\documentclass[pre,showpacs,longbibliography,twocolumn]{revtex4-1}
\usepackage{graphicx}
\usepackage{color}
\usepackage{amsmath}
\usepackage{subfigure}
\usepackage{epstopdf}
\usepackage{multirow,array}
\usepackage{makecell}
\usepackage{amssymb}
\usepackage{bbm}
\usepackage[mathscr]{eucal}
\newcommand{\RNum}[1]{\uppercase\expandafter{\romannumeral #1\relax}}
\usepackage[colorlinks,linkcolor=blue,citecolor=blue,urlcolor=blue,hyperindex,breaklinks]{hyperref}
\usepackage{appendix}

\begin{document}  

\title{Singular transport in non-equilibrium strongly internal-coupled 1D tilted field spin-1/2 chain}
\author{Yi-jia Yang, Yu-qiang Liu, Zheng Liu, and Chang-shui Yu}
\email{Electronic address: ycs@dlut.edu.cn}
\affiliation{School of Physics, Dalian University of Technology, Dalian 116024, China}
\date{\today}

\begin{abstract}
Non-equilibrium spin-chain systems have been attracting increasing interest in energy transport. This work studies a one-dimensional non-equilibrium Ising chain immersed in a tilted magnetic field, every spin contacts a Boson reservoir with the dissipative system-environment interaction. We analytically investigate the dynamics and the steady-state energy transport taking advantage of the Born-Markov-secular master equation. In the longitudinal field, one can find that the non-dissipative $N^\prime$ spins decompose the spin chain into $N^\prime+1$ independent subchains and block the heat currents from the hot end to the cool end. Moreover, for the non-dissipative $\mu$th spin, its nearest two bulk spins become the nodal spins in the subchains and have the corresponding energy correction of $\pm J_{\mu-1,\mu}$ and $\pm J_{\mu,\mu+1}$ depending on the excited/ground state of the $\mu$th spin. Therefore, a magnetically controlled heat modulator can be designed by adjusting the direction of the magnetic field in which the non-dissipative spin is located. For the transverse field case, the whole Hilbert space of the chain can always be divided into two independent subspaces regardless of whether the bulk spin is dissipative.
This work provides new insight into the dynamics and energy transport of the dissipative Ising model.
\end{abstract}
\maketitle

A cutting-edge field in non-equilibrium open quantum system, also named as driven-dissipation system, is the study of the quantum singular properties, which are different from the equilibrium states \cite{PhysRevLett.106.217206,PhysRevLett.109.240402,PhysRevLett.114.040402,PhysRevLett.115.200502,PhysRevLett.117.207201,PhysRevLett.122.127202,PhysRevLett.122.050501,RevModPhys.93.015008,0Emergence,PhysRevB.107.115118,PhysRevLett.132.120401}. 
The dissipative quantum system has been simulated in many platforms \cite{RevModPhys.90.041002,RevModPhys.94.045006,doi:10.1126/science.abb2823,Arrachea_2023}, such as trapped ions \cite{Barreiro:2011fge,PhysRevLett.128.080502}, superconducting circuit \cite{doi:10.1126/science.1231930,Wendin_2017,PhysRevLett.123.080501,Leger2019,maillet2020electric,RevModPhys.93.025005,PhysRevResearch.6.033107}, etc.
The dissipative Ising model is widely mentioned \cite{PhysRevA.84.031402,PhysRevA.88.063811,PhysRevB.92.174418,PhysRevE.94.052132,PhysRevA.95.042133,PhysRevLett.123.090401,PhysRevB.99.224432,PhysRevB.102.094204,PhysRevB.104.214301,PhysRevA.104.023713,PhysRevB.103.224210,Paz_2021,10.21468/SciPostPhys.15.3.096,PhysRevE.94.052132,PhysRevA.109.022420}, since it has been experimentally implemented in Rydberg atomic system \cite{PhysRevLett.111.113901,PhysRevLett.113.023006,PhysRevX.7.021020,PhysRevLett.119.190402,2021Quantum,PhysRevResearch.6.L032069,0Quantum}.  
In recent years, the tilted field Ising model, namely, the transverse field (TF) and longitudinal field (LF) exist simultaneously \cite{PhysRevB.57.8375,DONASCIMENTO2017224,2016Real,PhysRevE.99.012122,PhysRevA.100.022124,PhysRevB.99.180302,Białończyk_2020,2017A,PhysRevB.103.235117,PhysRevLett.126.200501,ncs41467-022-35301-6,PhysRevB.105.125413,PhysRevB.106.214311,PhysRevLett.131.190403,PhysRevB.109.104312,PhysRevB.110.054432,10.1007/s11128-024-04567-8}, has been attracted more attention because it has been implemented in a tilted field optical lattice \cite{nature09994}.
The tilted field Ising model has been combined with
bistability \cite{PhysRevA.84.031402,PhysRevX.7.021020}, 
multicritical behavior \cite{PhysRevA.95.042133,DONASCIMENTO2017224},   
quantum phase transition \cite{PhysRevLett.114.040402,PhysRevB.103.235117,PhysRevB.110.054432,PhysRevA.100.022124,PhysRevLett.131.190403},
energy transport \cite{PhysRevB.99.180302},
prethermalization \cite{ncs41467-022-35301-6},
quantum sensing \cite{PhysRevLett.126.200501},  
quantum integrability \cite{PhysRevB.106.214311,2016Real}, and so on.

In general, the interaction between the system and the environment is carried out through dissipation and decoherence in an open quantum system \cite{PhysRevLett.129.056402,PhysRevLett.132.266701,PhysRevB.110.064304,fazio2024manybodyopenquantumsystems,minganti2024openquantumsystems}, the complexity of the system's dynamics greatly increases compared to the closed system. 
There are many methods used to effectively simulate the dissipative Ising model, such as 
Monte Carlo simulation \cite{PhysRevB.81.104302,PhysRevB.103.245135,PhysRevB.109.024431}, 
density matrix renormalization group \cite{Humeniuk_2020,PhysRevB.107.104414,PhysRevResearch.5.043270}, 
tensor network method \cite{ncs41467-022-35301-6,2019Tensor,PhysRevA.109.022420,PhysRevLett.123.090401}, 
effective non-Hermitian Hamiltonian \cite{PhysRevB.99.224432,PhysRevB.109.104312}, 
full counting statistics \cite{PhysRevLett.113.023006,10.21468/SciPostPhys.15.3.096}, 
mean field approximation \cite{PhysRevX.6.031011,PhysRevE.94.052132,PhysRevB.92.174418,PhysRevB.104.214301,PhysRevB.106.115122}, 
quantum master equation \cite{PhysRevLett.109.240402,PhysRevB.98.241108,PhysRevB.102.115109}.
The Born-Markov-secular (BMS) master equation based on the density matrix is a competitive candidate to study energy transport because of its strict solvability \cite{breuer2002theory,10.1063/1.5115323}.
For the case of weak coupling inside the chain, the dissipative properties are described by phenomenally introducing the jump operator $\sigma^-=\frac{1}{2}(\sigma^x-\mathrm{i}\sigma^y)$, where $\sigma^x$ and $\sigma^y$ are Pauli matrices \cite{Levy_2014,Hofer_2017,PhysRevLett.120.200603,Cattaneo_2019,hewgill2021quantum,PhysRevE.105.024120}.  The longitudinal field Ising chain can be designed as a perfect diode because of the directional transfer of the heat current \cite{PhysRevE.99.032116}. 
As the coupling between the nearest neighbor sites increases, the dynamics of the spin chain, which must be considered as a whole, is quite complex since the system's dimension increases in power with the number of spins.
Therefore, a lot of works focus on the few-body system, such as the quantum thermal transistor based on three spins \cite{PhysRevLett.116.200601,PhysRevB.101.245402,PhysRevA.103.052613}. 
It is an urgent problem to study analytically the multi-body strongly internal-coupled system by using the BMS master equation.

\begin{figure}
	\centering
		\includegraphics[width=8.3cm]{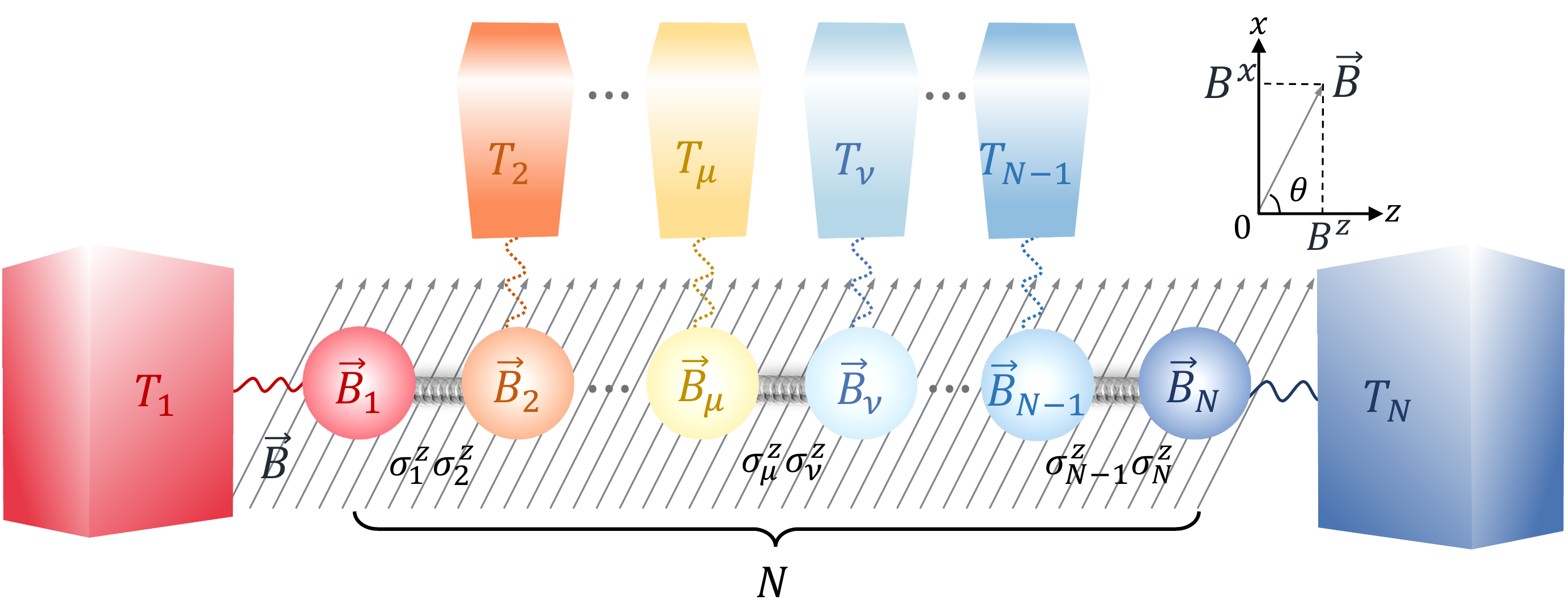}
	\caption{1D Ising model immersed in the direction-adjustable tilted magnetic field with fixed amplitude $\vec{B}=B^x\vec{e}_x+B^z\vec{e}_z$, $\vert\vec{B}\vert=\sqrt{{B^x}^2+{B^z}^2}$, where $B^x=\vert\vec{B}\vert\sin\theta$ and $B^z=\vert\vec{B}\vert\cos\theta$. Each spin is independently connected to a thermal reservoir, where the temperature of the $\mu$th reservoir is $T_\mu$.}
\label{model}
\end{figure}

In this work, we study the non-equilibrium Ising model immersed in a tilted field and each spin is connected to a local reservoir with the dissipative system-reservoir interaction, as shown in Fig. \ref{model}. We employ the BMS master equation to analytically describe the system's dynamics and deal with the steady state and steady-state heat currents. It is found that, if there exist $N^\prime$ non-dissipative spins, LF not only decomposes the chain into $N^\prime+1$ independent subchains but also has a certain energy correction for the two spins which are nearest neighbor to the non-dissipative spin. Besides, the heat currents cannot flow across the subchains but are present in the interior of each subchain. TF divides total Hilbert space into two independent subspaces despite whether bulk spins are in contact with the corresponding reservoirs. Based on these singular heat currents, we propose a theoretical scheme of magnetic-controlled heat current modulator. This modulator allows one to adjust the heat current from zero to a finite value by changing the magnetic field of the non-dissipative spin from LF to TF.

The structure of this paper is as follows. The non-equilibrium 1D tilted field Ising chain is modeled in Sec. \ref{section2}. Sec. \ref{section3} gives the dynamics of this open spin chain.
The steady-state heat current of the system is described in Sec. \ref{section4}. Based on the characteristic of steady-state heat current, we simulate the process of quantum magnetically controlled heat modulator in Sec. \ref{section5}. 
Section \ref{section7} summarizes this work.

\section{\label{section2}Dissipative tilted field Ising model}

The 1D Ising chain consists of $N$ spin-1/2 systems coupling with the nearest-neighbor interaction$\sigma^z_\mu\sigma^z_{\mu+1}$, $\mu=1,\cdots,N-1$, where each spin is connected to an independent heat reservoir. The first and last spins are nodes, and the rest are bulk spins. When this spin chain is immersed in a tilted magnetic field in $x$-0-$z$ plane, the Hamiltonian of the open quantum system is $H=\tilde{H}_S+H_E+H_{SE}$, which is modeled in Fig. \ref{model}. The system's  Hamiltonian is
\begin{align}
\tilde{H}_S=&\frac{1}{2}(\sum_{\mu=1}^N B^x_\mu\sigma_\mu^x+B^z_\mu\sigma_\mu^z+\sum_{\mu=1}^{N-1}J_{\mu,\mu+1}\sigma_\mu^z\sigma_{\mu+1}^z),\label{Hamiltonian}
\end{align}
where $B_\mu^{z(x)}=\vert\vec{B}_\mu\vert\cos\theta(\sin\theta)$ represents the Zeeman energy of the $\mu$th spin in $z(x)$ direction reduced by the magnetic field $\vec{B}_\mu$ with $\theta$ denoting the angle between the magnetic field and horizontal direction, and $J_{\mu,\mu+1}$ denotes the coupling strength between the nearest-neighbor spins. 

Each site is connected to an independent heat reservoir, which means that the "flip" operation of each spin can only be performed by the reservoir in contact with it. Each reservoir is composed of infinite dimensional harmonic oscillators, the model introduced by Caldeira and Leggett \cite{PhysRevLett.46.211,RevModPhys.59.1}, and the Hamiltonian of environment is 
\begin{align}
H_E=\sum_\mu\sum_k\omega_{\mu k}a^\dagger_{\mu k}a_{\mu k},
\end{align} 
where $a_{\mu k}^\dagger$ (or $a_{\mu k})$ denotes creating (or annihilating) an oscillator of frequency $\omega_{\mu k}$ in the $\mu$th reservoir.

In general, the interaction between the spin and the corresponding reservoir is expressed as
\begin{align}
H_{\mathrm{spin-res}}=\vec{\sigma}\cdot\sum_k \vec{f}_k(a_k^\dagger +a_k),\label{Hspinres}
\end{align}
where $\vec{\sigma}=(\sigma^x,\sigma^y,\sigma^z)^\mathrm{T}$ (or $\vec{f}_k=(f^x_k,f^y_k,f^z_k)^\mathrm{T}$) is the vector formed by the Pauli matrix (or coupling strength corresponding to the $k$th mode of the reservoir) along the three directions in the Bloch sphere \cite{Arrachea_2023}.
For the spin-$1/2$ system with the Hamiltonian $\frac{\omega}{2}\sigma^z$, the $\sigma^x$-type and $\sigma^z$-type system-reservoir interactions correspond to dissipative and dephasing channels. Therefore, these two types of interactions are defined as dissipative coupling and dephasing coupling, respectively \cite{Cattaneo_2019}.
Since the dissipative interaction occurs naturally in the superconducting circuit system \cite{doi:10.1126/science.1231930,Wendin_2017,RevModPhys.93.025005} and the focus of this paper is the energy transfer, the dissipative system-reservoir interaction, also known as the dipole interaction, is expressed as
\begin{align}
H_{SE}=\sum_\mu\sigma_\mu^x\sum_k f_{\mu k}(a_{\mu k}^\dagger+a_{\mu k}),\label{HSE}
\end{align} 
where $f_{\mu k}\equiv f_{\mu k}^x$ denotes the $x$-direction coupling strength between the $\mu$th spin and oscillator with frequency $\omega_{\mu k}$ in the corresponding reservoir.

\section{Dynamics of the dissipative tilted field Ising model}
\label{section3}
As an efficient method for analytically calculating the dynamics of open quantum systems, the application of the BMS master equation usually satisfies the following three conditions \cite{breuer2002theory,10.1063/1.5115323,RevModPhys.94.045006}.
First, the coupling strength between the system and the environment is so weak that the system does not affect the environment, which is the Born approximation. 
Second, the reservoir correlation time $\tau_E$ is far less than the relaxation time of the system $\tau_R$, so that the state of the system depends only on the present moment, corresponding to no memory effect, i.e. Markov approximation.  
Third, secular approximation, also known as the rotating wave approximation, requires that the difference between the transition frequencies is much larger than the inverse relaxation time, i.e. $\vert \omega^\prime-\omega\vert\gg\tau_R^{-1}=O(f^2)$ with $f$ denoting the coupling strength between the system and environment, which ensures that the fast oscillation terms are ignored \cite{Cattaneo_2019}. The secular approximation is valid for any well-spaced system, internal-strongly coupling or different magnetic fields for each cite is the fundamental settings.
The weak and strong coupling between subsystems correspond to local and global master equations \cite{doi:10.1142/S1230161217400108,Cattaneo_2019}. When the coupling is weak and the energy gap of each spin is almost unaffected, the local master equation is efficient, that is, the reservoir can induce only one spin connected to it to undergo the transition $\sigma^-=\frac{1}{2}(\sigma^x-\mathrm{i}\sigma^y)$. With the coupling increases, the reservoir will affect the whole system and the global master equation dominates the dynamics.

This work focuses on the dynamics of strong coupling between subsystems, and diagonalization representation is convenient.
The system's Hamiltonian Eq. (\ref{Hamiltonian}) can be re-expressed by eigen-decomposition as 
\begin{align}
H_S=\Lambda \tilde{H}_S\Lambda^\dagger=\sum_{i=1}^{2^N}\lambda_i\vert i\rangle\langle i\vert,
\end{align} 
where $\lambda_i$ and $\vert i\rangle=\sum_{j=1}^{2^N}\Lambda(i,j)\vert\tilde{j}\rangle$ are eigenvalue and the corresponding eigenstate with bare bases $\{\vert\tilde{i}\rangle,i\in[1,2^N]\}$ and $\Lambda(i,j)$ representing matrix element in the $i$th row and the $j$th column of transformation matrix $\Lambda$.
Without loss of generality, the bare bases are defined as $\vert\tilde{1}\rangle=\otimes_{\mu=1}^N\vert g\rangle_\mu,\cdots,\vert\widetilde{2^N}\rangle=\otimes_{\mu=1}^N\vert e\rangle_\mu$, where $\vert e\rangle_\mu=\left(\begin{smallmatrix}1\\0\end{smallmatrix}\right)$ and $\vert g\rangle_\mu=\left(\begin{smallmatrix}0\\1\end{smallmatrix}\right)$ are the excited and ground state of the $\mu$th spin. Thus, one can write the master equation for the reduced density matrix $\rho(t)=\mathrm{Tr}_E[\rho_{SE}(t)]$, in the schr$\ddot{\mathrm{o}}$dinger picture, as
\begin{align}
\dot{\rho}(t)=-\mathrm{i}[H_S,\rho(t)]+\sum_\mu\mathcal{L}_\mu[\rho(t)].
\end{align}
The dissipator $\mathcal{L}_\mu[\rho(t)]$ related to the $\mu$th reservoir reads
\begin{align}
\nonumber
\mathcal{L}_\mu[\rho(t)]=\sum_{\omega_{\mu}^{ij}}&J_\mu(-\omega_{\mu}^{ij})[{2V_\mu^{ij}\rho(t)V_\mu^{ij}}^\dagger-\{{V_\mu^{ij}}^\dagger V_\mu^{ij},\rho(t)\}]\\
+&J_\mu(+\omega_{\mu}^{ij})[{2V_\mu^{ij}}^\dagger\rho(t)V_\mu^{ij}-\{V_\mu^{ij}{V_\mu^{ij}}^\dagger,\rho(t)\}],\label{BMS_ME}
\end{align}
where $J_\mu(\pm\omega_\mu^{ij})=\pm\kappa(\omega_\mu^{ij})\bar{n}_\mu(\pm\omega_\mu^{ij})$ stands for spectral density, $\bar{n}_\mu(\omega_\mu^{ij})=(e^{\omega_\mu^{ij}/T_\mu}-1)^{-1}$ is average photon number of the frequency $\omega_\mu^{ij}$ in the $\mu$th reservoir and 
\begin{align}
\kappa(\omega_\mu^{ij})=\sum_k\pi\vert f_{\mu k}\vert ^2\delta(\omega_\mu^{ij}-\omega_{\mu k})\label{kappa_omega}
\end{align}
describes the dissipative strength between the system and the reservoir.
In general, this hybridization function is represented by a power-law function $\kappa(\omega_\mu^{ij})\propto(\omega_\mu^{ij})^s$, where $s<1$, $s=1$, and $s>1$ correspond to sub-ohmic, ohmic, and super-ohmic environment. Its coefficient is usually expressed as the dissipation rate \cite{Arrachea_2023}. Based on Eq. (\ref{BMS_ME}), it is the environment-induced transitions $V_\mu^{ij}(\omega_\mu^{ij})$ determines the dynamics of the system, not the specific form of the spectral density. Therefore, different dissipative spectrum only yields quantitative change rather than qualitative change in physical processes \cite{Upadhyay_2024}. For convenience, the flat spectrum is considered in this work, namely, the dissipative rate corresponding to any transition frequency is the same and can be expressed as $\kappa(\omega_\mu^{ij})\equiv\kappa_\mu$.
In Eq. (\ref{BMS_ME}), the eigen-operator $V_\mu^{ij}$ is defined as 
\begin{align}
\sum_{\omega_\mu^{ij}} V_\mu^{ij}+{V_\mu^{ij}}^\dagger=\sum_{i,j}\vert i\rangle\langle i\vert\sigma_\mu^x\vert j\rangle\langle j\vert,\label{V_define}
\end{align} 
and can be expressed as
\begin{align}
V_\mu^{ij}=\Lambda_\mu^{ij}\vert i\rangle\langle j\vert,\quad\mathrm{for}\quad i\in[1,2^N-1],j\in[i+1,2^N],\label{V_general}
\end{align}
which denotes the transition $\vert i\rangle\leftrightarrow\vert j\rangle$ with the frequency $\omega_\mu^{ij}\equiv\lambda_j-\lambda_i$ induced by the $\mu$th reservoir, and eigen-frequency $\omega_\mu^{ij}$ satisfy $[H_S,V_\mu^{ij}]=-\omega_\mu^{ij}V_\mu^{ij}$.
The coefficient $\Lambda_\mu^{ij}$ in Eq. (\ref{V_general}) is given by Eq. (\ref{Lambdax}). 
Hence, the number of the eigen-operator for each spin is $N_{\mathrm{tot}}\equiv\sum_{x=1}^{2^{N}-1}x=2^{2N-1}-2^{N-1}$, 
that is, each reservoir can induce the transition between any two levels. According to the similar definition of the eigen-operator in Eq. (\ref{V_define}) and simple algebra, the addition of $\sigma^z_\mu$-type interaction in Eq. (\ref{HSE}) only changes the coefficient $\Lambda_\mu^{ij}$ in the eigen-operators, but does not change the dissipative channels between the system and the environment. As a consequence, it is reasonable to consider only the dissipative interaction in Eq. (\ref{HSE}).

According to  Eq. (\ref{BMS_ME}), one can obtain the system's steady state and find that the dynamics of population and coherence, denoted by the diagonal and off-diagonal entries, are decoupled from each other. It is our duty to remind the reader that the well-spaced system is a necessary condition for the decoupling of population and coherence \cite{PhysRevE.76.031115}.
For example, if a reservoir can induce the transitions between one level to two degenerate levels simultaneously, the coherences between these two degenerate levels are entangled to the populations of the involved levels \cite{PhysRevE.107.064125}.

For the case where all eigen-frequencies are different, there is $\rho_{ij}^S=0, i\neq j,$ because its dynamics satisfies 
\begin{align}
\dot{\rho}_{ij}\equiv\langle i\vert\dot{\rho}(t)\vert j\rangle=-C_{ij}\rho_{ij}\label{dynamic_offdiagonal}
\end{align}
with $C_{ij}$ representing the positive constant related to the spectral density of the transitions involving the levels $\vert i\rangle$ and $\vert j\rangle$. 
When there are two transitions $\vert i\rangle\leftrightarrow\vert j\rangle$ and $\vert k\rangle\leftrightarrow\vert l\rangle$ induced by each reservoir have the same transition $\omega$ (then the eigen-operator is described by $V_\mu(\omega)=\Lambda_\mu^{ij}\vert i\rangle\langle j\vert+\Lambda_\mu^{kl}\vert k\rangle\langle l\vert, \mu=1,\cdots,N$), off-diagonal elements $\rho_{ij}$, $\rho_{il}$, $\rho_{jl}$, $\rho_{kl}$ and their conjugate terms obey the independent evolution as Eq. (\ref{dynamic_offdiagonal}), and $\rho_{ik}$ and $\rho_{jl}$ satisfy
\begin{align}
\nonumber
&\begin{pmatrix}
\dot{\rho}_{ik}\\\dot{\rho}_{jl}
\end{pmatrix}
=\mathcal{A}\begin{pmatrix}
\rho_{ik}\\\rho_{jl}
\end{pmatrix},\\
&\mathcal{A}=-\sum_\mu
\begin{pmatrix}
({\Lambda_\mu^{ij}}^2+{\Lambda_\mu^{kl}}^2)J_\mu(+\omega) & -2\Lambda_\mu^{ij}\Lambda_\mu^{kl}J_\mu(-\omega) \\
 -2\Lambda_\mu^{ij}\Lambda_\mu^{kl}J_\mu(+\omega) & ({\Lambda_\mu^{ij}}^2+{\Lambda_\mu^{kl}}^2)J_\mu(-\omega)
\end{pmatrix}.
\end{align}
By simple calculation, the condition of $\rho_{ik}^S=0$ and $\rho_{jl}^S=0$ is $\det[\mathcal{A}]\neq 0$, that is, $\Lambda_\mu^{ij}\neq\Lambda_\mu^{kl}, \mu=1,\cdots,N$. Otherwise, $\rho_{ik}^S$ and $\rho_{jl}^S$ have infinite solutions. Fortunately, this condition is always satisfied taking the 3-spin eigen-operators in Appendix \ref{AppendixE} as an example. In brief, the off-diagonal entries vanish in the steady state.

Besides, the dynamics of populations is
\begin{align}
\dot{\rho}_{ii}=-\sum_{\mu=1}^N\sum_{j=1,j\neq i}^{2^N}\Gamma^\mu_{ij},\quad i=1,\cdots,2^N,\label{pop}
\end{align}
where $\Gamma^\mu_{ij}=2(\Lambda_\mu^{ij})^2[J_\mu(+\omega_\mu^{ij})\rho_{ii}-J_\mu(-\omega_\mu^{ij})\rho_{jj}]$ denotes the net transition rate between energy levels $\vert i\rangle$ and $\vert j\rangle$.
The steady state means $\dot{\rho}^S(t)=0$, and Eq. (\ref{pop}) can be rewritten in the matrix form as
\begin{align}
\vert\dot{\rho}^S(t)\rangle=\sum_\mu\mathcal{M}_\mu\vert\rho^S\rangle=0,\label{Eq_matrixform}
\end{align}
where $\vert\rho^S\rangle=[\rho_{11}^S,\rho_{22}^S,\cdots,\rho_{2^N,2^N}^S]^{\mathrm{T}}$ is the column vector consisting of the populations. 
In Eq. (\ref{Eq_matrixform}), $\mathcal{M}_\mu$ is a $2^N\times 2^N$-dimensional coefficient matrix consisted to spectral density of the $\mu$th reservoir, the element in the $i$th row and $j$th column is
\begin{align}
\mathcal{M}_\mu(i,j)=\left\{
\begin{array}{rcl}
2(\Lambda_\mu^{ij})^2J_\mu(-\omega_\mu^{ij}),&\quad &i<j,\\
2(\Lambda_\mu^{ji})^2J_\mu(+\omega_\mu^{ji}),&\quad &i>j,\\
-\sum_{j\neq i}^{2^N}\mathcal{M}_\mu(i,j),&\quad &i=j.
\end{array}\right.\label{M_matrixelements}
\end{align}
$i<j$, $i>j$, and $i=j$ corresponding to the elements in the lower triangle, upper triangle, and diagonal regions of matrix $\mathcal{M}_\mu$.
One can check that the rank of matrix $\mathcal{M}=\sum_\mu\mathcal{M}_\mu$ is $2^N-1$, therefore, the steady state is unique with the normalization Condition $\mathrm{Tr}\rho=1$.
\begin{figure}
	\centering
		\includegraphics[width=8.3cm]{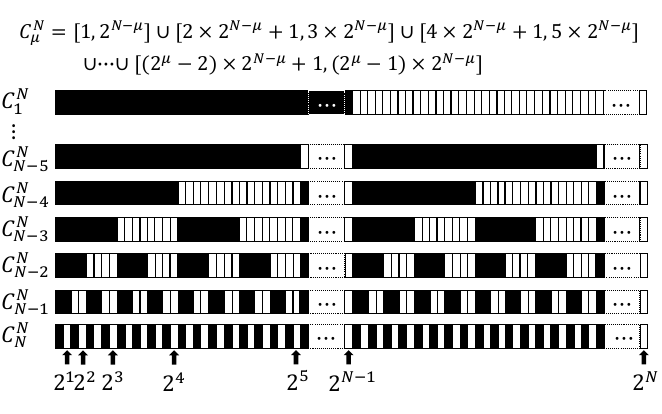}
	\caption{Sketch of the expression for coefficient $C_\mu^N$. In each row, the squares from left to right represent the numbers from $1$ to $2^N$, which is the dimension of the Hilbert space of the chain consisting of $N$ spins. The collection of numbers represented by the black squares in each row is $C_\mu^N$.}
\label{CmuN}
\end{figure}

\subsection{Longitudinal field case}
\label{section3a}

When only LF exists ($\theta=0$), the system's Hamiltonian is already diagonalized, i.e., $\tilde{H}_S^\parallel=H_S^\parallel$, and transformation matrix is an identity matrix with dimension $2^N$, i.e. $\Lambda^\parallel=\mathbbm{1}_{2^N}$.
Thus the number of the allowed transitions for each spin is $2^{N-1}$ and the coefficient is given in Eq. (\ref{LambdaxL}).
All transitions corresponding to the same frequency are merged into one eigen-operator.  A simple calculation indicates that the eigen-operator for each spin is $\sigma_\mu^-$, which is consistent with its counterpart in the local master equation. The difference is that the dissipation channels are also related to the states of the nearest-neighbor spins, that is, 
\begin{align}
\sigma_\mu^-\mapsto\otimes_{\nu=1}^{\mu-2}\mathbbm{1}_\nu\otimes\rho^{e(g)}_{\mu-1}\otimes\sigma_\mu^-\otimes\rho^{e(g)}_{\mu+1}\otimes_{\nu=\mu+2}^N\mathbbm{1}_\nu
\end{align} 
with $\rho^{e(g)}_{\mu}=\vert e(g)\rangle_\mu\langle e(g)\vert$. Thus, the nodal (or bulk) spin has 2 (or 4) eigen-operators, the specific forms are shown in Appendix \ref{AppendixA}.

The dynamics of diagonal entries of the density matrix is reduced as 
\begin{align}
\dot{\rho}_{ii}^\parallel=-\sum_{\mu=1}^N\Gamma^\mu_{i,i\bullet 2^{N-\mu}},i=1,\cdots, N,
\end{align} 
where $\bullet=+$ for $i\in C_\mu^N$ and $\bullet=-$ for others with $C_\mu^N$ defining in Fig. \ref{CmuN}. 
We study the dynamics of $1$-, $2$-, and $3$-spin chains in detail in Appendix \ref{AppendixA}.

When the $\mu$th spin is not in contact with the corresponding reservoir, the system can be divided into two independent subspaces due to the transition $\sigma_\mu^-$ is forbidden, where the first subspace is $\mathbb{S}_1^\parallel=\{\vert i\rangle,i\in C_\mu^N\}$ and the other energy levels constitute subspace $\mathbb{S}_2^\parallel$. The state of the $\mu$th spin in $\mathbb{S}_1^\parallel$ (or $\mathbb{S}_2^\parallel$) is $\rho_\mu^g$ (or $\rho_\mu^e$). 
In $\mathbb{S}_1^\parallel$, for the $(\mu-1)$th or $(\mu+1)$th bulk spin, which is the nearest neighbor to the $\mu$th spin, the number of eigen-operator is reduced from four to two, expressed as $\otimes_{\nu=1}^{\mu-3}\mathbbm{1}_\nu\otimes\rho^{e(g)}_{\mu-2}\otimes\sigma_{\mu-1}^-\otimes\rho^{g}_{\mu}\otimes_{\nu=\mu+1}^N\mathbbm{1}_\nu$ or $\otimes_{\nu=1}^{\mu-1}\mathbbm{1}_\nu\otimes\rho^{g}_{\mu}\otimes\sigma_{\mu+1}^-\otimes\rho^{e(g)}_{\mu+2}\otimes_{\nu=\mu+3}^N\mathbbm{1}_\nu$. The corresponding eigen-frequencies are $(B_{\mu-1}-J_{\mu-1,\mu})\pm J_{\mu-2,\mu-1}$ and $(B_{\mu+1}-J_{\mu,\mu+1})\pm J_{\mu+1,\mu+2}$.
Similarly, in subspace $\mathbb{S}_2^\parallel$, the eigen-frequencies for the $(\mu-1)$th and $(\mu+1)$th spins are $(B_{\mu-1}+J_{\mu-1,\mu})\pm J_{\mu-2,\mu-1}$ and $(B_{\mu+1}+J_{\mu+1,\mu+2})\pm J_{\mu,\mu+1}$, since the $\mu$th spin is in the excited state. 
Likewise, when there are $N^\prime$ spins that are not connected to reservoirs, the system is divided into $2^{N^\prime}$ independent subspaces $\mathbb{S}$. For example, subspace $\mathbb{S}=\mathbb{S}_{1^\prime_e2^\prime_e\cdots N^\prime_e}$ denotes that the states of all non-dissipative spins are spin-up. 
No matter which subspace, the non-dissipative $\mu$th bulk spin in the excited/ground state produces an energy correction for its nearest-neighbor spins, specifically, the new nodal spin $\mu-1$ (or $\mu+1$) is imposed by an effective magnetic field $B_{\mu-1}\pm J_{\mu-1,\mu}$ (or $B_{\mu+1}\pm J_{\mu,\mu+1}$). 
\begin{figure}
	\centering
		\includegraphics[width=8.3cm]{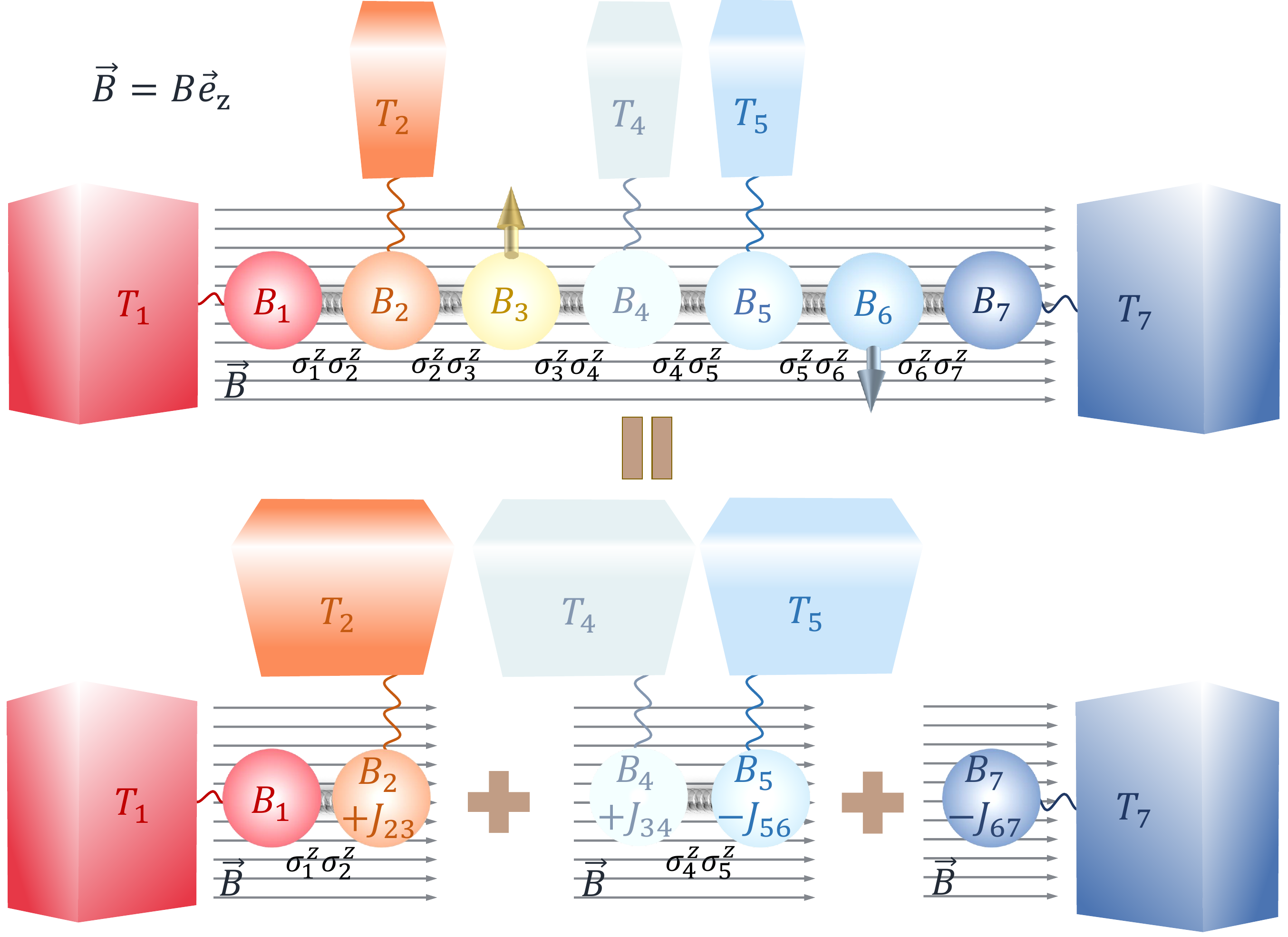}
	\caption{A schematic illustration shows that the longitudinal field not only induces the non-dissipative spin splitting spin chain but also has an energy correction to its nearest-neighbor spins that depends on the state of the non-dissipative spin. In this $7$-spin chain, the third and sixth spins are not connected to the corresponding reservoirs. Without loss of generality, provided that the $3$rd spin is in the up state and the $6$th spin is in the down state.}
\label{model_7qubit}
\end{figure}

For the dynamics, the coefficient matrix and steady state in Eq. (\ref{Eq_matrixform}) satisfy 
\begin{align}
\mathcal{M}^\parallel=\oplus_\mathbb{S}\mathcal{M}^{\parallel,\mathbb{S}},\quad
\rho^{\parallel}=\sum_{m=1}^{2^{N^\prime}}p_m^\parallel\rho^{\parallel,\mathbb{S}}\otimes\rho^{\mathbb{S}},
\end{align}
where $\rho^{\mathbb{S}}=\otimes_{\nu=1^\prime}^{N^\prime}\rho_{\nu}^{e(g)}$ with $\rho_{\nu}^{e(g)}$ denoting the state of the $\nu$th spin disconnecting to the reservoir and $p_m^\parallel$ is the fraction of the $m$th subspace $\mathbb{S}$ in the initial state. 
In any $\mathbb{S}$, the coefficient matrix is 
\begin{align}
\nonumber
\mathcal{M}^{\parallel,\mathbb{S}}=&\mathcal{M}^{\parallel,\mathbb{S}}_{1,\cdots,1^\prime-1}\otimes\mathcal{M}^{\parallel,\mathbb{S}}_{1^\prime+1,\cdots,2^\prime-1}\otimes\mathcal{M}^{\parallel,\mathbb{S}}_{2^\prime+1,\cdots,3^\prime-1}\\
&\otimes\cdots\otimes\mathcal{M}^{\parallel,\mathbb{S}}_{N^\prime+1,\cdots, N},
\end{align}
and the steady state of the system can be expressed as 
\begin{align}
\nonumber
\vert\rho^{\parallel,\mathbb{S}}\rangle=&\vert\rho^{\parallel,\mathbb{S}}_{1,\cdots,1^\prime-1}\rangle\otimes\vert\rho^{\parallel,\mathbb{S}}_{1^\prime+1,\cdots,2^\prime-1}\rangle\otimes\vert\rho^{\parallel,\mathbb{S}}_{2^\prime+1,\cdots,3^\prime-1}\rangle\\
&\otimes\cdots\otimes\vert\rho^{\parallel,\mathbb{S}}_{N^\prime+1,\cdots, N}\rangle.
\end{align}
Therefore, the spins of the unconnected heat reservoir can decompose the spin chain into $N^\prime+1$ independent subchains in each subspace $\mathbb{S}$.

Let's take a 7-spin chain as an example, as shown in Fig. \ref{model_7qubit}.  When the 3rd and 6th spins are not connected to the local reservoirs, the system can be divided into four subspaces $\mathbb{S}_{3_e6_e}$, $\mathbb{S}_{3_e6_g}$, $\mathbb{S}_{3_g6_e}$, and $\mathbb{S}_{3_g6_g}$. In subspace $\mathbb{S}_{3_{e(g)}6_{e(g)}}$, the coefficient matrix can be represented as $\mathcal{M}^{\parallel,\mathbb{S}}=\mathcal{M}^{\parallel,\mathbb{S}}_{12}\otimes\rho^{e(g)}_3\otimes\mathcal{M}^{\parallel,\mathbb{S}}_{45}\otimes\rho^{e(g)}_6\otimes\mathcal{M}^{\parallel,\mathbb{S}}_{7}$, so the spin chain can be decomposed into three independent parts. For the subspace $\mathbb{S}_{3_e6_g}$, the eigen-frequencies for the second and fourth spins are $(B_{2}+J_{23})\pm J_{12}$ and $(B_{4}+J_{34})\pm J_{45}$, and the eigen-frequencies for the fifth and seventh spins are $(B_{5}-J_{56})\pm J_{45}$ and $B_{7}-J_{67}$. 
The detailed calculation process is given in Appendix \ref{AppendixB}.

\subsection{Transverse field case}
\label{section3b}

Where only TF exists ($\theta=\pi/2$), the system's Hamiltonian can be diagonalized as $H_S^\perp=\Lambda^\perp\tilde{H}_S^\perp{(\Lambda^\perp)}^{-1}$, and transition matrix entries satisfy 
\begin{align}
\Lambda^\perp(i,j)=\left\{
\begin{array}{rcl}
+\Lambda^\perp(i,2^N+1-j),&\quad &i\in[1,2^{N-1}],\\
-\Lambda^\perp(i,2^N+1-j),&\quad &i\in[2^{N-1}+1,2^N].
\end{array}\right.
\end{align}
The system can be divided into two subsystems according to the symmetry of eigenstates, which are the rows of matrix $\Lambda^\perp$.
The transition operator shows that the transitions between the energy levels in subsystem $\mathbb{S}_1^\perp=\{\vert i\rangle,i\in[1,2^{N-1}]\}$ is entirely independent of those in subsystem $\mathbb{S}_2^\perp=\{\vert i\rangle,i\in[2^{N-1}+1,2^N]\}$. In other words, there is no cross-transition between the two subsystems. 
Therefore, each subspace has $N_{\mathrm{TF}}\equiv\sum_{x=1}^{2^{N-1}-1}x=2^{2N-3}-2^{N-2}$ eigen-operators, and the coefficients of the transition operators in two subspaces are given in Eq. (\ref{Vx_TF}). We should emphasize that these two independent subspaces generated in TF don't depend on whether spins are isolated from their local reservoirs, which differs from LF case.

\begin{figure}
	\centering
		\includegraphics[width=8.3cm]{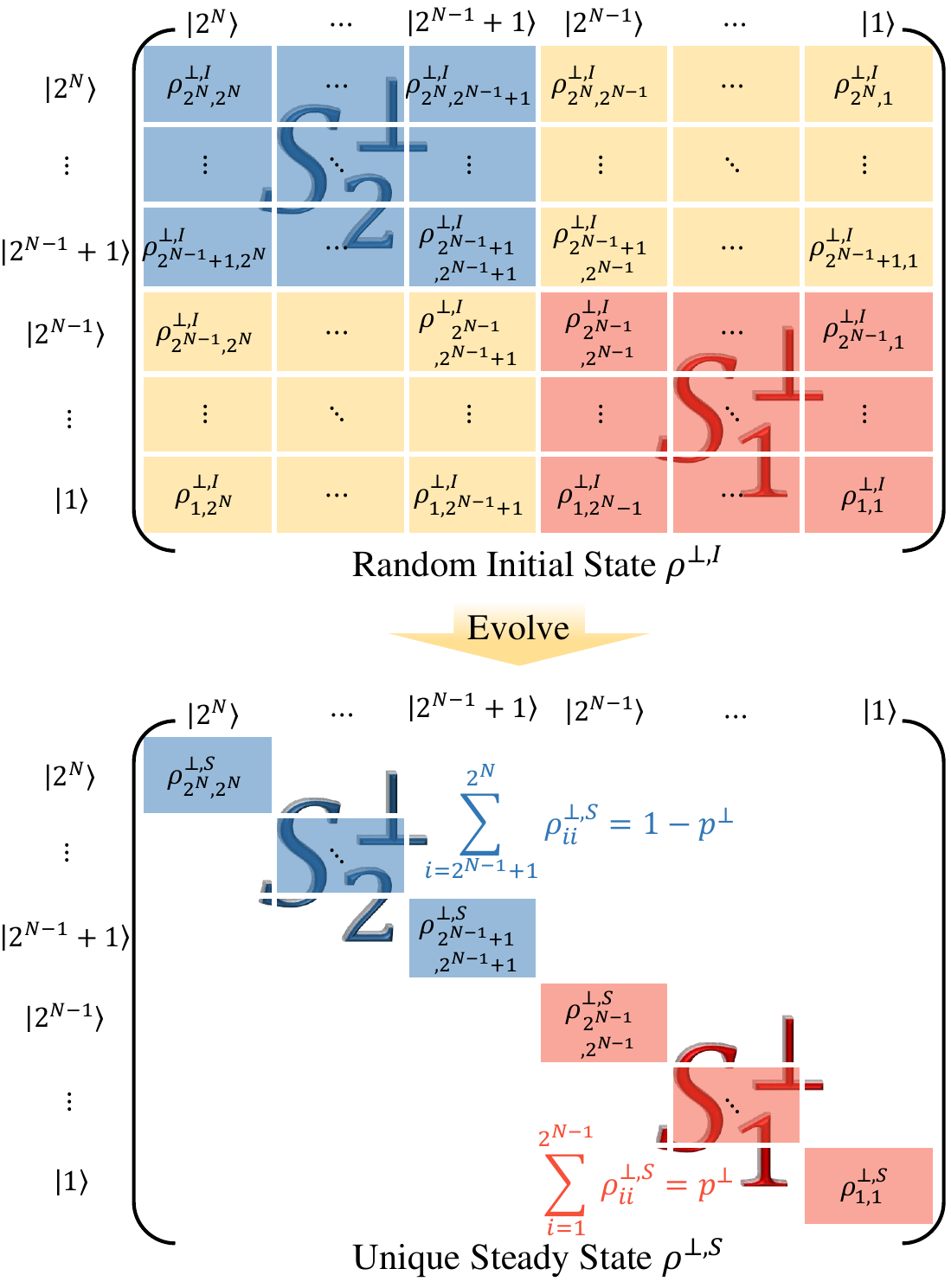}
	\caption{The dynamics of every element of the density matrix when the spin chain is immersed in the transverse field. The red and blue shades correspond to the elements of the subspaces composed of two sets of eigenstates with different symmetries. The yellow shade corresponds to the cross area of the two subspaces. The upper plane is a random initial state where all elements exist, and the lower plane represents a unique steady state, where all off-diagonal elements vanish and are represented in the blank area. The symbol "$\perp$" has been omitted in the eigenlevels for simplicity. }
\label{infinal}
\end{figure}
Based on the above analysis, we can get the dynamics in two subspaces, respectively. The dynamics in two subspaces are
\begin{align}
\nonumber
\dot{\rho}_{ii}^{\perp_1}&=-\sum_\mu\sum_{j=1,j\neq i}^{2^{N-1}}\Gamma_{ij}^{\perp,\mu}, &\quad &i=1,\cdots,2^{N-1},\\
\dot{\rho}_{ii}^{\perp_2}&=-\sum_\mu\sum_{j=2^{N-1}+1,j\neq i}^{2^N}\Gamma_{ij}^{\perp,\mu}, &\quad& i=2^{N-1}+1,\cdots,2^N.
\end{align}
The coefficient matrices in these two subspaces have the same form as Eq. (\ref{M_matrixelements}). The fraction of each subspace is determined by the system's initial state, which doesn't change with time. Let $p^\perp$ be the fraction of subspace $\mathbb{S}_1^\perp$, then the steady state is
\begin{align}
\vert\rho^{\perp,S}\rangle=p^\perp\vert\rho_1^{\perp,S}\rangle+(1-p^\perp)\vert\rho_2^{\perp,S}\rangle,
\end{align}
where $\vert\rho_1^{\perp,S}\rangle=[\rho_{11}^{\perp,S},\cdots,\rho_{2^{N-1},2^{N-1}}^{\perp,S}]^{\mathrm{T}}$, and $\vert\rho_2^{\perp,S}\rangle=[\rho_{2^{N-1}+1,2^{N-1}+1}^{\perp,S},\cdots,\rho_{2^N,2^N}^{\perp,S}]^{\mathrm{T}}$.

To more clearly describe the dynamic independence of these two subspaces, we give an illustration in Fig. \ref{infinal}. In the $N$-spin system, the density matrix elements of a random initial state $\rho^{\perp, I}$ are represented in the up plane in Fig. \ref{infinal}. 
In the initial state, the fraction of subspace $\mathbb{S}^\perp_1$ is $p^\perp$. In the process of evolution, the transitions in subspace $\mathbb{S}^\perp_1$ are independent of transitions in subspace $\mathbb{S}^\perp_2$, i.e., the dynamics of each subspace evolve independently. When the system arrives at the steady state, all off-diagonal elements decay to zero, and the sum of the populations in subspace $\mathbb{S}^\perp_1$ is still $p^\perp$.

It is worth mentioning that when the dephasing interaction is added in Eq. (\ref{HSE}), the remaining $N_{TF}^z=N_{\mathrm{tot}}-2N_{\mathrm{TF}}$ transitions that are prohibited in the transverse field pure dissipative model are re-induced. Thus, for more complex interactions, the characteristic that Hilbert space can be divided into two independent subspaces is no longer maintained.


\section{Steady-state heat current}
\label{section4}
According to the first law of thermodynamics, the increase of internal energy of a system is equal to the work done by the surroundings and the heat absorbed from the environment. Since the dissipative Ising model in this work has no external work done to it, we are only concerned with the thermal transport of the system. Heat current is defined by the change rate of the system's energy as
\begin{align}
\dot{Q}_\mu^N=\mathrm{Tr}\{H_S\mathcal{L}_\mu[\rho(t)]\}=-\sum_{i=1}^{2^N-1}\sum_{j=i+1}^{2^N}\omega_\mu^{ij}\Gamma_\mu^{ij}.\label{Q_define}
\end{align}
$\dot{Q}_\mu^N>0$ means that the heat flows from the $\mu$th reservoir to the chain, and vice versus. $\sum_{\mu=1}^N\dot{Q}_\mu^N=0$ is an inevitable consequence of the law of conservation of energy. 
The second equality in Eq. (\ref{Q_define}) is derived in Appendix \ref{AppendixC}. 
\begin{figure}
	\centering
		\includegraphics[width=8.3cm]{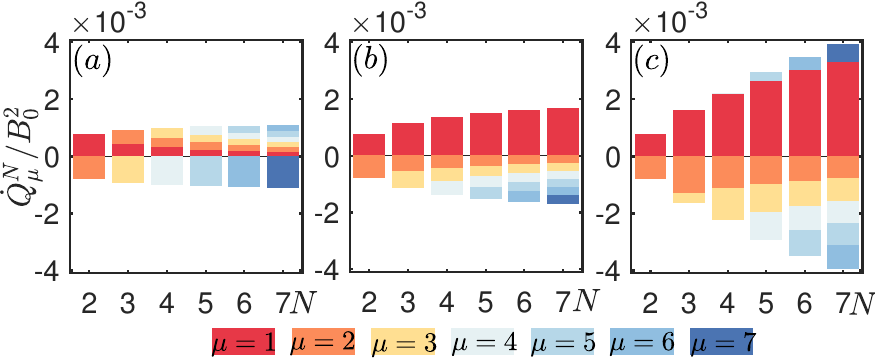}
	\caption{The heat current from the $\mu$th reservoir into the $N$-spin chain $\dot{Q}_\mu^N$. The temperatures of reservoirs connected to bulk spins are $T_b=10B_0,5B_0,2B_0$ in (a), (b), (c), respectively. $B_0=1$, $B_\mu=5B_0$, $J_{\mu,\mu+1}=0.1B_0$, $\kappa_\mu=0.001B_0$, $T_1=10B_0$, $T_N=5B_0$, and $\theta=\pi/4$.}
\label{heat_7}
\end{figure}

In Fig. \ref{heat_7}, we analyze the heat current from each reservoir into the $N$-spin chain under the condition that LF and TF have equal weights ($\theta=\pi/4$). Without loss of generality, it is assumed that all reservoirs connected to the bulk spin have the same temperature $T_b$ and the same dissipation rate $\kappa_b$. All the imposed magnetic fields and the nearest-neighbor coupling strengths between spins are the same, i.e., $B_\mu=B$ and $J_{\mu,\mu+1}=J$. However, the temperatures of reservoirs connected with the nodal spins are fixed as $T_1=10B_0$ and $T_N=5B_0$. We find that $T_b$ affects the heat current quite significantly. When $T_1=T_b>T_N$, as shown in Fig. \ref{heat_7}(a), the heat current flows into the chain from the 1st to the $(N-1)$th spins, and then all flows out to the $N$th spin. When $T_1>T_b=T_N$, as shown in Fig. \ref{heat_7}(b), the heat current flows from the 1st spin into the chain and out of the other reservoirs. When $T_b$ is small, heat current flows into the system from two nodal spins and exits through the bulk spins, shown in Fig. \ref{heat_7}(c). 
However, $T_b$ is invalid for heat current when $\kappa_b=0$, the heat current flows into the chain from the first spin, through all the bulk spins, and out of the $N$th spin. As the dissipation $\kappa_b$ between the bulk spin and the corresponding reservoir gradually decreases, the influence of $T_b$ decreases. 
The effect of $T_b$ on heat current is detailed in Appendix \ref{AppendixD}.

When only LF is present, we can decompose the spin chain into $N^\prime+1$ independent subchains if $N^\prime$ spins are not connected to reservoirs. Thus, the heat current cannot flow from the chain's left end to the right end under $T_1>T_b>T_N$. This phenomenon is attributed to the fact that the state of the non-dissipative spin in LF does not change during evolution, since the net energy transfer is achieved by the spin flip induced by the reservoir. 
However, heat transfer is still present in the interior of each subchain. What's more, for a completely symmetric two-spin system, that is, the magnetic field of both spins and the temperature of the heat reservoir connected to them are the same, the system in thermal equilibrium can be driven to non-equilibrium by connecting one of the spins to an additional spin, which is proven in Appendix \ref{AppendixB}. 

For TL case, the total Hilbert space can always be divided into two decoupled subspaces, each subspace has the non-vanishing steady-state heat current by a simple calculation.

\begin{figure}
	\centering
		\includegraphics[width=8.3cm]{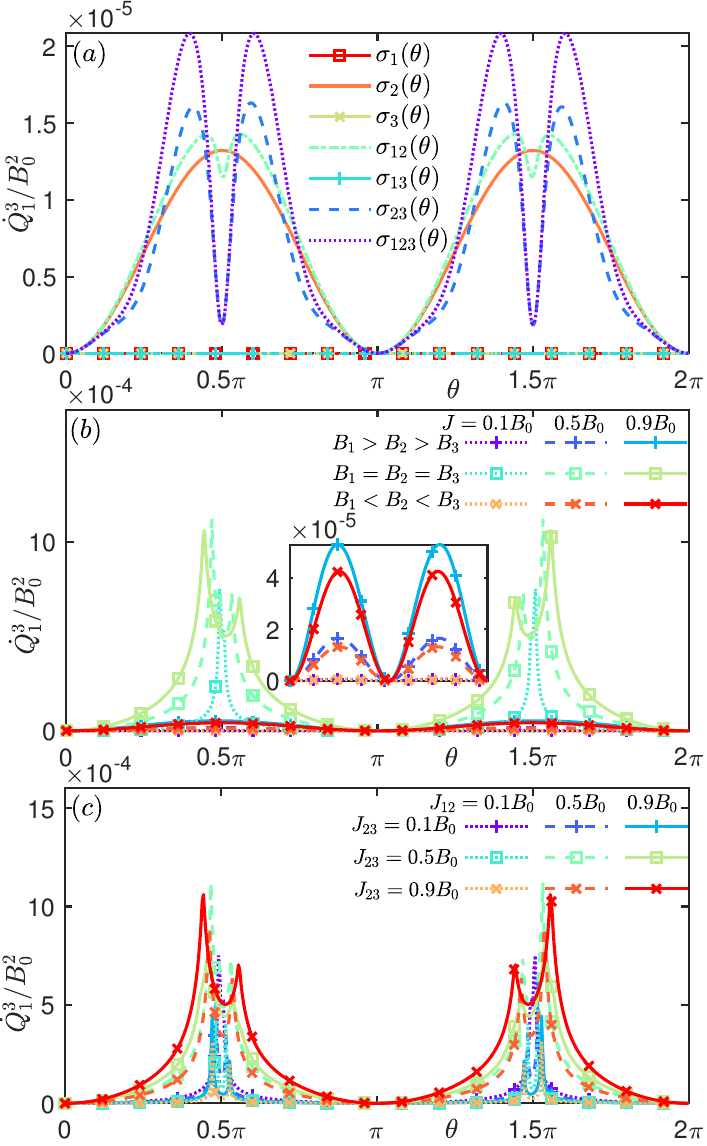}
	\caption{Plane (a) describes modulating heat current $\dot{Q}_1^3=-\dot{Q}_3^3$ by adjusting the direction of the magnetic field where the different spins are immersed. $\sigma_{12}(\theta)$ indicates that the magnetic fields of the first and second spins are modulated simultaneously. Otherwise, the process is similar. $B_1=2B_0$, $B_2=5B_0$, $B_3=8B_0$, $J_{12}=J_{23}=0.5B_0$. Planes (b)  and (c) simulate the process of $\sigma_2(\theta)$ modulating heat current under different parameters. In (b), $[B_1,B_2,B_3]\in[2B_0,5B_0,8B_0]$, $J_{12}=J_{23}\equiv J\in[0.1B_0,0.5B_0,0.9B_0]$. In (c), $B_1=B_2=B_3=5B_0$. Here, $B_0=1$, $\kappa_1=\kappa_3=0.001B_0$, $\kappa_2=0$, $T_1=10B_0$, $T_2=3B_0$, and $T_3=5B_0$.}
\label{Q3_theta}
\end{figure}
\section{Magnetically controlled quantum heat modulator}
\label{section5}

We designed a magnetic-controlled heat modulator based on the heat currents of spin chains as an example of a 3-spin chain. Figure \ref{Q3_theta}(a) simulates the modulation of heat currents by adjusting the direction of the magnetic field. The subscript marks the spin with the adjustable magnetic field, and the unmarked spins are in LF. For example, $\sigma_{12}(\theta)$ corresponds to the case where magnetic fields of the first and second spins are adjusted simultaneously, and the third spin is immersed in LF, i.e., the Hamiltonian of the free system is 
\begin{align}
\sigma_{12}(\theta)=\frac{1}{2}\sum_{\mu=1}^2B_\mu(\sin\theta\sigma_\mu^x+\cos\theta\sigma_\mu^x)+B_3\sigma_3^z.
\end{align}
According to Fig. \ref{Q3_theta} (a), one can excitingly find that the heat current is blocked once the non-dissipative spin is immersed in LF. 
A detailed analysis is shown in the Table \ref{table1}. Regardless of the magnetic field, the transition channels of the first and third spins are kept, and the dynamics of the system at $\kappa_2=0$ can be divided into two independent parts according to the eigen-operator. When the system reaches the steady state, the first and third spins are in thermal equilibrium with their local reservoirs, respectively, where $\rho^{\mathrm{th}}_\mu=p_\mu^e\vert e\rangle_\mu\langle e\vert+p_\mu^g\vert g\rangle_\mu\langle g\vert$ denotes the thermal equilibrium state with $p_\mu^e=\frac{\bar{n}_\mu(\omega)}{2\bar{n}_\mu(\omega)+1}$ and $p_\mu^g=\frac{\bar{n}_\mu(\omega)+1}{2\bar{n}_\mu(\omega)+1}$. 
We must emphasize that each model corresponds to a unique eigen-system. Therefore, for the different models, the nodal spins in thermal equilibrium have different frequencies --- their corresponding eigen-frequencies. The eigen-system and transition operations in each model are given in Appendix \ref{AppendixE}.
\begin{table}[!h] 
\renewcommand{\arraystretch}{1.5}
\centering
\caption{Transitions and dynamics of the 3-spin chain under different magnetic fields when the $2$nd spin is in the longitudinal field.} \label{table1}
\scalebox{0.9}{
\begin{tabular}{c|c|c|c|c|c}
  \hline
  \hline
    \multicolumn{2}{c|}{Magnetic field in turn} & LLL & LLT & TLL & TLT \\
  \hline
    \multirow{3}{*}{\makecell[c]{Transition\\operators}} 
       & 1st & \multicolumn{4}{c}{$V_1^{15},V_1^{26},V_1^{37},V_1^{48}$} \\
       \cline{2-6}
       & 2nd & \makecell[c]{$V_2^{13},V_2^{24},$\\$V_2^{57},V_2^{68}$} 
             & \makecell[c]{$V_2^{13},V_2^{24},$\\$V_2^{57},V_2^{68},$\\$V_2^{14},V_2^{23},$\\$V_2^{58},V_2^{57}$} 
             & \makecell[c]{$V_2^{13},V_2^{24},$\\$V_2^{57},V_2^{68},$\\$V_2^{17},V_2^{28},$\\$V_2^{35},V_2^{46}$} 
             & \makecell[c]{$V_2^{13},V_2^{24},$\\$V_2^{57},V_2^{68},$\\$V_2^{14},V_2^{23},$\\$V_2^{58},V_2^{67},$\\$V_2^{17},V_2^{28},$\\$V_2^{35},V_2^{46},$\\$V_2^{18},V_2^{27},$\\$V_2^{36},V_2^{45}$} \\
       \cline{2-6}
       & 3rd & \multicolumn{4}{c}{$V_3^{12},V_3^{34},V_3^{56},V_3^{78}$}\\
  \hline
    \multirow{2}{*}{\makecell[c]{Dynamics\\with\\$\kappa_2=0$}}  
       & $\mathbb{S}_{2_g}$ & \multicolumn{4}{c}{\makecell[c]{$\dot{\rho}_{11}=-\Gamma_{15}^1-\Gamma_{12}^3,\quad\dot{\rho}_{22}=-\Gamma_{26}^1+\Gamma_{12}^3,$\\$\dot{\rho}_{55}=+\Gamma_{15}^1-\Gamma_{56}^3,\quad\dot{\rho}_{66}=+\Gamma_{26}^1+\Gamma_{56}^3$}} \\
       \cline{2-6}
       & $\mathbb{S}_{2_e}$ & \multicolumn{4}{c}{\makecell[c]{$\dot{\rho}_{33}=-\Gamma_{37}^1-\Gamma_{34}^3,\quad\dot{\rho}_{44}=-\Gamma_{48}^1+\Gamma_{34}^3,$\\$\dot{\rho}_{77}=+\Gamma_{37}^1-\Gamma_{78}^3,\quad\dot{\rho}_{88}=+\Gamma_{48}^1+\Gamma_{78}^3$}}\\
  \hline
      \multirow{2}{*}{\makecell[c]{Steady\\state}}  
       & $\mathbb{S}_{2_g}$ & \multicolumn{4}{c}{$\rho^{\mathrm{th}}_1\otimes\vert e\rangle_2\langle e\vert\otimes\rho^{\mathrm{th}}_3$}\\
       \cline{2-6}
       & $\mathbb{S}_{2_e}$ & \multicolumn{4}{c}{$\rho^{\mathrm{th}}_1\otimes\vert g\rangle_2\langle g\vert\otimes\rho^{\mathrm{th}}_3$}\\
  \hline
    \multicolumn{2}{c|}{Conduction} & \multicolumn{4}{c}{No}\\
 \hline
 \hline
 \end{tabular}}
 \end{table}

Since the 2nd spin can block the heat current if immersed in LF, we can effectively use the magnetic field to modulate the heat current. Heat current can be monotonically modulated by individually changing the magnetic field on the 2nd spin from LF to TF, which corresponds to the orange unmarked solid line in Fig. \ref{Q3_theta}(a). However, this monotonic feature could not be general. Figures \ref{Q3_theta}(b) and (c) show that heat current manipulation appears the non-monotonic behavior at $B_1=B_2=B_3$. However, regardless of the local non-monotonicity, the heat current can always be modulated from some maximum to minimum values in a large range. Besides, one can also find that the heat current is enhanced with the increase of coupling strength.

\begin{figure}
	\centering
		\includegraphics[width=8.3cm]{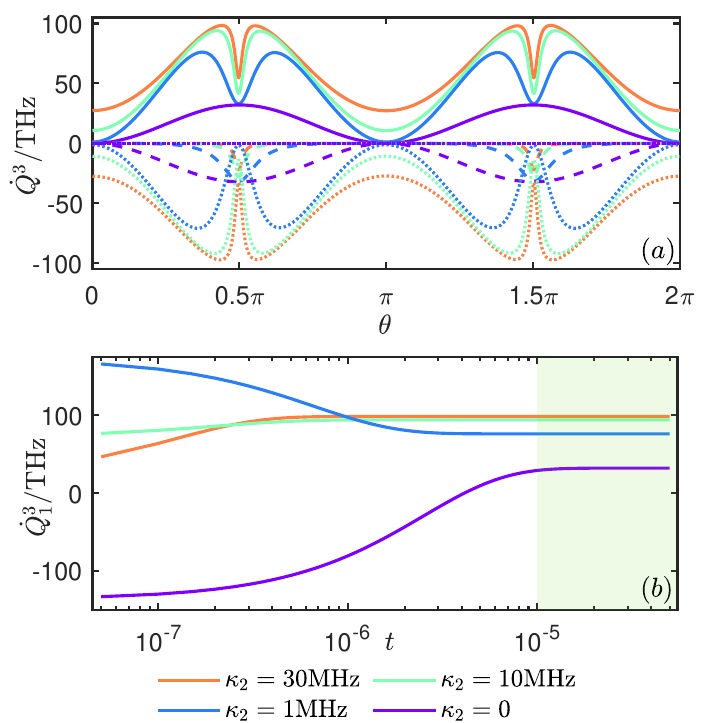}
	\caption{Numerical simulation of heat currents modulated by adjusting the direction of the magnetic field in (a). The variation of heat current between the left reservoir and the system with time is shown in (b), the green background indicate that the system is in the steady state. The solid, dotted, and dashed lines correspond to the heat currents between the first, second, and third atom and the corresponding reservoir. Here, $E_C^1=1.3\mathrm{GHz}$, $E_C^2=2.3\mathrm{GHz}$, $E_C^3=3.3\mathrm{GHz}$, $E_J^i=5E_C^i, i=1,2,3$, $n_g=0.2$, $J_{12}=J_{23}=125\mathrm{MHz}$, $\kappa_1=\kappa_3=30\mathrm{MHz}$, $\kappa_2=1\mathrm{MHz}$, $T_1=0.2\mathrm{K}$, $T_2=0.1\mathrm{K}$, and $T_3=0.1\mathrm{K}$.}
\label{superconducting_cricuit}
\end{figure}

Superconducting circuits have been used to design a wide variety of quantum thermal devices, such as quantum transistors \cite{PhysRevB.101.184510,Yamamoto_2021,Gubaydullin2022}, diodes \cite{Sanchez_2017,10.1063/1.4991516,senior2020heat,Diaz_2021,10.1063/5.0160675}.
In a superconducting circuit, the qubit is composed of Josephson junctions and capacitances, whose state can be controlled by adjusting the magnetic flux and the gate voltage \cite{GU20171,RevModPhys.93.025005,RevModPhys.73.357}. 
Here, a spin-$1/2$ atom can be designed by the Copper pair box \cite{doi:10.1126/science.285.5430.1036,PhysRevA.76.042319,PhysRevApplied.17.064022,Arrachea_2023}, which consists of two superconductors with a Josephson junction, and the Hamiltonian can be expressed as 
\begin{align}
H_{\mathrm{qubit}}=-B^z\sigma^z-B^x\sigma^x=-B(\cos\theta\sigma^z+\sin\theta\sigma^x),
\end{align} 
where $B^z=2E_C(1-2n_g)$, $B^x=E_J\cos\phi$, $B=\sqrt{{B^x}^2+{B^z}^2}$, and $\arctan\theta=B^x/B^z$. 
$E_C=e^2/{2(C_\Sigma)}$ is the superconductor's charging energy, where $C_\Sigma=C_g+C_J$ is the capacitance of the two superconductors themselves and between them. $E_J=\Phi_0 I_C/(2\pi)$ is the Josephson coupling energy, where $\Phi_0$ is the flux quantum and $I_C$ is critical current.
$n_g=C_gV_g/(2e)$ is a dimensionless gate charge that can be regulated by controlling the gate voltage $V_g$. 
The longitudinal magnetic field strength can be achieved by adjusting $n_g$. 
The capacitive coupling between two CPBs can be achieved by a superconducting quantum interference device (SQUID) \cite{Kerman_2019,PhysRevApplied.13.034037,10.1063/5.0069530,10.1063/5.0160675,PhysRevApplied.17.064022}. 
The coupling between qubit and environment is usually achieved by introducing a microwave resonator, the leakage rate of the cavity describes the dissipation \cite{PhysRevLett.94.090501,GU20171,Harrington2022}.
In order to better evaluate quantum thermal management, Fig. \ref{superconducting_cricuit}(a) show the numerical processes for heat current modulation of the non-equilibrium 3-spin chain in different dissipation rate. Only the magnetic field where the second atom is immersed is changed, and the first and third sites are in the longitudinal field. As the dissipation of the second atom is continuously reduced, the heat current in the longitudinal field gradually approaches $0$.
Furthermore, we studied  the time to reach the maximum regulatable heat current in a certain modulation process in Fig. \ref{superconducting_cricuit}(b). Theoretically, the time for a non-equilibrium open quantum system to reach steady state is infinite. Here, we define the system as being in a steady state when the heat current's rate of change approaches zero (indicated by green background). Under this criterion, the heat current stabilizes at its maximum value. Obviously, after $10^{-5}\mathrm{s}$, the heat current can be stably modulated to the maximum value.

\section{Conclusion}
\label{section7}

In this paper, the steady-state heat current of the non-equilibrium strongly internal-coupled spin chain immersed in the tilted magnetic field is obtained analytically by the BMS master equation. The singular heat transport behaviors are observed, demonstrating that in LF, the non-dissipative spin can decompose the chain into several independent subchains. Meanwhile, the energy of the nodal spins in the subchain is corrected by their couplings with their nearest non-dissipative spin based on their states. Interestingly, in TF, the system can always be divided into two subsystems regardless of whether non-dissipative spins exist, which provides the potential to control the steady-state behaviors by initial states. Based on the characteristics of heat currents, the magnetic-controlled heat modulator is designed, which provides a more feasible scheme for developing a quantum thermal device. We want to emphasize that our findings could have potential effects on magnetic order \cite{PhysRevB.103.094424,PhysRevLett.132.266701} and non-equilibrium phase transitions \cite{PhysRevLett.127.043902,PhysRevLett.126.116401,PhysRevA.105.053311,PhysRevLett.129.183602,PhysRevApplied.19.L031001} in Ising models, and so on, which deserves our forthcoming investigations.

\section*{Acknowledgments}
This work was supported by the National Natural Science Foundation of China under Grant Nos. 12175029, 12011530014, and 11775040.


\appendix
\begin{widetext}
\section{Expression of the coefficient of the transition operator}
\label{Appendix0}  
 
The eigen-operator between the energy levels $\vert i\rangle$ and $\vert j\rangle$ can be expressed as $V^{ij}_\mu=\Lambda_\mu^{ij}\vert i\rangle\langle j\vert$, where the coefficient $\Lambda_\mu^{ij}$ has different expression under different magnetic field.
When the transverse and longitudinal fields are present at the same time,
\begin{align}
\Lambda_\mu^{ij}=\sum_{k\in C_\mu^N}\Lambda(i,k)\Lambda(j,k+2^{N-\mu})+\Lambda(i,k+2^{N-\mu})\Lambda(j,k),\label{Lambdax}
\end{align} 
where $C_\mu^N$ is defined in Fig. \ref{CmuN}. 
When only the longitudinal field is present, the coefficient is 
\begin{align}
\Lambda_\mu^{ij,\parallel}=\left\{
\begin{array}{rcl}
1,&\quad &i\in C_\mu^N,j=i+2^{N-\mu},\\
0,&\quad &i,j\in\mathrm{others}.
\end{array}\right.\label{LambdaxL}
\end{align}
When only the transverse field is present, the transition coefficients of the $1$st spin in $\mathbb{S}_1^\perp$ and $\mathbb{S}_2^\perp$ are $\Lambda_1^{ij,\perp_1}=\Lambda_1^{ij,\perp},i\in[1,2^{N-1}-1],j\in[i+1,2^{N-1}]$ and $\Lambda_1^{ij,\perp_2}=-\Lambda_1^{ij,\perp},i\in[2^{N-1}+1,2^N-1],j\in[i+1,2^N]$, respectively. For the other spins, the coefficient is $\Lambda_{\mu}^{ij,\perp}$, which can be expressed as
\begin{align}
\Lambda_\mu^{ij,\perp}=\left\{
\begin{array}{rcl}
&2\sum_{k\in C_1^N}\Lambda(i,k)\Lambda(j,2^{N-1}+1-k), \hspace{3.4cm}\quad \mu=1,\\
&2\sum_{k\in C_{\mu-1}^{N-1}}\Lambda(i,k)\Lambda(j,k+2^{N-\mu})+\Lambda(i,k+2^{N-\mu})\Lambda(j,k),\quad \mu\neq 1.
\end{array}\right.\label{Vx_TF}
\end{align}


\section{Global analytic solution of 1D longitudinal field Ising chain}
\label{AppendixA} 
 
In the longitudinal field (LF) Ising model, the system is diagonalized. Compared with the eigen-operators obtained from the local master equation, the eigen-operator of each spin obtained by the global approach is still a down operator. The difference is that it depends on the state of its nearest-neighbor spins. Therefore, there are two (or four) eigen-operators per spin for the nodal (or bulk) spins. In the absence of a specific statement, the symbol "$\parallel$" has been omitted for simplicity. Specifically, the eigen-operators and the corresponding eigen-frequencies for the nodal spins are
\begin{align}
\nonumber
V_{11}&=\sigma_1^-\otimes \rho_2^g\otimes\mathbbm{1}_{3,\cdots,N},\quad&\omega_{11}&=B_1-J_{12},\\
\nonumber
V_{12}&=\sigma_1^-\otimes \rho_2^e\otimes\mathbbm{1}_{3,\cdots,N},\quad&\omega_{12}&=B_1+J_{12},\\
\nonumber
V_{N1}&=\mathbbm{1}_{1,\cdots,N-2}\otimes \rho_{N-1}^g\otimes\sigma_N^-,\quad&\omega_{N1}&=B_N-J_{N-1,N},\\
V_{N2}&=\mathbbm{1}_{1,\cdots,N-2}\otimes \rho_{N-1}^e\otimes\sigma_N^-,\quad&\omega_{N2}&=B_N+J_{N-1,N},
\end{align}
where $V_{\mu l}$ denotes the $l$th eigen-operator of the $\mu$th spin, $\mathbbm{1}_{1,\cdots,\mu}\equiv\mathbbm{1}_1\otimes\mathbbm{1}_2\otimes\cdots\otimes\mathbbm{1}_\mu$,  $\sigma_\mu^-=\vert g\rangle_\mu\langle e\vert$ (or $\sigma_\mu^e=\vert g\rangle_\mu\langle g\vert$) is the down (or up) operator, and $\rho_\mu^{g(e)}=\vert {g(e)}\rangle_\mu\langle {g(e)}\vert$ denotes the excited (ground) state of the $\mu$th spin. 
For the $\mu$th bulk spin, the eigen-operators and the corresponding eigen-frequencies are 
\begin{align}
\nonumber
V_{\mu 1}&=\mathbbm{1}_{1,\cdots,\mu-2}\otimes \rho_{\mu-1}^g\otimes \sigma_\mu^-\otimes\rho_{\mu+1}^g\otimes\mathbbm{1}_{\mu+2,\cdots,N},\quad&\omega_{\mu 1}&=B_\mu-J_{\mu-1,\mu}-J_{\mu,\mu+1},\\
\nonumber
V_{\mu 2}&=\mathbbm{1}_{1,\cdots,\mu-2}\otimes \rho_{\mu-1}^g\otimes \sigma_\mu^-\otimes \rho_{\mu+1}^e\otimes\mathbbm{1}_{\mu+2,\cdots,N},\quad&\omega_{\mu 2}&=B_\mu-J_{\mu-1,\mu}+J_{\mu,\mu+1},\\
\nonumber
V_{\mu 3}&=\mathbbm{1}_{1,\cdots,\mu-2}\otimes \rho_{\mu-1}^e\otimes \sigma_\mu^-\otimes \rho_{\mu+1}^g\otimes\mathbbm{1}_{\mu+2,\cdots,N},\quad&\omega_{\mu 3}&=B_\mu+J_{\mu-1,\mu}-J_{\mu,\mu+1},\\
V_{\mu 4}&=\mathbbm{1}_{1,\cdots,\mu-2}\otimes \rho_{\mu-1}^e\otimes \sigma_\mu^-
\otimes \rho_{\mu+1}^e\otimes\mathbbm{1}_{\mu+2,\cdots,N},\quad&\omega_{\mu 4}&=B_\mu+J_{\mu-1,\mu}+J_{\mu,\mu+1}.
\end{align}

The Lindblad quantum master equation can be rewritten as the matrix form $\vert\dot{\rho}_N(t)\rangle=\sum_\mu\mathcal{M}_\mu\vert\rho_N(t)\rangle$, $\vert\rho_N(t)\rangle=[\rho_{11},\rho_{22},\rho_{33},\rho_{44},\cdots,\rho_{2^N2^N}]^T$ is the column vector consisting of the populations, and the coefficient matrices satisfy
\begin{align}
\nonumber
\mathcal{M}_1&=\mathrm{M}_{11}\otimes \rho_2^g\otimes\mathbbm{1}_{3,\cdots,N}+\mathrm{M}_{12}\otimes \rho_2^e\otimes\mathbbm{1}_{3,\cdots,N},\\
\nonumber
\mathcal{M}_\mu&=\mathbbm{1}_{1,\cdots,\mu-2}\otimes \rho_{\mu-1}^g\otimes\mathrm{M}_{\mu 1}\otimes \rho_{\mu+1}^g\otimes\mathbbm{1}_{\mu+2,\cdots,N}+\mathbbm{1}_{1,\cdots,\mu-2}\otimes \rho_{\mu-1}^g\otimes\mathrm{M}_{\mu 2}\otimes \rho_{\mu+1}^e\otimes\mathbbm{1}_{\mu+2,\cdots,N}\\
\nonumber
&+\mathbbm{1}_{1,\cdots,\mu-2}\otimes \rho_{\mu-1}^e\otimes\mathrm{M}_{\mu 3}\otimes \rho_{\mu+1}^g\otimes\mathbbm{1}_{\mu+2,\cdots,N}+\mathbbm{1}_{1,\cdots,\mu-2}\otimes \rho_{\mu-1}^e\otimes\mathrm{M}_{\mu 4}\otimes \rho_{\mu+1}^e\otimes\mathbbm{1}_{\mu+2,\cdots,N},\\
\mathcal{M}_N&=\mathbbm{1}_{1,\cdots,N-2}\otimes \rho_{N-1}^g\otimes\mathrm{M}_{N1}+\mathbbm{1}_{1,\cdots,N-2}\otimes \rho_{N-1}^e\otimes\mathrm{M}_{N2},
\end{align}
where $\mathrm{M}_{\mu l}=2\begin{pmatrix}-J_\mu(+\omega_{\mu l})&J_\mu(-\omega_{\mu l})\\J_\mu(+\omega_{\mu l})&-J_\mu(-\omega_{\mu l})\end{pmatrix}$ and $\vert\rho_N(t)\rangle=[\rho_{1_g2_g\cdots(N-1)_gN_g},\rho_{1_g2_g\cdots(N-1)_gN_e},$ $\rho_{1_g2_g\cdot(N-1)_eN_g},\rho_{1_g2_g\cdots(N-1)_eN_e},\cdots,\rho_{1_e2_e\cdots(N-1)_eN_e}]^{\mathrm{T}}$. $\rho_{1_g2_g\cdots(N-1)_gN_g}$ represents the population of the energy level that all spins are in the ground state and so on.

When a single spin is connected to a heat reservoir with the temperature $T$, the steady-state dynamic is $\mathcal{M}_1\vert\rho_1^S\rangle=0$. The coefficient matrix is $\mathcal{M}_1=\mathrm{M}=2\begin{pmatrix}
-J(+B) & J(-B)\\J(+B) & -J(-B)
\end{pmatrix}$, where $J(\pm B)=\pm\kappa n(\pm B)$ with $\kappa$ denoting the dissipation rate and $n(B)=(\mathrm{exp}[B/T]-1)^{-1}$ denoting the average photon number. 
Therefore, the vector consisted of the steady-state populations is $\vert\rho_1^S\rangle=[\rho_{1_g},\rho_{1_e}]^T$, where 
\begin{align}
\rho_{1_g}=\frac{n(B)+1}{2n(B)+1},\quad\rho_{1_e}=\frac{n(B)}{2n(B)+1}.
\end{align}
This single spin is in thermal equilibrium with the local reservoir, the net transition rate is 
\begin{align}
\Gamma_{1_\updownarrow}=2[J(+B)\rho_{1_g}-J(-B)\rho_{1_e}]=0.
\end{align}
Therefore, when the single spin is connected to a reservoir, there is no steady-state net transition between the energy levels and no steady-state heat current.

The dynamic of the 2-spin chain is $\mathcal{M}_{12}\vert\rho_{12}^S\rangle=0$, where 
\begin{align}
\nonumber
\mathcal{M}_{12}&=\mathrm{M}_{11}\otimes \rho^g_2+\mathrm{M}_{12}\otimes \rho^e_2+\rho^g_1\otimes\mathrm{M}_{21}+\rho^e_1\otimes\mathrm{M}_{22},\\
\vert\rho_{12}^S\rangle&=[\rho_{1_g2_g},\rho_{1_g2_e},\rho_{1_e2_g},\rho_{1_e2_e}]^T
\end{align}
with $\rho_{1_{g(e)}2_{g(e)}}=\frac{1}{N_{12}}\breve{\rho}_{1_{g(e)}2_{g(e)}}$, and the non-normalized populations are
\begin{align}
\nonumber
\breve{\rho}_{1_g2_g}=J_{11}^-J_{12}^-J_{21}^-+J_{11}^-J_{12}^+J_{22}^-+J_{11}^-J_{21}^-J_{22}^-+J_{12}^-J_{21}^-J_{22}^+,\\
\nonumber
\breve{\rho}_{1_g2_e}=J_{11}^-J_{12}^-J_{21}^++J_{11}^+J_{12}^-J_{22}^++J_{11}^-J_{21}^+J_{22}^-+J_{12}^-J_{21}^+J_{22}^+,\\
\nonumber
\breve{\rho}_{1_e2_g}=J_{11}^+J_{12}^-J_{21}^-+J_{11}^+J_{12}^+J_{22}^-+J_{11}^+J_{21}^-J_{22}^-+J_{12}^+J_{21}^+J_{22}^-,\\
\breve{\rho}_{1_e2_e}=J_{11}^-J_{12}^+J_{21}^++J_{11}^+J_{12}^+J_{22}^++J_{11}^+J_{21}^-J_{22}^++J_{12}^+J_{21}^+J_{22}^+,
\end{align}
where $J_{\mu l}^{\pm}\equiv J_\mu(\pm\omega_{\mu l})=\pm\kappa_\mu n_\mu(\pm\omega_{\mu l})$ stands for spectral density.
For the 2-spin system, by simple calculation, the four transitions induced by two heat reservoirs have the same net transition rate as
\begin{align}
\Gamma_{1_\updownarrow 2_g}=\frac{2}{N_{12}}(J_{11}^+J_{12}^-J_{21}^-J_{22}^+-J_{11}^-J_{12}^+J_{21}^+J_{22}^-)=-\Gamma_{1_\updownarrow 2_e}=-\Gamma_{1_g 2_\updownarrow}=\Gamma_{1_e 2_\updownarrow}\equiv\Gamma_{12}.
\end{align}
Therefore, the steady-state currents are
\begin{align}
\dot{Q}^2_1=-2J_{12}\Gamma_{12}=-\dot{Q}^2_2.
\end{align}
Suppose these two spins with the same energy are connected to the reservoirs with the same temperature. In that case, the steady-state heat currents are blocked since the system in this condition is entirely symmetric.

For the dynamics of the 3-spin system $\mathcal{M}_{123}\vert\rho_{123}^S\rangle=0$, where
\begin{align}
\nonumber
\mathcal{M}_{123}&=\mathrm{M}_{11}\otimes \rho_2^g\otimes\mathbbm{1}_3+\mathrm{M}_{12}\otimes \rho_2^e\otimes\mathbbm{1}_3\\
\nonumber
&+\rho_1^g\otimes\mathrm{M}_{21}\otimes \rho_3^g+\rho_1^g\otimes\mathrm{M}_{22}\otimes \rho_3^e+\rho_1^e\otimes\mathrm{M}_{23}\otimes \rho_3^g+\rho_1^e\otimes\mathrm{M}_{24}\otimes \rho_3^e\\
\nonumber
&+\mathbbm{1}_1\otimes \rho_2^g\otimes\mathrm{M}_{31}+\mathbbm{1}_1\otimes \rho_2^e\otimes\mathrm{M}_{32},\\
\vert\rho_{123}^S\rangle&=[\rho_{1_g2_g3_g},\rho_{1_g2_g3_e},\rho_{1_g2_e3_g},\rho_{1_g2_e3_e},\rho_{1_e2_g3_g},\rho_{1_e2_g3_e},\rho_{1_e2_e3_g},\rho_{1_e2_e3_e}]^T.
\end{align}
Similar calculations can be used to obtain the analytical expressions of the steady state and the corresponding heat currents. However, they are not expressed here because the form is too complicated. 
For the special case of $T_1=T_3$, $B_1=B_2=B_3\equiv B$, and $J_{12}=J_{23}\equiv J$, the eigen-frequencies are $\omega_{1,1(2)}=\omega_{3,1(2)}=B+(-)J\equiv\omega_{n,1(2)}$, $\omega_{21}=B+2J\equiv\omega_{b1}$, $\omega_{22}=B\equiv\omega_{b2}$, and $\omega_{23}=B+2J\equiv\omega_{b3}$. Taking advantage of the normalization condition $\mathrm{Tr}\rho_{123}=1$, the steady state density matrix can be obtained. The non-normalized populations are
\begin{align}
\nonumber
\breve{\rho}_{1_g2_g3_g}=&{J_{b1}^-} (({J_{b2}^-} {J_{n1}^-} + {J_{b2}^+} {J_{n2}^-}) ({J_{b3}^-} {J_{n1}^-} + {J_{b3}^+} {J_{n2}^-}) + 
    {J_{b3}^-} {J_{n1}^-}^2 ({J_{n2}^+} + {J_{n2}^-}) + {J_{b3}^+} {J_{n2}^-}^2 ({J_{n1}^+} + {J_{n1}^-})) \\
\nonumber
+& 2 {J_{n1}^-} ({J_{b1}^-} {J_{n2}^-} ({J_{b2}^-} {J_{n1}^-} + {J_{b2}^+} {J_{n2}^-}) + 
    {J_{b2}^-} {J_{n2}^+} ({J_{b3}^-} {J_{n1}^-} + {J_{b3}^+} {J_{n2}^-}) + 
    {J_{n1}^-} ({J_{b1}^-} {J_{n2}^-}^2 + {J_{b3}^-} {J_{n2}^+}^2 + 2 {J_{b2}^-} {J_{n2}^+} {J_{n2}^-})), \\
\nonumber
\breve{\rho}_{1_g2_g3_e}=&{J_{b2}^-} ({J_{b1}^-} {J_{n1}^+} + {J_{b1}^+} {J_{n2}^+}) ({J_{b3}^-} {J_{n1}^-} + {J_{b3}^+} {J_{n2}^-}) + 
 {J_{b1}^-} {J_{n1}^+} {J_{n2}^-} ( {J_{b3}^-} {J_{n1}^-} + {J_{b3}^+} {J_{n2}^-}) + 
 {J_{b3}^-} {J_{n1}^-} {J_{n2}^+} ({J_{b1}^-} {J_{n1}^+} + {J_{b1}^+} {J_{n2}^+}) \\
\nonumber
+& 2 {J_{b2}^-} ({J_{n1}^-} {J_{n1}^+} ({J_{b1}^-} {J_{n2}^-} + {J_{b3}^-}  {J_{n2}^+}) + 
    {J_{n2}^-} {J_{n2}^+} ({J_{b1}^+} {J_{n1}^-} + {J_{b3}^+} {J_{n1}^+} )) + 
 2  {J_{n1}^-} {J_{n1}^+} ({J_{b3}^-} {J_{n2}^+}^2 + {J_{b1}^-} {J_{n2}^-}^2 + 2 {J_{b2}^-} {J_{n2}^-} {J_{n2}^+})\\
 \nonumber
 =&\breve{\rho}_{1_e2_g3_g},\\
\nonumber
\breve{\rho}_{1_g2_e3_g}=&{J_{b1}^+} (({J_{b2}^-} {J_{n1}^-} + {J_{b2}^+} {J_{n2}^-}) ({J_{b3}^-} {J_{n1}^-} + {J_{b3}^+} {J_{n2}^-}) + 
    {J_{b3}^-} {J_{n1}^-}^2 ({J_{n2}^+} + {J_{n2}^-}) + {J_{b3}^+} {J_{n2}^-}^2 ({J_{n1}^+} + {J_{n1}^-})) \\
\nonumber
+& 2 {J_{n2}^-} ({J_{b1}^+} {J_{n1}^-} ({J_{b2}^-} {J_{n1}^-} + {J_{b2}^+} {J_{n2}^-}) + 
    {J_{b2}^+} {J_{n1}^+} ({J_{b3}^-} {J_{n1}^-} + {J_{b3}^+} {J_{n2}^-}) + 
    {J_{n2}^-} ({J_{b1}^+} {J_{n1}^-}^2 + {J_{b3}^+} {J_{n1}^+}^2 + 2 {J_{b2}^+} {J_{n1}^+} {J_{n1}^-})),\\
\nonumber
\breve{\rho}_{1_g2_e3_e}=&{J_{b2}^+} ({J_{b1}^-} {J_{n1}^+} + {J_{b1}^+} {J_{n2}^+}) ({J_{b3}^-} {J_{n1}^-} + {J_{b3}^+} {J_{n2}^-}) + 
 {J_{b1}^+} {J_{n1}^-} {J_{n2}^+} ( {J_{b3}^-} {J_{n1}^-} + {J_{b3}^+} {J_{n2}^-}) + 
 {J_{b3}^+} {J_{n1}^+} {J_{n2}^-} ({J_{b1}^-} {J_{n1}^+} + {J_{b1}^+} {J_{n2}^+}) \\
\nonumber
+& 2 {J_{b2}^+} ({J_{n1}^-} {J_{n2}^-} ({J_{b1}^-} {J_{n1}^+} + {J_{b1}^+} {J_{n2}^+}) + 
    {J_{n1}^+} {J_{n2}^+} ({J_{b3}^+} {J_{n2}^-} + {J_{b3}^-} {J_{n1}^-} )) + 
 2  {J_{n2}^+} {J_{n2}^-} ({J_{b1}^+} {J_{n1}^-}^2 + {J_{b3}^+} {J_{n1}^+}^2 + 2 {J_{b2}^+} {J_{n1}^+} {J_{n1}^-})\\
 \nonumber
 =&\breve{\rho}_{1_e2_e3_g}, \\
\nonumber
\breve{\rho}_{1_e2_g3_e}=&{J_{b3}^-} (({J_{b1}^-} {J_{n1}^+} + {J_{b1}^+} {J_{n2}^+}) ({J_{b2}^-} {J_{n1}^+} + {J_{b2}^+} {J_{n2}^+}) + 
    {J_{b1}^-} {J_{n1}^+}^2 ({J_{n2}^-} + {J_{n2}^+}) + {J_{b1}^+} {J_{n2}^+}^2 ({J_{n1}^+} + {J_{n1}^-})) \\
\nonumber
+& 2 {J_{n1}^+} ({J_{b2}^-} {J_{n2}^-} ({J_{b1}^-} {J_{n1}^+} + {J_{b1}^+} {J_{n2}^+}) + 
    {J_{b3}^-} {J_{n2}^+} ({J_{b2}^-} {J_{n1}^+} + {J_{b2}^+} {J_{n2}^+}) + 
    {J_{n1}^+} ({J_{b1}^-} {J_{n2}^-}^2 + {J_{b3}^-} {J_{n2}^+}^2 + 2 {J_{b2}^-} {J_{n2}^+} {J_{n2}^-})),\\
\nonumber
\breve{\rho}_{1_e2_e3_e}=& {J_{b3}^+} (({J_{b1}^-} {J_{n1}^+} + {J_{b1}^+} {J_{n2}^+}) ({J_{b2}^-} {J_{n1}^+} + {J_{b2}^+} {J_{n2}^+}) + 
    {J_{b1}^-} {J_{n1}^+}^2 ({J_{n2}^-} + {J_{n2}^+}) + {J_{b1}^+} {J_{n2}^+}^2 ({J_{n1}^+} + {J_{n1}^-})) \\
+& 2 {J_{n2}^+} ({J_{b2}^+} {J_{n1}^-} ({J_{b1}^-} {J_{n1}^+} + {J_{b1}^+} {J_{n2}^+}) + 
    {J_{b3}^+} {J_{n1}^+} ({J_{b2}^-} {J_{n1}^+} + {J_{b2}^+} {J_{n2}^+}) + 
    {J_{n2}^+} ({J_{b1}^+} {J_{n1}^-}^2 + {J_{b3}^+} {J_{n1}^+}^2 + 2 {J_{b2}^+} {J_{n1}^+} {J_{n1}^-})),
\end{align}
where $\rho_{1_{g(e)}2_{g(e)}3_{g(e)}}=\frac{1}{N_{123}}\breve{\rho}_{1_{g(e)}2_{g(e)}3_{g(e)}}$.
The net transition rates are 
\begin{align}
\nonumber
\Gamma_{1_\updownarrow2_g3_g}&=\frac{2}{N_{123}}\big(({J_{b3}^-} {J_{n1}^-} + {J_{b3}^+} {J_{n2}^-} + 2 {J_{n1}^-} {J_{n2}^-}) ({J_{b1}^-} {J_{b2}^+} {J_{n1}^+} {J_{n2}^-} - {J_{b1}^+} {J_{b2}^-} {J_{n1}^-} {J_{n2}^+})+ {J_{b1}^-} {J_{b3}^+} {J_{n1}^+}^2 {J_{n2}^-}^2 - {J_{b1}^+} {J_{b3}^-} {J_{n1}^-}^2 {J_{n2}^+}^2\big)\\
\nonumber
&=-\Gamma_{1_\updownarrow2_e3_g}=\Gamma_{1_g2_g3_\updownarrow}=-\Gamma_{1_g2_e3_\updownarrow}=-\frac{1}{2}\Gamma_{1_g2_\updownarrow3_g}\equiv\Gamma^1_{123},\\
\nonumber
\Gamma_{1_\updownarrow2_g3_e}&=\frac{2}{N_{123}}\big( ({J_{b1}^+} {J_{n2}^+} + {J_{b1}^-} {J_{n1}^+} +  2 {J_{n1}^+} {J_{n2}^+})({J_{b2}^-} {J_{b3}^+} {J_{n1}^+} {J_{n2}^-} - {J_{b2}^+} {J_{b3}^-} {J_{n1}^-} {J_{n2}^+}) + {J_{b1}^-} {J_{b3}^+} {J_{n1}^+}^2 {J_{n2}^-}^2 - {J_{b1}^+} {J_{b3}^-} {J_{n1}^-}^2 {J_{n2}^+}^2\big)\\
\nonumber
&=-\Gamma_{1_\updownarrow2_e3_e}=\Gamma_{1_e2_g3_\updownarrow}=-\Gamma_{1_e2_e3_\updownarrow}=+\frac{1}{2}\Gamma_{1_e2_\updownarrow3_e}\equiv\Gamma^2_{123},\\
\Gamma_{1_g2_\updownarrow3_e}&=\Gamma_{1_e2_\updownarrow3_g}=\Gamma^1_{123}-\Gamma^2_{123}\equiv\Gamma^3_{123}.
\end{align}
Therefore, the steady-state heat currents are
\begin{align}
\dot{Q}^3_1=\dot{Q}_3^3=-\frac{1}{2}\dot{Q}^3_2&=-2J(\Gamma_{123}^1+\Gamma_{123}^2).\label{Q3_13}
\end{align}
Equation (\ref{Q3_13}) also shows that when the system is completely symmetric, the heat currents that flow into the system from two nodal spins are equal and flow out together through the middle spin.

If the second spin is not connected to the local reservoir, i.e., $\kappa_2=0$, $J_{2}(\pm\omega_{2l})=0$, and $\dot{Q}_3^2=0$. In this case, there are two subspaces $S^{2_{g}}$ and $S^{2_{e}}$. Subspace $S^{2_{g(e)}}$ consists of energy levels $\vert 1_g2_{g(e)}3_g\rangle$, $\vert 1_g2_{g(e)}3_e\rangle$, $\vert 1_e2_{g(e)}3_g\rangle$, and $\vert 1_e2_{g(e)}3_e\rangle$, where $\vert 1_g2_g3_e\rangle\equiv\vert g\rangle_1\otimes\vert g\rangle_2\otimes\vert g\rangle_3$ and so on. The steady-state populations are 
\begin{align}
\nonumber
\rho_{1_g2_{g(e)}3_g}&=\frac{n_1(\omega_{1,1(2)})+1}{2n_1(\omega_{1,1(2)})+1}\cdot\frac{n_3(\omega_{3,1(2)})+1}{2n_3(\omega_{3,1(2)})+1},&\quad\rho_{1_g2_{g(e)}3_e}&=\frac{n_1(\omega_{1,1(2)})+1}{2n_1(\omega_{1,1(2)})+1}\cdot\frac{n_3(\omega_{3,1(2)})}{2n_3(\omega_{3,1(2)})+1},\\
\rho_{1_e2_{g(e)}3_g}&=\frac{n_1(\omega_{1,1(2)})}{2n_1(\omega_{1,1(2)})+1}\cdot\frac{n_3(\omega_{3,1(2)})+1}{2n_3(\omega_{3,1(2)})+1},&\quad\rho_{1_e2_{g(e)}3_e}&=\frac{n_1(\omega_{1,1(2)})}{2n_1(\omega_{1,1(2)})+1}\cdot\frac{n_3(\omega_{3,1(2)})}{2n_3(\omega_{3,1(2)})+1}.
\end{align}
The steady state is the thermal equilibrium state. However, it must be noted that when the second spin is in the ground state (or excited state), the frequencies of the nodal spins in the thermal state are $\omega_{1,1(2)}=B_1-(+)J_{12}$ and $\omega_{3,1(2)}=B_3-(+)J_{23}$. Therefore, the heat current is blocked when the second spin has no dissipation in the LF 3-spin Ising chain.

\section{Dynamics and Steady-state heat currents of 1D longitudinal feild 7-spin Ising chain}
\label{AppendixB}
 
In the chain consisting of seven spins, if the third and the sixth spins are not connected to the reservoirs, the transitions $\mathbbm{1}_1\otimes \rho_2^{g(e)}\otimes\sigma_3^-\otimes \rho_4^{g(e)}\otimes\mathbbm{1}_5\otimes\mathbbm{1}_6\otimes\mathbbm{1}_7$ and $\mathbbm{1}_1\otimes\mathbbm{1}_2\otimes\mathbbm{1}_3\otimes\mathbbm{1}_4\otimes \rho_5^{g(e)}\otimes\sigma_6^-\otimes \rho_7^{g(e)}$ are forbidden, which is shown in Fig. \ref{model_7qubit}. The system's steady-state dynamic is $\mathcal{M}_{1234567}\vert\rho_{1234567}^S\rangle=0$, the coefficient matrix is 
\begin{align}
\nonumber
\mathcal{M}_{1234567}&=[(\mathrm{M}_{11}\otimes \rho_2^{g}+\mathrm{M}_{12}\otimes \rho_2^{e}+\rho_1^{g}\otimes\mathrm{M}_{21}+\rho_1^{e}\otimes\mathrm{M}_{23})\otimes(\mathrm{M}_{41}\otimes \rho_5^{g}+\mathrm{M}_{42}\otimes \rho_5^{e}+\rho_4^{g}\otimes\mathrm{M}_{51}+\rho_4^{e}\otimes\mathrm{M}_{53})\otimes\mathrm{M}_{71}]\otimes \rho_3^{g}\rho_6^{g}\\
\nonumber
&+[(\mathrm{M}_{11}\otimes \rho_2^{g}+\mathrm{M}_{12}\otimes \rho_2^{e}+\rho_1^{g}\otimes\mathrm{M}_{21}+\rho_1^{e}\otimes\mathrm{M}_{23})\otimes(\mathrm{M}_{41}\otimes \rho_5^{g}+\mathrm{M}_{42}\otimes \rho_5^{e}+\rho_4^{g}\otimes\mathrm{M}_{52}+\rho_4^{e}\otimes\mathrm{M}_{54})\otimes\mathrm{M}_{72}]\otimes \rho_3^{g}\rho_6^{e}\\
\nonumber
&+[(\mathrm{M}_{11}\otimes \rho_2^{g}+\mathrm{M}_{12}\otimes \rho_2^{e}+\rho_1^{g}\otimes\mathrm{M}_{22}+\rho_1^{e}\otimes\mathrm{M}_{24})\otimes(\mathrm{M}_{43}\otimes \rho_5^{g}+\mathrm{M}_{44}\otimes \rho_5^{e}+\rho_4^{g}\otimes\mathrm{M}_{51}+\rho_4^{e}\otimes\mathrm{M}_{53})\otimes\mathrm{M}_{71}]\otimes \rho_3^{e}\rho_6^{g}\\
\nonumber
&+[(\mathrm{M}_{11}\otimes \rho_2^{g}+\mathrm{M}_{12}\otimes \rho_2^{e}+\rho_1^{g}\otimes\mathrm{M}_{22}+\rho_1^{e}\otimes\mathrm{M}_{24})\otimes(\mathrm{M}_{43}\otimes \rho_5^{g}+\mathrm{M}_{44}\otimes \rho_5^{e}+\rho_4^{g}\otimes\mathrm{M}_{52}+\rho_4^{e}\otimes\mathrm{M}_{54})\otimes\mathrm{M}_{72}]\otimes \rho_3^{e}\rho_6^{e}\\
&\equiv(\mathcal{M}^{\mathbb{S}_{3_g6_g}}_{12457}\otimes \rho_3^{g}\rho_6^{g})+(\mathcal{M}^{\mathbb{S}_{3_g6_e}}_{12457}\otimes \rho_3^{g}\rho_6^{e})+(\mathcal{M}^{\mathbb{S}_{3_g6_g}}_{12457}\otimes \rho_3^{e}\rho_6^{g})+(\mathcal{M}^{\mathbb{S}_{3_e6_e}}_{12457}\otimes \rho_3^{e}\rho_6^{e}),
\end{align}
and the vector consisted of the populations is
\begin{align}
\vert\rho_{1234567}\rangle=(\vert\rho_{12457}^{\mathbb{S}_{3_g6_g}}\rangle\otimes\vert 3_g6_g\rangle)+(\vert\rho_{12457}^{\mathbb{S}_{3_g6_e}}\rangle\otimes\vert 3_g6_e\rangle)+(\vert\rho_{12457}^{\mathbb{S}_{3_e6_g}}\rangle\otimes\vert 3_e6_g\rangle)+(\vert\rho_{12457}^{\mathbb{S}_{3_e6_e}}\rangle\otimes\vert 3_e6_e\rangle).
\end{align}
The space formed by the third and the sixth spins is an invariant subspace. The dynamics of spins, which the other five local reservoirs connect, can be divided into three parts: spins 1 and 2, spins 4 and 5, and spin 7. The total Hilbert space can be expressed as a direct sum of four subspaces, which are $\mathbb{S}_{3_e6_e}$, $\mathbb{S}_{3_e6_g}$, $\mathbb{S}_{3_g6_e}$, and $\mathbb{S}_{3_g6_g}$, based on the states of the third and the sixth spins.

In the subspace $\mathbb{S}_{3_g6_g}$, for example, the dynamics for the remaining five spins are $\mathcal{M}_{12457}^{3_g6_g}\vert\rho_{12457}^{3_g6_g}\rangle=0$, where
\begin{align}
\mathcal{M}_{12457}^{\mathbb{S}_{3_g6_g}}=\mathcal{M}_{12}^{\mathbb{S}_{3_g6_g}}\otimes\mathcal{M}_{45}^{\mathbb{S}_{3_g6_g}}\otimes\mathcal{M}_{7}^{\mathbb{S}_{3_g6_g}},\quad\vert\rho_{12457}^{\mathbb{S}_{3_g6_g}}\rangle=\vert\rho_{12}^{\mathbb{S}_{3_g6_g}}\rangle\otimes\vert\rho_{45}^{\mathbb{S}_{3_g6_g}}\rangle\otimes\vert\rho_{7}^{\mathbb{S}_{3_g6_g}}\rangle.
\end{align}
The dynamics of these spins are divided into three independent parts: the first part contains the first and second spins, the fourth and fifth spins constitute the second part, and the seventh spin will be in thermal equilibrium with the heat reservoir connected to it at the steady state.
Therefore, the steady-state populations are $\rho^{\mathbb{S}_{3_g6_g}}_{1_{g(e)}2_{g(e)}4_{g(e)}5_{g(e)}7_{g(e)}}=\rho^{\mathbb{S}_{3_g6_g}}_{1_{g(e)}2_{g(e)}}\rho^{{\mathbb{S}_{3_g6_g}}}_{4_{g(e)}5_{g(e)}}\rho^{\mathbb{S}_{3_g6_g}}_{7_{g(e)}}$. 

It is convenient and necessary to obtain the dynamics of each subsystem. 
In the previous section, we describe the dynamics of a 2-spin chain in detail. Thus, in the first part, the net transition rates are respectively
\begin{align}
      \Gamma_{1_\updownarrow2_{e(g)}4_{e(g)}5_{e(g)}7_{e(g)}}^{\mathbb{S}_{3_g6_g}}=\Gamma_{1_\updownarrow2_{e(g)}}^{\mathbb{S}_{3_g6_g}}\rho_{4_{e(g)}5_{e(g)}}^{\mathbb{S}_{3_g6_g}}\rho_{7_{e(g)}}^{\mathbb{S}_{3_g6_g}},\quad
         \Gamma_{1_{e(g)}2_\updownarrow4_{e(g)}5_{e(g)}7_{e(g)}}^{\mathbb{S}_{3_g6_g}}=\Gamma_{1_{e(g)}2_\updownarrow}^{\mathbb{S}_{3_g6_g}}\rho_{4_{e(g)}5_{e(g)}}^{\mathbb{S}_{3_g6_g}}\rho_{7_{e(g)}}^{\mathbb{S}_{3_g6_g}},
\end{align}
where
\begin{align}
\Gamma^{\mathbb{S}_{3_g6_g}}_{1_\updownarrow2_g}=\frac{1}{N_{12}^{\mathbb{S}_{3_g6_g}}}(J_{11}^+J_{12}^-J_{21}^-J_{23}^+-J_{11}^-J_{12}^+J_{21}^+J_{23}^-)=-\Gamma^{\mathbb{S}_{3_g6_g}}_{1_\updownarrow2_e}=-\Gamma^{\mathbb{S}_{3_g6_g}}_{1_g2_\updownarrow}=\Gamma^{\mathbb{S}_{3_g6_g}}_{1_e2_\updownarrow}\equiv\Gamma^{\mathbb{S}_{3_g6_g}}_{12}.
\end{align}
It's worth noting that the transition frequencies induced by the second reservoir are $\omega_{2,1(3)}=(B_2-J_{23})-(+)J_{12}$, in $J_{2,1(3)}^\pm=\pm\kappa_2n_2(\pm\omega_{2,1(3)})$. 
Therefore, in the LF spin chain model, the $\mu$th spin of the unconnected heat reservoir not only divides the dynamics of the system but also produces an energy correction $J_{\mu-1,\mu}$ or $J_{\mu,\mu+1}$ for the transition frequency of its nearest-neighbor spins, whether the correction is positive or negative depends on the state of the $\mu$th spin.
If the $\mu$th spin is in the ground (or excited) state, the transition frequencies of the $(\mu-1)$th and $(\mu+1)$th spins decrease (or increase) $J_{\mu-1,\mu}$ and $J_{\mu,\mu+1}$.
Hence, the steady-state heat currents are
\begin{align}
\nonumber
\dot{Q}_1^{7,{\mathbb{S}_{3_g6_g}}}=&-2J_{12}\Gamma_{12}^{\mathbb{S}_{3_g6_g}}(\rho_{4_g5_g}^{\mathbb{S}_{3_g6_g}}+\rho_{4_g5_e}^{\mathbb{S}_{3_g6_g}}+\rho_{4_e5_g}^{\mathbb{S}_{3_g6_g}}+\rho_{4_e5_e}^{\mathbb{S}_{3_g6_g}})(\rho_{7_g}^{\mathbb{S}_{3_g6_g}}+\rho_{7_e}^{\mathbb{S}_{3_g6_g}})\\
=&-2J_{12}\Gamma_{12}^{\mathbb{S}_{3_g6_g}}=-\dot{Q}_2^{7,{\mathbb{S}_{3_g6_g}}}.\label{Q11}
\end{align}
Using the same calculation, one can get 
\begin{align}
\nonumber
\dot{Q}_4^{7,{\mathbb{S}_{3_g6_g}}}&=-2J_{45}\Gamma_{45}^{\mathbb{S}_{3_g6_g}}=-\dot{Q}_5^{7,{\mathbb{S}_{3_g6_g}}},\\ \dot{Q}_7^{7,{\mathbb{S}_{3_g6_g}}}&=0,\label{Q100100}
\end{align}
where $\Gamma_{45}^{\mathbb{S}_{3_g6_g}}=\frac{1}{N_{45}^{\mathbb{S}_{3_g6_g}}}(J_{41}^+J_{42}^-J_{51}^-J_{53}^+-J_{41}^-J_{42}^+J_{51}^+J_{53}^-)$, $\omega_{4,1(2)}=(B_4-J_{34})-(+)J_{45}$, and $\omega_{5,1(3)}=(B_5-J_{56})-(+)J_{45}$.

In the other three subspaces $S^{\mathbb{S}_{3_g6_e}}$, $S^{\mathbb{S}_{3_e6_g}}$, and $S^{\mathbb{S}_{3_e6_e}}$, the steady-state heat currents have the same form as Eqs. (\ref{Q11},\ref{Q100100}),
where
\begin{small}
\begin{align}
\nonumber
\Gamma_{12}^{\mathbb{S}_{3_g6_e}}&=\Gamma_{12}^{\mathbb{S}_{3_g6_g}}, &\Gamma_{45}^{\mathbb{S}_{3_g6_e}}&=\frac{1}{N_{45}^{\mathbb{S}_{3_g6_e}}}(J_{41}^+J_{42}^-J_{52}^-J_{54}^+-J_{41}^-J_{42}^+J_{52}^+J_{54}^-),\\
\nonumber
\Gamma_{12}^{\mathbb{S}_{3_e6_g}}&=\frac{1}{N_{12}^{\mathbb{S}_{3_e6_g}}}(J_{11}^+J_{12}^-J_{22}^-J_{24}^+-J_{11}^-J_{12}^+J_{22}^+J_{24}^-), &\Gamma_{45}^{\mathbb{S}_{3_e6_g}}&=\frac{1}{N_{45}^{\mathbb{S}_{3_e6_g}}}(J_{43}^+J_{44}^-J_{51}^-J_{53}^+-J_{43}^-J_{44}^+J_{51}^+J_{53}^-),\\
\Gamma_{12}^{\mathbb{S}_{3_e6_e}}&=\Gamma_{12}^{\mathbb{S}_{3_e6_g}}, &\Gamma_{45}^{\mathbb{S}_{3_e6_e}}&=\frac{1}{N_{45}^{\mathbb{S}_{3_e6_e}}}(J_{43}^+J_{44}^-J_{52}^-J_{54}^+-J_{43}^-J_{44}^+J_{52}^+J_{54}^-).
\end{align}
\end{small}

For the special case $B_\mu=B$, $J_{\mu,\mu+1}=J$, $T_1=T_N\equiv T_n$, and $T_2=T_3=\cdots T_{N-1}=T_b$. The eigen-frequencies for the nodal and bulk spins are
\begin{align}
\omega_{n1}=B-J,\quad\omega_{n2}=B+J,\quad\omega_{b1}=B-2J,\quad\omega_{b2}=B,\quad\omega_{b3}=B+2J.
\end{align}
Since the transition frequencies of the nodal and the bulk spins induced by the corresponding reservoir are different, the net transition rates in the four spaces are
\begin{align}
\nonumber
\Gamma_{12}^{\mathbb{S}_{3_g6_g}}&=\Gamma_{12}^{\mathbb{S}_{3_g6_e}}=\frac{1}{N_{12}^{\mathbb{S}_{3_g6_g}}}(J_{n1}^+J_{n2}^-J_{b1}^-J_{b2}^+-J_{n1}^-J_{n2}^+J_{b1}^+J_{b2}^-),\\
\nonumber
\Gamma_{12}^{\mathbb{S}_{3_e6_g}}&=\Gamma_{12}^{\mathbb{S}_{3_e6_e}}=\frac{1}{N_{12}^{\mathbb{S}_{3_e6_g}}}(J_{n1}^+J_{n2}^-J_{b2}^-J_{b3}^+-J_{n1}^-J_{n2}^+J_{b2}^+J_{b3}^-),\\
\nonumber
\Gamma_{45}^{\mathbb{S}_{3_g6_g}}&=\Gamma_{45}^{\mathbb{S}_{3_e6_e}}=0,\\
\Gamma_{45}^{\mathbb{S}_{3_e6_g}}&=-\Gamma_{45}^{\mathbb{S}_{3_g6_e}}=-\frac{1}{N_{45}^{\mathbb{S}_{3_g6_e}}}(J_{b1}^+J_{b2}^-J_{b2}^-J_{b3}^+-J_{b1}^-J_{b2}^+J_{b2}^+J_{b3}^-).
\end{align}
The heat currents are
\begin{align}
\nonumber
\dot{Q}_1^{\mathbb{S}_{3_g6_g}}&=\dot{Q}_1^{\mathbb{S}_{3_g6_e}}=-\dot{Q}_2^{\mathbb{S}_{3_g6_g}}=-\dot{Q}_2^{\mathbb{S}_{3_g6_e}},\\
\nonumber
\dot{Q}_1^{\mathbb{S}_{3_e6_g}}&=\dot{Q}_1^{\mathbb{S}_{3_e6_e}}=-\dot{Q}_2^{\mathbb{S}_{3_e6_g}}=-\dot{Q}_2^{\mathbb{S}_{3_e6_e}},\\
\nonumber
\dot{Q}_4^{\mathbb{S}_{3_g6_g}}&=-\dot{Q}_4^{\mathbb{S}_{3_e6_e}}=-\dot{Q}_5^{\mathbb{S}_{3_g6_g}}=\dot{Q}_5^{\mathbb{S}_{3_e6_e}}=0,\\
\dot{Q}_4^{\mathbb{S}_{3_g6_e}}&=-\dot{Q}_4^{\mathbb{S}_{3_e6_g}}=-\dot{Q}_5^{\mathbb{S}_{3_g6_e}}=\dot{Q}_5^{\mathbb{S}_{3_e6_g}}.
\end{align}
It is easy to understand that for a subchain containing a nodal spin, the heat currents are not the same and depend on the state of the third spin since the transitions induced by the two heat reservoirs are different. However, for the system composed entirely of bulk spins, when the third and the sixth spins are in the same state, the transitions induced by the fourth and the fifth reservoirs are the same; that is, the net transition rate in this subchain is zero. When the third and the sixth have opposite states, taking $\mathbb{S}_{3_e6_g}$ as an example, the transition channel between the fourth reservoir and the system contains two channels with frequencies $B+2J$ and $B$, while the two frequencies induced by fifth reservoir are $B-2J$ and $B$, respectively, the net transition rates are disappeared, and the steady-state heat currents exit. When the spin's states on both sides of this subchain are exchanged, i.e., in the subspace $\mathbb{S}_{3_g6_e}$, the steady-state heat currents are reversed. 
Surprisingly, in a completely symmetric system consisting of two spins, namely, the magnetic field Zeeman energy of each spin is the same and the temperature of the connected local reservoir is the same. When one of the spins is connected with a perturbed spin, this system, without heat transfer at a steady state, will generate a heat current.

\section{Derivation of heat current expression}
\label{AppendixC}

Quantum master equation is an effective method to describe the dynamics of the reduced system $\rho_S(t)=\mathrm{Tr}_E\rho(t)$ in open quantum systems. According to the Liouville-von Neumann equation $\dot{\rho}(t)=-\mathrm{i}[H,\rho(t)]$,
\begin{align}
\frac{d\rho_S(t)}{dt}=-\int_0^\infty ds\mathrm{Tr}_B\left[H_I(t),[H_I(t-s),\rho_S(t)\otimes\rho_B]\right]\label{dotrho}
\end{align}
can be obtained by iterating back to the Liouville equation by formally integrating the time in the interaction picture using the Born-Markov approximation \cite{breuer2002theory}. The heat current is defined as $\dot{Q}=\frac{d\langle H_S\rangle}{dt}=\mathrm{Tr}_S[H_S\frac{d\rho_S(t)}{dt}]$, then
\begin{align}
\dot{Q}&=\mathrm{Tr}\int_0^\infty ds[[H_I(t),H_S],H_I(t-s)]\rho_S(t)\otimes\rho_B.
\end{align}
The interaction between the system and the environment can be expressed as $H_I(t)=\sum_{\mu=1}^N\sum_\omega e^{-\mathrm{i}\omega t}A_\mu(\omega)\otimes B_\mu(t)=\sum_{\mu=1}^N\sum_\omega e^{+\mathrm{i}\omega t} A^\dagger_\mu(\omega)\otimes B^\dagger_\mu(t)$ with $B_\mu(t)=e^{\mathrm{i}H_Bt}B_\mu e^{-\mathrm{i}H_Bt}$, where the eigen-operator and the corresponding eigen-frequency satisfy $A_\mu(\omega)=\sum_{\omega=\lambda^\prime-\lambda}\Pi(\lambda)A_\mu\Pi(\lambda^\prime)$ with $\Pi(\lambda)=\vert\lambda\rangle\langle\lambda\vert$ denoting the projection operator.
Therefore, the heat current can be expressed as
\begin{align}
\nonumber
\dot{Q}&=\sum_{\mu,\nu}\sum_{\omega,\omega^\prime}e^{\mathrm{i}(\omega^\prime-\omega)t}\Gamma_{\mu\nu}(\omega)\mathrm{Tr}_S[(A_\mu^\dagger(\omega^\prime)H_SA_\nu(\omega)-H_SA_\mu^\dagger(\omega^\prime)A_\nu(\omega))\rho_S(t)]\\
&+\sum_{\mu,\nu}\sum_{\omega,\omega^\prime}e^{-\mathrm{i}(\omega^\prime-\omega)t}\Gamma_{\nu\mu}^\ast(\omega)\mathrm{Tr}_S[(A_\nu^\dagger(\omega^\prime)H_SA_\mu(\omega)-A_\nu^\dagger(\omega^\prime)A_\mu(\omega)H_S)\rho_S(t)],
\end{align}
where $\Gamma_{\mu\nu}(\omega)=\int_0^\infty ds e^{\mathrm{i}\omega s}\langle B_\mu^\dagger(t)B_\nu(t-s)\rangle$ is the Fourier transform with the reservoir correlation function $\mathrm{Tr}_E[B_\mu^\dagger (t)B_\nu(t-s)\rho_B]\equiv\langle B_\mu^\dagger(t)B_\nu(t-s)\rangle=\langle B_\mu^\dagger(s)B_\nu(0)\rangle$. Using the secular approximation, there is
\begin{align}
\nonumber
\dot{Q}=\sum_{\mu,\nu}\sum_{\omega}&\Gamma_{\mu\nu}(\omega)\left(\langle A_\mu^\dagger(\omega)H_SA_\nu(\omega)\rangle-\langle H_SA_\mu^\dagger(\omega)A_\nu(\omega)\rangle\right)\\
+&\Gamma_{\nu\mu}^\ast(\omega)\left(\langle A_\nu^\dagger(\omega)H_SA_\mu(\omega)\rangle-\langle A_\nu^\dagger(\omega)A_\mu(\omega)H_S\rangle\right).
\end{align}
In terms of the correlation function, we define $\gamma_{\mu\nu}(\omega)=\Gamma_{\mu\nu}(\omega)+\Gamma_{\nu\mu}^\ast(\omega)$ and $S_{\mu\nu}(\omega)=\frac{1}{2\mathrm{i}}(\Gamma_{\mu\nu}(\omega)-\Gamma_{\nu\mu}^\ast(\omega))$, then the heat current is
\begin{small}
\begin{align}
\nonumber
\dot{Q}&=\sum_{\mu,\nu}\sum_{\omega}\frac{1}{2}\gamma_{\mu\nu}(\omega)\left(\left\langle \left[A_\mu^\dagger(\omega),H_S\right]A_\nu(\omega)\right\rangle+\left\langle A_\mu^\dagger(\omega)\left[H_S,A_\nu(\omega)\right]\right\rangle\right)\\
\nonumber
&\quad+\mathrm{i}\sum_{\mu,\nu}\sum_{\omega}S_{\mu\nu}(\omega)\left(\left\langle \left[A_\mu^\dagger(\omega),H_S\right]A_\nu(\omega)\right\rangle-\left\langle A_\mu^\dagger(\omega)\left[H_S,A_\nu(\omega)\right]\right\rangle\right)\\
&=\sum_{\mu,\nu}\sum_{\omega}\frac{1}{2}\gamma_{\mu\nu}(\omega)\left(2\langle A_\mu^\dagger(\omega)H_SA_\nu(\omega)\rangle-\langle \{H_S,A_\mu^\dagger(\omega)A_\nu(\omega)\}\rangle\right)-\mathrm{i}S_{\mu\nu}(\omega)\langle [H_S,A_\mu^\dagger(\omega)A_\nu(\omega)]\rangle.
\end{align}
\end{small}
With conditions $[H_S,A_\mu(\omega)]=-\omega A_\mu(\omega)$ and $[H_S,A_\mu^\dagger(\omega)]=\omega A_\mu^\dagger(\omega)$, there is
\begin{align}
\dot{Q}=-\sum_{\mu,\nu}\sum_{\omega}\omega\gamma_{\mu\nu}(\omega)\langle A_\mu^\dagger(\omega)A_\nu(\omega)\rangle,
\end{align}
where 
\begin{align}
\nonumber
\langle A_\mu^\dagger(\omega)A_\nu(\omega)\rangle&=\sum_\lambda\langle\lambda\vert\left[\sum_{\omega=\lambda^\prime-\lambda}\Pi(\lambda^{\prime})A_\mu^\dagger\Pi(\lambda) \sum_{\omega=\lambda^{\prime\prime\prime}-\lambda^{\prime\prime}}\Pi(\lambda^{\prime\prime})A_\nu\Pi(\lambda^{\prime\prime\prime})\rho_S(t)\right]\vert\lambda\rangle\\
\nonumber
&=\sum_\lambda\langle\lambda\vert\left[\sum_{\lambda^\prime,\lambda^{\prime\prime\prime}}\Pi(\lambda^\prime)A_\mu^\dagger A_\nu\Pi(\lambda^{\prime\prime\prime})\rho_S(t)\right]\vert\lambda\rangle\\
&=\rho_{\lambda^\prime}(\omega)
\end{align}
where $\rho_{\lambda^\prime}(\omega)$ represents the population of the excited state contained in a transition of frequency $\omega$, therefore, the heat current is
\begin{align}
\dot{Q}=-\sum_{\mu,\nu}\sum_{\omega}\omega\gamma_{\mu\nu}(\omega)\rho_{\lambda^\prime}(\omega).
\end{align}

For the Lindblad quantum master equation Eq. (\ref{BMS_ME}), the heat currents have a clearer physical form as
\begin{align}
\nonumber
\dot{Q}_\mu&=\sum_l J_\mu(-\omega_{\mu l})\mathrm{Tr_S}\big[(V_{\mu l}^\dagger H_S V_{\mu l}-\frac{1}{2}\{H_S,V_{\mu l}^\dagger V_{\mu l}\})\rho_S(t)\big]+ J_\mu(+\omega_{\mu l})\mathrm{Tr_S}\big[(V_{\mu l} H_S V_{\mu l}^\dagger-\frac{1}{2}\{H_S,V_{\mu l} V_{\mu l}^\dagger\})\rho_S(t)\big]\\
\nonumber
&=\sum_l\frac{1}{2}\big[J_\mu(-\omega_{\mu l})\big(\langle V_{\mu l}^\dagger[H_S,V_{\mu l}]\rangle-\langle[H_S,V_{\mu l}^\dagger] V_{\mu l}\rangle\big)+J_\mu(+\omega_{\mu l})\big(\langle V_{\mu l}[H_S,V_{\mu l}^\dagger]\rangle-\langle[H_S,V_{\mu l}] V_{\mu l}^\dagger\rangle\big)\big]\\
\nonumber
&=\sum_l\omega_{\mu l}[J_\mu(+\omega_{\mu l})\langle V_{\mu l}V_{\mu l}^\dagger\rangle-J_\mu(-\omega_{\mu l})\langle V_{\mu l}^\dagger V_{\mu l}\rangle]\\
\nonumber
&=-\sum_l\omega_{\mu l}[J_\mu(-\omega_{\mu l})\rho^{\mu l}_{ee}-J_\mu(+\omega_{\mu l})\rho^{\mu l}_{gg}]\\
&=-\sum_l\omega_{\mu l}\Gamma^{\mu l}_{ge},
\end{align}
where $\rho^{\mu l}_{gg}=\langle V_{\mu l}V_{\mu l}^\dagger\rangle$ and $\rho^{\mu l}_{ee}=\langle V_{\mu l}^\dagger V_{\mu l}\rangle$ denote the populations of the ground and the excited levels involved in the certain transition.

\section{Dependence of steady-state heat current on temperature $T_b$}
\label{AppendixD}

\begin{figure}
	\centering
		\includegraphics[width=18cm]{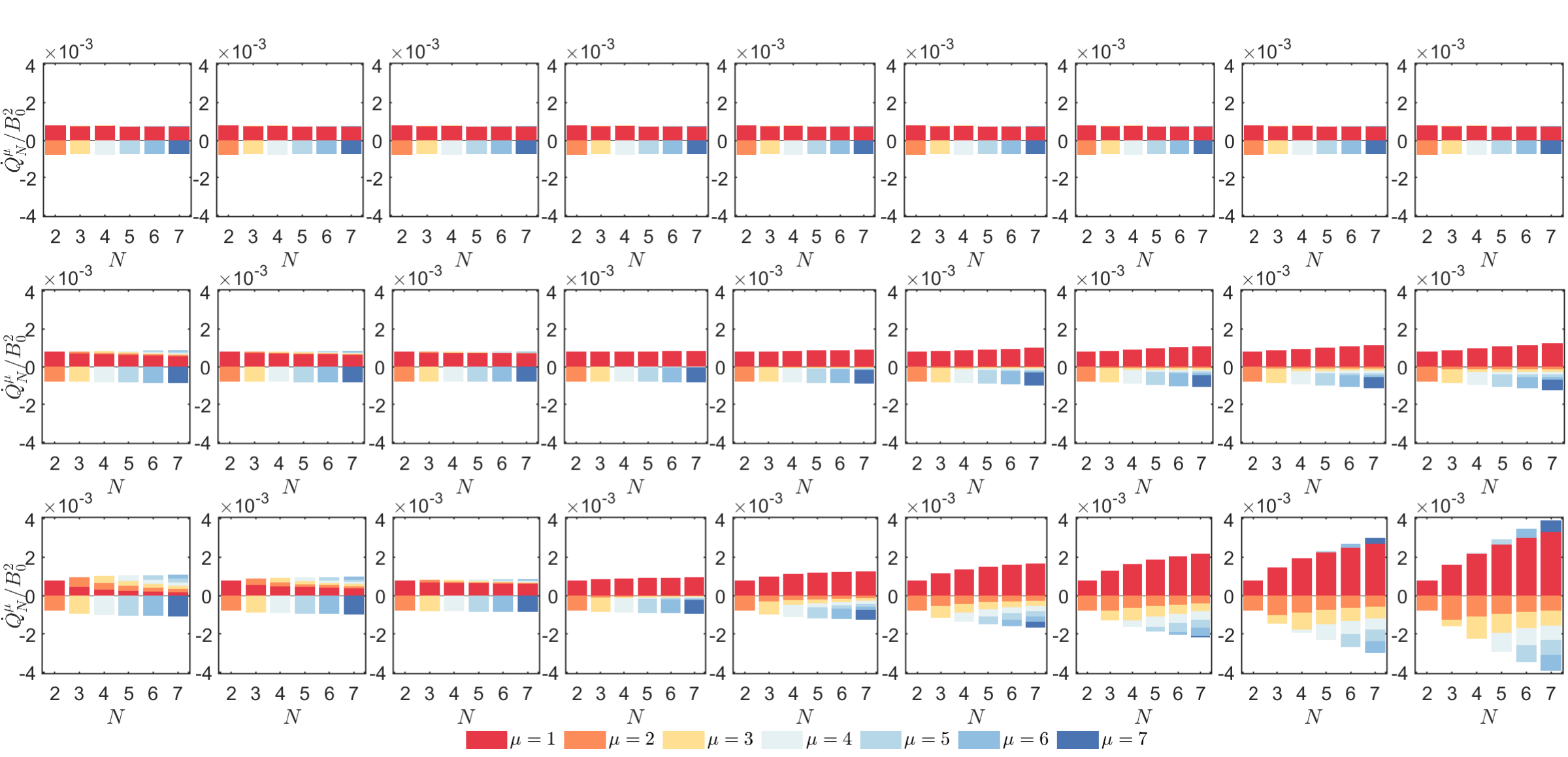}
	\caption{The heat current from the $\mu$th heat reservoir into the system at $N$-spin open chain. The left, middle, and right rows correspond to the dissipation rate of the bulk spin $\kappa_{b}=0, 0.1\kappa_1, \kappa_1$. The temperatures of the heat reservoirs connected by the bulk spins in the first to the ninth column are $T_b=10B_0,9B_0,8B_0,7B_0,6B_0,5B_0,4B_0,3B_0,2B_0$, respectively. $B_0=1$, $B_\mu=5B_0$, $J_{\mu,\mu+1}=0.1B_0$, $\kappa_1=\kappa_N=0.001B_0$, $T_1=10B_0$, $T_N=5B_0$, and $\theta=\pi/4$.}
\label{Q_N_large}
\end{figure}

This section complementally describes the influence of the heat reservoir's temperature connected to the bulk spin on the heat transfer of the whole spin chain. In Fig. \ref{Q_N_large}, the temperatures of the reservoirs connected to the nodal spins are $T_1=10B_0$ and $T_N=10B_0$, respectively. From left to right, the temperature of the heat reservoir connected to the bulk spin is $T_b=10B_0,9B_0,8B_0,7B_0,6B_0,5B_0,4B_0,3B_0,2B_0$. From top to bottom, the dissipation rates between the bulk spins and the corresponding reservoir are $\kappa_\mu=0,0.1\kappa_1,\kappa_1$, where $\kappa_1=\kappa_N=0.001B_0$ denotes the dissipation rate of the nodal spin, in order.
When all bulk spins are not dissipated, it is evident that $T_b$ doesn't affect heat current because of $\dot{Q}_b=0$.  As the dissipation of the bulk spin increases, $T_b$ becomes very important in the heat current flow from the left to the right. 
When $T_1$ and $T_b$ are both large, the heat current that flows out through the $N$th spin flows into the chain through the remaining spins. As the difference between $T_1$ and $T_b$ increases, the heat current from the first spin into the open chain gradually increases. When the temperature of the heat reservoir connected to the bulk spin is small, the heat currents flow in from both ends and out through the other bulk spin, invalid when the bulk spins are weakly dissipated.

\section{Transitions and dynamics of 3-spin chain with different magnetic fields}
\label{AppendixE}
 
In this section, we add the eigen-systems of a 3-spin chain under different magnetic fields. "Model LLL" indicates that the magnetic fields of the three spins are LF, LF, and LF, and so on.

\textit{Model LLL}-
When the chain is entirely immersed in LF, the free Hamiltonian is $H_{S0}^{LLL}=\frac{1}{2}(B_1\sigma_1^z+B_2\sigma_2^z+B_3\sigma_3^z)$ and the eigen-system can be written as
\begin{align}
\nonumber
\vert 1\rangle&=\vert\tilde{1}\rangle, \lambda_1=\frac{1}{2}(+B_1+B_2+B_3+J_{12}+J_{23}),& \vert 2\rangle&=\vert\tilde{2}\rangle, \lambda_2=\frac{1}{2}(+B_1+B_2-B_3+J_{12}-J_{23}),\\
\nonumber
\vert 3\rangle&=\vert\tilde{3}\rangle, \lambda_3=\frac{1}{2}(+B_1-B_2+B_3-J_{12}-J_{23}),& \vert 4\rangle&=\vert\tilde{4}\rangle, \lambda_4=\frac{1}{2}(+B_1-B_2-B_3-J_{12}+J_{23}),\\
\nonumber
\vert 5\rangle&=\vert\tilde{5}\rangle, \lambda_5=\frac{1}{2}(-B_1+B_2+B_3-J_{12}+J_{23}),& \vert 6\rangle&=\vert\tilde{6}\rangle, \lambda_6=\frac{1}{2}(-B_1+B_2-B_3-J_{12}-J_{23}),\\
\vert 7\rangle&=\vert\tilde{7}\rangle, \lambda_7=\frac{1}{2}(-B_1-B_2+B_3+J_{12}-J_{23}),& \vert 8\rangle&=\vert\tilde{8}\rangle, \lambda_8=\frac{1}{2}(-B_1-B_2-B_3+J_{12}+J_{23}).
\end{align}
The eigen-operators and the corresponding frequencies are
\begin{align}
\nonumber
&V_{11}=\vert 5\rangle\langle 1\vert+\vert 6\rangle\langle 2\vert, &\omega_{11}&=B_1+J_{12},
\quad&V_{12}&=\vert 7\rangle\langle 3\vert+\vert 8\rangle\langle 4\vert, &\omega_{12}&=B_1-J_{12},\\
\nonumber
&V_{21}=\vert 3\rangle\langle 1\vert, &\omega_{21}&=B_2+J_{12}+J_{23},
\quad&V_{22}&=\vert 4\rangle\langle 2\vert,\quad&\omega_{22}&=B_2+J_{12}-J_{23},\\
\nonumber
&V_{23}=\vert 7\rangle\langle 5\vert, &\omega_{23}&=B_2-J_{12}+J_{23},  
 &V_{24}&=\vert 8\rangle\langle 6\vert, &\omega_{24}&=B_2-J_{12}-J_{23}, \\
&V_{31}=\vert 2\rangle\langle 1\vert+\vert 6\rangle\langle 5\vert, &\omega_{31}&=B_3+J_{23},
 &V_{32}&=\vert 4\rangle\langle 3\vert+\vert 8\rangle\langle 7\vert, &\omega_{31}&=B_3-J_{23}.
\end{align}
%

\textit{Model TLL}-When the first spin is in TF, the eigen-system is
\begin{align}
\nonumber
&\vert 1\rangle=+\Lambda_1\vert\tilde{1}\rangle+\Lambda_2\vert\tilde{5}\rangle,&\lambda_1=\frac{1}{2}(+\Omega_++B_2+B_3), &\vert 2\rangle=+\Lambda_1\vert\tilde{2}\rangle+\Lambda_2\vert\tilde{6}\rangle,&\lambda_2=\frac{1}{2}(+\Omega_-+B_2-B_3),\\
\nonumber
&\vert 3\rangle=+\Lambda_2\vert\tilde{3}\rangle+\Lambda_1\vert\tilde{7}\rangle,&\lambda_3=\frac{1}{2}(+\Omega_--B_2+B_3), &\vert 4\rangle=+\Lambda_2\vert\tilde{4}\rangle+\Lambda_1\vert\tilde{8}\rangle,&\lambda_4=\frac{1}{2}(+\Omega_+-B_2-B_3),\\
\nonumber
&\vert 5\rangle=-\Lambda_2\vert\tilde{1}\rangle+\Lambda_1\vert\tilde{5}\rangle,&\lambda_5=\frac{1}{2}(-\Omega_-+B_2+B_3), &\vert 6\rangle=-\Lambda_2\vert\tilde{2}\rangle+\Lambda_1\vert\tilde{6}\rangle,&\lambda_6=\frac{1}{2}(-\Omega_++B_2-B_3),\\
&\vert 7\rangle=-\Lambda_1\vert\tilde{3}\rangle+\Lambda_2\vert\tilde{7}\rangle,&\lambda_7=\frac{1}{2}(-\Omega_+-B_2+B_3), &\vert 8\rangle=-\Lambda_1\vert\tilde{4}\rangle+\Lambda_2\vert\tilde{8}\rangle,&\lambda_8=\frac{1}{2}(-\Omega_--B_2-B_3),
\end{align}
where $\Omega_\pm=\sqrt{B_1^2+J_{12}^2}\pm J_{23}$, $\Lambda_1=\frac{\sqrt{B_1^2+J_{12}^2}+J_{12}}{\sqrt{B_1^2+(\sqrt{B_1^2+J_{12}^2}+J_{12})^2}}$, and $\Lambda_2=\sqrt{1-\Lambda_1^2}$.
The eigen-operators and the corresponding frequencies are
\begin{align}
\nonumber
&V_1=A_2(\vert 5\rangle\langle 1\vert+\vert 6\rangle\langle 2\vert-\vert 7\rangle\langle 3\vert-\vert 8\rangle\langle 4\vert), &\omega_1&=\sqrt{B_1^2+J_{12}^2},\\
\nonumber
&V_{21}=A_1(\vert 3\rangle\langle 1\vert+\vert 7\rangle\langle 5\vert), &\omega_{21}&=B_2+J_{23}
&V_{22}&=A_1(\vert 4\rangle\langle 2\vert+\vert 8\rangle\langle 6\vert), &\omega_{22}&=B_2-J_{23},\\
\nonumber
&V_{23}=-A_2\vert 7\rangle\langle 1\vert, &\omega_{23}&=\Omega_++B_2,
&V_{24}&=A_2\vert 6\rangle\langle 4\vert, &\omega_{24}&=\Omega_+-B_2,\\
\nonumber
&V_{25}=-A_2\vert 8\rangle\langle 2\vert, &\omega_{25}&=\Omega_-+B_2,
&V_{26}&=A_2\vert 5\rangle\langle 3\vert, &\omega_{26}&=\Omega_--B_2,\\\
&V_{31}=\vert 2\rangle\langle 1\vert+\vert 6\rangle\langle 5\vert, &\omega_{31}&=B_3+J_{23},
&V_{32}&=\vert 4\rangle\langle 3\vert+\vert 8\rangle\langle 7\vert, &\omega_{32}&=B_3-J_{23},
\end{align}
where $A_1=2\Lambda_1\Lambda_2$ and $A_2=\Lambda_1^2-\Lambda_2^2$.

\textit{Model LTL}-When the second spin immersed in TF, the eigenvalues and the corresponding eigenlevels are
\begin{align}
\nonumber
\vert 1\rangle=+\Lambda_{11}\vert\tilde{1}\rangle+\Lambda_{12}\vert\tilde{3}\rangle, \lambda_1=\frac{1}{2}(+B_1+\Omega_++B_3), \vert 2\rangle=+\Lambda_{21}\vert\tilde{2}\rangle+\Lambda_{22}\vert\tilde{4}\rangle, \lambda_2=\frac{1}{2}(+B_1+\Omega_--B_3),\\
\nonumber
\vert 3\rangle=-\Lambda_{12}\vert\tilde{1}\rangle+\Lambda_{11}\vert\tilde{3}\rangle, \lambda_3=\frac{1}{2}(+B_1-\Omega_++B_3), \vert 4\rangle=-\Lambda_{22}\vert\tilde{2}\rangle+\Lambda_{21}\vert\tilde{4}\rangle, \lambda_4=\frac{1}{2}(+B_1-\Omega_--B_3),\\
\nonumber
\vert 5\rangle=+\Lambda_{22}\vert\tilde{5}\rangle+\Lambda_{21}\vert\tilde{7}\rangle, \lambda_5=\frac{1}{2}(-B_1+\Omega_-+B_3), \vert 6\rangle=+\Lambda_{12}\vert\tilde{6}\rangle+\Lambda_{11}\vert\tilde{8}\rangle, \lambda_6=\frac{1}{2}(-B_1+\Omega_+-B_3),\\
\vert 7\rangle=-\Lambda_{21}\vert\tilde{5}\rangle+\Lambda_{22}\vert\tilde{7}\rangle, \lambda_7=\frac{1}{2}(-B_1-\Omega_-+B_3), \vert 8\rangle=-\Lambda_{11}\vert\tilde{6}\rangle+\Lambda_{12}\vert\tilde{8}\rangle, \lambda_8=\frac{1}{2}(-B_1-\Omega_+-B_3),
\end{align}
where $\Omega_\pm=\sqrt{B_2^2+(J_{12}\pm J_{23})^2}$, $\Lambda_{11}=\frac{\Omega_++J_{12}+J_{23}}{\sqrt{B_2^2+(\Omega_++J12_+J_{23})^2}}$, $\Lambda_{12}=\sqrt{1-\Lambda_{11}^2}$, $\Lambda_{21}=\frac{\Omega_-+J_{12}-J_{23}}{\sqrt{B_2^2+(\Omega_-+J12_-J_{23})^2}}$, and $\Lambda_{22}=\sqrt{1-\Lambda_{21}^2}$.
The eigen-operators and the corresponding eigen-frequencies are
\begin{align}
\nonumber
&V_{11}=+A_1(\vert 5\rangle\langle 1\vert+\vert 8\rangle\langle 4\vert),&\omega_{11}&=\frac{1}{2}(2B_1-\Omega_-+\Omega_+),\\
\nonumber
&V_{12}=-A_4(\vert 7\rangle\langle 1\vert+\vert 8\rangle\langle 2\vert),&\omega_{12}&=\frac{1}{2}(2B_1+\Omega_-+\Omega_+),\\
\nonumber
&V_{13}=+A_1(\vert 6\rangle\langle 2\vert+\vert 7\rangle\langle 3\vert),&\omega_{13}&=\frac{1}{2}(2B_1+\Omega_--\Omega_+),\\
\nonumber
&V_{14}=+A_4(\vert 5\rangle\langle 3\vert+\vert 6\rangle\langle 4\vert),&\omega_{14}&=\frac{1}{2}(2B_1-\Omega_--\Omega_+),\\
\nonumber
&V_{21}=+A_5(\vert 3\rangle\langle 1\vert-\vert 8\rangle\langle 6\vert),&\omega_{21}&=\Omega_+,\\
\nonumber
&V_{22}=+A_6(\vert 4\rangle\langle 2\vert-\vert 7\rangle\langle 5\vert),&\omega_{22}&=\Omega_-,\\
\nonumber
&V_{31}=+A_3(\vert 2\rangle\langle 1\vert+\vert 8\rangle\langle 7\vert),&\omega_{31}&=\frac{1}{2}(+2B_3-\Omega_-+\Omega_+),\\
\nonumber
&V_{32}=-A_2(\vert 4\rangle\langle 1\vert+\vert 8\rangle\langle 5\vert),&\omega_{32}&=\frac{1}{2}(+2B_3+\Omega_-+\Omega_+),\\
\nonumber
&V_{33}=+A_2(\vert 3\rangle\langle 2\vert+\vert 7\rangle\langle 6\vert),&\omega_{33}&=\frac{1}{2}(-2B_3+\Omega_-+\Omega_+),\\
&V_{34}=+A_3(\vert 4\rangle\langle 3\vert+\vert 6\rangle\langle 5\vert),&\omega_{34}&=\frac{1}{2}(+2B_3+\Omega_--\Omega_+),
\end{align}
where $A_1=\Lambda_{11}\Lambda_{22}+\Lambda_{12}\Lambda_{21}$, $A_2=\Lambda_{11}\Lambda_{22}-\Lambda_{12}\Lambda_{21}$, $A_3=\Lambda_{11}\Lambda_{21}+\Lambda_{12}\Lambda_{22}$, $A_4=\Lambda_{11}\Lambda_{21}-\Lambda_{12}\Lambda_{22}$, $A_5=\Lambda_{11}^2-\Lambda_{12}^2$, and $A_6=\Lambda_{21}^2-\Lambda_{22}^2$.

\textit{Model LLT}-When the third spin is immersed in TF, the eigen-system is  
\begin{align}
\nonumber
&\vert 1\rangle=+\Lambda_1\vert\tilde{1}\rangle+\Lambda_2\vert\tilde{2}\rangle,&\lambda_1=\frac{1}{2}(+B_1+B_2+\Omega_+),\quad&\vert 2\rangle=-\Lambda_2\vert\tilde{1}\rangle+\Lambda_1\vert\tilde{2}\rangle,&\lambda_2=\frac{1}{2}(+B_1+B_2-\Omega_-),\\
\nonumber
&\vert 3\rangle=+\Lambda_2\vert\tilde{3}\rangle+\Lambda_1\vert\tilde{4}\rangle,&\lambda_3=\frac{1}{2}(+B_1-B_2+\Omega_-),\quad&\vert 4\rangle=-\Lambda_1\vert\tilde{3}\rangle+\Lambda_2\vert\tilde{4}\rangle,&\lambda_4=\frac{1}{2}(+B_1-B_2-\Omega_+),\\
\nonumber
&\vert 5\rangle=+\Lambda_1\vert\tilde{5}\rangle+\Lambda_2\vert\tilde{6}\rangle,&\lambda_5=\frac{1}{2}(-B_1+B_2+\Omega_-),\quad&\vert 6\rangle=-\Lambda_2\vert\tilde{5}\rangle+\Lambda_1\vert\tilde{6}\rangle,&\lambda_6=\frac{1}{2}(-B_1+B_2-\Omega_+),\\
&\vert 7\rangle=+\Lambda_2\vert\tilde{7}\rangle+\Lambda_1\vert\tilde{8}\rangle,&\lambda_7=\frac{1}{2}(-B_1-B_2+\Omega_+),\quad&\vert 8\rangle=-\Lambda_1\vert\tilde{7}\rangle+\Lambda_2\vert\tilde{8}\rangle,&\lambda_8=\frac{1}{2}(-B_1-B_2-\Omega_-),
\end{align}
where $\Omega_\pm=\sqrt{B_3^2+J_{23}^2}\pm J_{12}$, $\Lambda_1=\frac{\sqrt{B_3^2+J_{23}^2}+J_{23}}{\sqrt{B_3^2+(\sqrt{B_3^2+J_{23}^2}+J_{23})^2}}$, and $\Lambda_2=\sqrt{1-\Lambda_1^2}$.
The eigen-operators and the corresponding frequencies are
\begin{footnotesize}
\begin{align}
\nonumber
&V_{11}=\vert 5\rangle\langle 1\vert+\vert 6\rangle\langle 2\vert, &\omega_{11}&=B_1+J_{12},
&V_{12}&=\vert 7\rangle\langle 3\vert+\vert 8\rangle\langle 4\vert, &\omega_{12}&=B_1-J_{12},\\
\nonumber
&V_{21}=+A_1(\vert 3\rangle\langle 1\vert+\vert 4\rangle\langle 2\vert), &\omega_{21}&=B_2+J_{12},
&V_{22}&=A_1(\vert 7\rangle\langle 5\vert+\vert 8\rangle\langle 6\vert), &\omega_{22}&=B_2-J_{12},\\
\nonumber
&V_{23}=-A_2\vert 4\rangle\langle 1\vert, &\omega_{23}&=B_2+\Omega_+,
&V_{24}&=A_2\vert 3\rangle\langle 2\vert, &\omega_{24}&=B_2+\Omega_-,\\
\nonumber
&V_{25}=-A_2\vert 8\rangle\langle 5\vert, &\omega_{25}&=B_2-\Omega_-,
&V_{26}&=A_2\vert 7\rangle\langle 6\vert, &\omega_{26}&=B_2-\Omega_+,\\
&V_3=+A_2(\vert 2\rangle\langle 1\vert-\vert 4\rangle\langle 3\vert+\vert 6\rangle\langle 5\vert-\vert 8\rangle\langle 7\vert), &\omega_3&=\sqrt{B_3^2+J_{23}^2},
\end{align}
\end{footnotesize}
where $A_1=2\Lambda_1\Lambda_2$ and $A_2=\Lambda_1^2-\Lambda_2^2$.

\textit{Model TTL}-When the first and second spins are in TF, the eigen-system is
\begin{align}
\nonumber
\vert 1\rangle&=+\Lambda_{11}\vert\tilde{1}\rangle+\Lambda_{12}\vert\tilde{3}\rangle+\Lambda_{13}\vert\tilde{5}\rangle+\Lambda_{14}\vert\tilde{7}\rangle,\lambda_1=\frac{1}{2}(+B_3+\Omega_+),\\
\nonumber
\vert 2\rangle&=-\Lambda_{21}\vert\tilde{1}\rangle-\Lambda_{22}\vert\tilde{3}\rangle+\Lambda_{23}\vert\tilde{5}\rangle+\Lambda_{24}\vert\tilde{7}\rangle,\lambda_2=\frac{1}{2}(+B_3+\Omega_-),\\
\nonumber
\vert 3\rangle&=-\Lambda_{22}\vert\tilde{1}\rangle+\Lambda_{21}\vert\tilde{3}\rangle-\Lambda_{24}\vert\tilde{5}\rangle+\Lambda_{23}\vert\tilde{7}\rangle,\lambda_3=\frac{1}{2}(+B_3-\Omega_-),\\
\nonumber
\vert 4\rangle&=+\Lambda_{12}\vert\tilde{1}\rangle-\Lambda_{11}\vert\tilde{3}\rangle-\Lambda_{14}\vert\tilde{5}\rangle+\Lambda_{13}\vert\tilde{7}\rangle,\lambda_4=\frac{1}{2}(+B_3-\Omega_+),\\
\nonumber
\vert 5\rangle&=+\Lambda_{14}\vert\tilde{2}\rangle+\Lambda_{13}\vert\tilde{4}\rangle+\Lambda_{12}\vert\tilde{6}\rangle+\Lambda_{11}\vert\tilde{8}\rangle,\lambda_5=\frac{1}{2}(-B_3+\Omega_+),\\
\nonumber
\vert 6\rangle&=-\Lambda_{24}\vert\tilde{2}\rangle-\Lambda_{23}\vert\tilde{4}\rangle+\Lambda_{22}\vert\tilde{6}\rangle+\Lambda_{21}\vert\tilde{8}\rangle,\lambda_6=\frac{1}{2}(-B_3+\Omega_-),\\
\nonumber
\vert 7\rangle&=-\Lambda_{23}\vert\tilde{2}\rangle+\Lambda_{24}\vert\tilde{4}\rangle-\Lambda_{21}\vert\tilde{6}\rangle+\Lambda_{22}\vert\tilde{8}\rangle,\lambda_7=\frac{1}{2}(-B_3-\Omega_-),\\
\vert 8\rangle&=+\Lambda_{13}\vert\tilde{2}\rangle-\Lambda_{14}\vert\tilde{4}\rangle-\Lambda_{11}\vert\tilde{6}\rangle+\Lambda_{12}\vert\tilde{8}\rangle,\lambda_8=\frac{1}{2}(-B_3-\Omega_+),
\end{align}
where $\Omega_\pm=\sqrt{B_1^2+B_2^2+J_{12}^2+J_{23}^2\pm 2\sqrt{B_1^2B_2^2+J_{23}^2(B_1^2+J_{12}^2)}}$. The specific expressions of $\Lambda_{11}$ and others are too verbose and are not given here. In this case, the eigen-operators and the corresponding eigen-frequencies are
\begin{footnotesize}
\begin{align}
\nonumber
&V_{11}=(+\Lambda_{11}\Lambda_{23}+\Lambda_{12}\Lambda_{24}-\Lambda_{13}\Lambda_{21}-\Lambda_{14}\Lambda_{22})(\vert 2\rangle\langle 1\vert-\vert 4\rangle\langle 3\vert-\vert 6\rangle\langle 5\vert+\vert 8\rangle\langle 7\vert), &\omega_{11}&=\frac{1}{2}(\Omega_+-\Omega_-),\\
\nonumber
&V_{12}=(-\Lambda_{11}\Lambda_{24}+\Lambda_{12}\Lambda_{23}-\Lambda_{13}\Lambda_{22}+\Lambda_{14}\Lambda_{21})(\vert 3\rangle\langle 1\vert+\vert 4\rangle\langle 2\vert-\vert 7\rangle\langle 5\vert-\vert 8\rangle\langle 6\vert), &\omega_{12}&=\frac{1}{2}(\Omega_++\Omega_-),\\
\nonumber
&V_{13}=2(-\Lambda_{11}\Lambda_{14}+\Lambda_{12}\Lambda_{13})(\vert 4\rangle\langle 1\vert+\vert 8\rangle\langle 5\vert), &\omega_{13}&=\Omega_+,\\
\nonumber
&V_{14}=2(+\Lambda_{21}\Lambda_{24}-\Lambda_{22}\Lambda_{23})(\vert 3\rangle\langle 2\vert+\vert 7\rangle\langle 6\vert), &\omega_{14}&=\Omega_-,\\
\nonumber
&V_{21}=(+\Lambda_{11}\Lambda_{22}-\Lambda_{12}\Lambda_{21}+\Lambda_{13}\Lambda_{24}+\Lambda_{14}\Lambda_{23})(\vert 2\rangle\langle 1\vert-\vert 4\rangle\langle 3\vert-\vert 6\rangle\langle 5\vert+\vert 8\rangle\langle 7\vert), &\omega_{21}&=\frac{1}{2}(\Omega_+-\Omega_-),\\
\nonumber
&V_{22}=(+\Lambda_{11}\Lambda_{21}-\Lambda_{12}\Lambda_{22}+\Lambda_{13}\Lambda_{23}-\Lambda_{14}\Lambda_{24})(\vert 3\rangle\langle 1\vert+\vert 4\rangle\langle 2\vert-\vert 7\rangle\langle 5\vert-\vert 8\rangle\langle 6\vert), &\omega_{22}&=\frac{1}{2}(\Omega_++\Omega_-),\\
\nonumber
&V_{23}=(-\Lambda_{11}^2+\Lambda_{12}^2+\Lambda_{13}^2-\Lambda_{14}^2)(\vert 4\rangle\langle 1\vert+\vert 8\rangle\langle 5\vert), &\omega_{23}&=\Omega_+,\\
\nonumber
&V_{24}=(-\Lambda_{21}^2+\Lambda_{22}^2+\Lambda_{23}^2-\Lambda_{24}^2)(\vert 3\rangle\langle 2\vert+\vert 7\rangle\langle 6\vert), &\omega_{24}&=\Omega_-,\\
\nonumber
&V_{31}=(-\Lambda_{11}\Lambda_{24}-\Lambda_{12}\Lambda_{23}+\Lambda_{13}\Lambda_{22}+\Lambda_{14}\Lambda_{21})(+\vert 6\rangle\langle 1\vert-\vert 8\rangle\langle 3\vert), &\omega_{31}&=B_3+\frac{1}{2}(\Omega_+-\Omega_-),\\
\nonumber
&V_{32}=(+\Lambda_{11}\Lambda_{24}+\Lambda_{12}\Lambda_{23}-\Lambda_{13}\Lambda_{22}-\Lambda_{14}\Lambda_{21})(+\vert 5\rangle\langle 2\vert-\vert 7\rangle\langle 4\vert), &\omega_{32}&=B_3-\frac{1}{2}(\Omega_+-\Omega_-),\\
\nonumber
&V_{33}=(-\Lambda_{11}\Lambda_{23}+\Lambda_{12}\Lambda_{24}-\Lambda_{13}\Lambda_{21}+\Lambda_{14}\Lambda_{22})(+\vert 7\rangle\langle 1\vert-\vert 8\rangle\langle 2\vert), &\omega_{33}&=B_3+\frac{1}{2}(\Omega_++\Omega_-),\\
\nonumber
&V_{34}=(+\Lambda_{11}\Lambda_{23}-\Lambda_{12}\Lambda_{24}+\Lambda_{13}\Lambda_{21}-\Lambda_{14}\Lambda_{22})(+\vert 5\rangle\langle 3\vert+\vert 6\rangle\langle 4\vert), &\omega_{34}&=B_3-\frac{1}{2}(\Omega_++\Omega_-),\\
&V_{35}=2[(\Lambda_{11}\Lambda_{14}+\Lambda_{12}\Lambda_{13})(\vert 5\rangle\langle 1\vert+\vert 8\rangle\langle 4\vert)+(\Lambda_{21}\Lambda_{24}+\Lambda_{22}\Lambda_{23})(\vert 6\rangle\langle 2\vert+\vert 7\rangle\langle 3\vert)], &\omega_{35}&=B_3.
\end{align}
\end{footnotesize}
It's worth noting that the transition channels induced by the first and second heat reservoirs are precisely the same.

\textit{Model $TLT$}-When the first and third spins are in TF, the eigen-system is
\begin{align}
\nonumber
&\vert 1\rangle=+\Lambda_1\vert\tilde{1}\rangle+\Lambda_2\vert\tilde{2}\rangle+\Lambda_3\vert\tilde{5}\rangle+\Lambda_4\vert\tilde{6}\rangle,\lambda_1=\frac{1}{2}(+B_2+\Omega_+),\\
\nonumber
&\vert 2\rangle=+\Lambda_2\vert\tilde{1}\rangle-\Lambda_2\vert\tilde{2}\rangle+\Lambda_4\vert\tilde{5}\rangle-\Lambda_3\vert\tilde{6}\rangle,\lambda_2=\frac{1}{2}(+B_2-\Omega_-),\\
\nonumber
&\vert 3\rangle=+\Lambda_4\vert\tilde{3}\rangle+\Lambda_3\vert\tilde{4}\rangle+\Lambda_2\vert\tilde{7}\rangle+\Lambda_1\vert\tilde{8}\rangle,\lambda_3=\frac{1}{2}(-B_2+\Omega_+),\\
\nonumber
&\vert 4\rangle=+\Lambda_3\vert\tilde{3}\rangle-\Lambda_4\vert\tilde{4}\rangle+\Lambda_1\vert\tilde{7}\rangle-\Lambda_2\vert\tilde{8}\rangle,\lambda_4=\frac{1}{2}(-B_2-\Omega_-),\\
\nonumber
&\vert 5\rangle=+\Lambda_3\vert\tilde{1}\rangle+\Lambda_4\vert\tilde{2}\rangle-\Lambda_1\vert\tilde{5}\rangle-\Lambda_2\vert\tilde{6}\rangle,\lambda_5=\frac{1}{2}(+B_2+\Omega_-),\\
\nonumber
&\vert 6\rangle=+\Lambda_4\vert\tilde{1}\rangle-\Lambda_3\vert\tilde{2}\rangle-\Lambda_2\vert\tilde{5}\rangle+\Lambda_1\vert\tilde{6}\rangle,\lambda_6=\frac{1}{2}(+B_2-\Omega_+),\\
\nonumber
&\vert 7\rangle=+\Lambda_2\vert\tilde{3}\rangle+\Lambda_1\vert\tilde{4}\rangle-\Lambda_4\vert\tilde{7}\rangle-\Lambda_3\vert\tilde{8}\rangle,\lambda_7=\frac{1}{2}(-B_2+\Omega_-),\\
&\vert 8\rangle=+\Lambda_1\vert\tilde{3}\rangle-\Lambda_2\vert\tilde{4}\rangle-\Lambda_3\vert\tilde{7}\rangle+\Lambda_4\vert\tilde{8}\rangle,\lambda_8=\frac{1}{2}(-B_2-\Omega_+),
\end{align}
where $\Omega_\pm=\sqrt{B_1^2+B_3^2+J_{12}^2+J_{23}^2\pm 2\sqrt{(B_1^2+J_{12}^2)(B_3^2+J_{23}^2)}}$, and $\Lambda_i,i=1,\cdots,4$ is quite difficult and isn't expressed here. The eigen-operators and the corresponding frequencies are
\begin{align}
\nonumber
&V_1=-A_1(\vert 5\rangle\langle 1\vert+\vert 6\rangle\langle 2\vert-\vert 7\rangle\langle 3\vert-\vert 8\rangle\langle 4\vert),\quad &\omega_1&=\frac{1}{2}(\Omega_+-\Omega_-),\\
\nonumber
&V_{21}=+A_4(\vert 3\rangle\langle 1\vert+\vert 4\rangle\langle 2\vert+\vert 7\rangle\langle 5\vert+\vert 8\rangle\langle 6\vert),\quad &\omega_{21}&=B_2,\\
\nonumber
&V_{22}=+A_5(\vert 4\rangle\langle 1\vert+\vert 8\rangle\langle 5\vert),\quad &\omega_{22}&=+B_2+\frac{1}{2}(\Omega_++\Omega_-),\\
\nonumber
&V_{23}=+A_2\vert 8\rangle\langle 1\vert,\quad&\omega_{23}&=B_2+\Omega_+,\\
\nonumber
&V_{24}=-A_5(\vert 3\rangle\langle 2\vert+\vert 7\rangle\langle 6\vert),\quad &\omega_{24}&=+B_2-\frac{1}{2}(\Omega_++\Omega_-),\\
\nonumber
& V_{25}=-A_2\vert 7\rangle\langle 2\vert,\quad&\omega_{25}&=B_2-\Omega_-,\\
\nonumber
&V_{26}=+A_6(\vert 1\rangle\langle 7\vert+\vert 2\rangle\langle 8\vert),\quad &\omega_{26}&=+B_2+\frac{1}{2}(\Omega_+-\Omega_-),\\
\nonumber
&V_{27}=+A_2\vert 6\rangle\langle 3\vert,\quad&\omega_{27}&=-B_2+\Omega_+,\\
\nonumber
&V_{28}=-A_6(\vert 5\rangle\langle 3\vert+\vert 6\rangle\langle 4\vert),\quad &\omega_{28}&=-B_2+\frac{1}{2}(\Omega_+-\Omega_-),\\
\nonumber
&V_{29}=-A_2\vert 5\rangle\langle 4\vert,\quad&\omega_{29}&=-B_2-\Omega_-,\\
&V_3=-A_3(\vert 2\rangle\langle 1\vert-\vert 4\rangle\langle 3\vert+\vert 6\rangle\langle 5\vert+\vert 8\rangle\langle 7\vert),\quad &\omega_3&=\frac{1}{2}(\Omega_++\Omega_-),
\end{align}
where $A_1=\Lambda_1^2+\Lambda_2^2-\Lambda_3^2-\Lambda_4^2$, $A_2=\Lambda_1^2-\Lambda_2^2-\Lambda_3^2+\Lambda_4^2$, $A_3=\Lambda_1^2-\Lambda_2^2+\Lambda_3^2-\Lambda_4^2$, $A_4=2(\Lambda_1\Lambda_4+\Lambda_2\Lambda_3)$,  $A_5=2(\Lambda_1\Lambda_3-\Lambda_2\Lambda_4)$, and $A_6=2(\Lambda_1\Lambda_2+\Lambda_3\Lambda_4)$.

\textit{Model $LTT$}-When the second and the third spins are immersed in TF, the eigen-system is
\begin{align}
\nonumber
\vert 1\rangle&=+\Lambda_{11}\vert\tilde{1}\rangle+\Lambda_{12}\vert\tilde{2}\rangle+\Lambda_{13}\vert\tilde{3}\rangle+\Lambda_{14}\vert\tilde{4}\rangle,\lambda_1=\frac{1}{2}(+B_1+\Omega_+),\\
\nonumber
\vert 2\rangle&=-\Lambda_{21}\vert\tilde{1}\rangle-\Lambda_{22}\vert\tilde{2}\rangle+\Lambda_{23}\vert\tilde{3}\rangle+\Lambda_{24}\vert\tilde{4}\rangle,\lambda_2=\frac{1}{2}(+B_1+\Omega_-),\\
\nonumber
\vert 3\rangle&=-\Lambda_{23}\vert\tilde{1}\rangle+\Lambda_{24}\vert\tilde{2}\rangle-\Lambda_{21}\vert\tilde{3}\rangle+\Lambda_{22}\vert\tilde{4}\rangle,\lambda_3=\frac{1}{2}(+B_1-\Omega_-),\\
\nonumber
\vert 4\rangle&=+\Lambda_{13}\vert\tilde{1}\rangle-\Lambda_{14}\vert\tilde{2}\rangle-\Lambda_{11}\vert\tilde{3}\rangle+\Lambda_{12}\vert\tilde{4}\rangle,\lambda_4=\frac{1}{2}(+B_1-\Omega_+),\\
\nonumber
\vert 5\rangle&=+\Lambda_{14}\vert\tilde{5}\rangle+\Lambda_{13}\vert\tilde{6}\rangle+\Lambda_{12}\vert\tilde{7}\rangle+\Lambda_{11}\vert\tilde{8}\rangle,\lambda_5=\frac{1}{2}(-B_1+\Omega_+),\\
\nonumber
\vert 6\rangle&=-\Lambda_{24}\vert\tilde{5}\rangle-\Lambda_{23}\vert\tilde{6}\rangle+\Lambda_{22}\vert\tilde{7}\rangle+\Lambda_{21}\vert\tilde{8}\rangle,\lambda_6=\frac{1}{2}(-B_1+\Omega_-),\\
\nonumber
\vert 7\rangle&=-\Lambda_{22}\vert\tilde{5}\rangle+\Lambda_{21}\vert\tilde{6}\rangle-\Lambda_{24}\vert\tilde{7}\rangle+\Lambda_{23}\vert\tilde{8}\rangle,\lambda_7=\frac{1}{2}(-B_1-\Omega_-),\\
\vert 8\rangle&=+\Lambda_{12}\vert\tilde{5}\rangle-\Lambda_{11}\vert\tilde{6}\rangle-\Lambda_{14}\vert\tilde{7}\rangle+\Lambda_{13}\vert\tilde{8}\rangle,\lambda_8=\frac{1}{2}(-B_1-\Omega_+),
\end{align}
where $\Omega_\pm=\sqrt{B_2^2+B_3^2+J_{12}^2+J_{23}^2\pm 2\sqrt{B_2^2B_3^2+J_{12}^2(B_3^2+J_{23}^2)}}$. The eigen-operators and the corresponding eigen-frequencies are
\begin{align}
\nonumber
&V_{11}=2[(\Lambda_{11}\Lambda_{14}+\Lambda_{12}\Lambda_{13})(\vert 5\rangle\langle 1\vert-\vert 8\rangle\langle 4\vert)+(\Lambda_{21}\Lambda_{24}+\Lambda_{22}\Lambda_{23})(\vert 6\rangle\langle 2\vert-\vert 7\rangle\langle 3\vert)], &\omega_{11}&=B_1,\\
\nonumber
&V_{12}=(-\Lambda_{11}\Lambda_{24}-\Lambda_{12}\Lambda_{23}+\Lambda_{13}\Lambda_{22}+\Lambda_{14}\Lambda_{21})(\vert 6\rangle\langle 1\vert+\vert 8\rangle\langle 3\vert), &\omega_{12}&=B_1+\frac{1}{2}(\Omega_+-\Omega_-),\\
\nonumber
&V_{13}=(+\Lambda_{11}\Lambda_{24}+\Lambda_{12}\Lambda_{23}-\Lambda_{13}\Lambda_{22}-\Lambda_{14}\Lambda_{21})(\vert 5\rangle\langle 2\vert+\vert 7\rangle\langle 4\vert), &\omega_{13}&=B_1-\frac{1}{2}(\Omega_+-\Omega_-),\\
\nonumber
&V_{14}=(-\Lambda_{11}\Lambda_{22}+\Lambda_{12}\Lambda_{21}-\Lambda_{13}\Lambda_{24}+\Lambda_{14}\Lambda_{22})(\vert 7\rangle\langle 1\vert-\vert 8\rangle\langle 2\vert), &\omega_{14}&=B_1+\frac{1}{2}(\Omega_++\Omega_-),\\
\nonumber
&V_{15}=(+\Lambda_{11}\Lambda_{22}-\Lambda_{12}\Lambda_{21}+\Lambda_{13}\Lambda_{24}-\Lambda_{14}\Lambda_{22})(\vert 5\rangle\langle 3\vert-\vert 6\rangle\langle 4\vert), &\omega_{15}&=B_1-\frac{1}{2}(\Omega_++\Omega_-),\\
\nonumber
&V_{21}=(+\Lambda_{11}\Lambda_{23}+\Lambda_{12}\Lambda_{24}-\Lambda_{13}\Lambda_{21}-\Lambda_{14}\Lambda_{22})(\vert 2\rangle\langle 1\vert+\vert 4\rangle\langle 3\vert-\vert 6\rangle\langle 5\vert-\vert 8\rangle\langle 7\vert), &\omega_{21}&=\frac{1}{2}(\Omega_+-\Omega_-),\\
\nonumber
&V_{22}=(-\Lambda_{11}\Lambda_{21}+\Lambda_{12}\Lambda_{22}-\Lambda_{13}\Lambda_{23}+\Lambda_{14}\Lambda_{24})(\vert 3\rangle\langle 1\vert-\vert 4\rangle\langle 2\vert-\vert 7\rangle\langle 5\vert+\vert 8\rangle\langle 6\vert), &\omega_{22}&=\frac{1}{2}(\Omega_++\Omega_-),\\
\nonumber
&V_{23}=(-\Lambda_{11}^2+\Lambda_{12}^2+\Lambda_{13}^2-\Lambda_{14}^2)(\vert 4\rangle\langle 1\vert+\vert 8\rangle\langle 5\vert), &\omega_{23}&=\Omega_+,\\
\nonumber
&V_{24}=(+\Lambda_{21}^2-\Lambda_{22}^2-\Lambda_{23}^2+\Lambda_{24}^2)(\vert 3\rangle\langle 2\vert+\vert 7\rangle\langle 6\vert), &\omega_{24}&=\Omega_-,\\
\nonumber
&V_{31}=(-\Lambda_{11}\Lambda_{22}-\Lambda_{12}\Lambda_{21}+\Lambda_{13}\Lambda_{24}+\Lambda_{14}\Lambda_{23})(\vert 2\rangle\langle 1\vert+\vert 4\rangle\langle 3\vert-\vert 6\rangle\langle 5\vert-\vert 8\rangle\langle 7\vert), &\omega_{31}&=\frac{1}{2}(\Omega_+-\Omega_-),\\
\nonumber
&V_{32}=(+\Lambda_{11}\Lambda_{24}-\Lambda_{12}\Lambda_{23}+\Lambda_{13}\Lambda_{22}-\Lambda_{14}\Lambda_{21})(\vert 3\rangle\langle 1\vert-\vert 4\rangle\langle 2\vert-\vert 7\rangle\langle 5\vert+\vert 8\rangle\langle 6\vert), &\omega_{32}&=\frac{1}{2}(\Omega_++\Omega_-),\\
\nonumber
&V_{33}=2(-\Lambda_{11}\Lambda_{14}+\Lambda_{12}\Lambda_{13})(\vert 4\rangle\langle 1\vert+\vert 8\rangle\langle 5\vert), &\omega_{33}&=\Omega_+,\\
&V_{34}=2(-\Lambda_{21}\Lambda_{24}+\Lambda_{22}\Lambda_{23})(\vert 3\rangle\langle 2\vert+\vert 7\rangle\langle 6\vert), &\omega_{34}&=\Omega_-.
\end{align}
Interestingly, the transition channels induced by the second and third heat reservoirs are the same, which is the same as model $TTL$.

\begin{table}[!h]
\renewcommand{\arraystretch}{2}
\centering
\caption{Transitions and dynamics of the 3-spin chain under different magnetic fields when the 2nd spin is in the transverse field. For model TTL (or LTT), the superscript of the net transition rate in the dynamics with $\kappa_2=0$ corresponds to the part outside (or inside) brackets.} \label{tableS1}
\scalebox{1}{
\begin{tabular}{c|c|c|c|c|c|c|c}
  \hline
    \multicolumn{2}{c|}{ } & LTL & TTL & \multicolumn{2}{c|}{LTT} & \multicolumn{2}{c}{TTT} \\
  \hline
    \multirow{3}{*}{\makecell[c]{Transition\\operators}} 
       & 1st & \makecell[c]{$V_1^{15},V_1^{26},V_1^{37},V_1^{48},$\\$V_1^{17},V_1^{28},V_1^{35},V_1^{46}$}
             & \multirow{2}{*}{\makecell[c]{$V_\mu^{13},V_\mu^{24},V_\mu^{57},V_\mu^{68},$\\$V_\mu^{12},V_\mu^{34},V_\mu^{56},V_\mu^{78},$\\$V_\mu^{14},V_\mu^{23},V_\mu^{58},V_\mu^{67}$}}
             & \multicolumn{2}{c|}{\makecell[c]{$V_1^{15},V_1^{26},V_1^{37},V_1^{48},$\\$V_1^{16},V_1^{25},V_1^{38},V_1^{47},$\\$V_1^{17},V_1^{28},V_1^{35},V_1^{46}$}}
             &\multicolumn{2}{c}{\multirow{3}{*}{\makecell[c]{$V_\mu^{12},V_\mu^{13},V_\mu^{14},V_\mu^{23},$\\$V_\mu^{24},V_\mu^{34},V_\mu^{56},V_\mu^{57},$\\$V_\mu^{58},V_\mu^{67},V_\mu^{68},V_\mu^{78}$}}}\\
             \cline{2-3} \cline{5-6}
       & 2nd & $V_2^{13},V_2^{24},V_2^{57},V_2^{68}$
             &  
             & \multicolumn{2}{c|}{\multirow{2}{*}{\makecell[c]{$V_\mu^{12},V_\mu^{34},V_\mu^{56},V_\mu^{78},$\\$V_\mu^{13},V_\mu^{24},V_\mu^{57},V_\mu^{68},$\\$V_\mu^{14},V_\mu^{23},V_\mu^{58},V_\mu^{67}$}}}
             & \multicolumn{2}{c}{ }\\ 
             \cline{2-4}
       & 3rd & \makecell[c]{$V_3^{12},V_3^{34},V_3^{56},V_3^{78},$\\$V_3^{14},V_3^{23},V_3^{58},V_3^{67}$} 
             & \makecell[c]{$V_3^{15},V_3^{26},V_3^{37},V_3^{48},$\\$V_3^{16},V_3^{25},V_3^{38},V_3^{47},$\\$V_3^{17},V_3^{28},V_3^{35},V_3^{46}$} 
             & \multicolumn{2}{c|}{}
             & \multicolumn{2}{c}{}
              \\
       \hline
       \multicolumn{2}{c|}{\multirow{2}*{\makecell[c]{Dynamics\\with\\$\kappa_2=0$}}}
       &\multirow{2}{*}{\makecell[c]{$\dot{\rho}_{11}=-\Gamma_{15}^1-\Gamma_{17}^1-\Gamma_{12}^3-\Gamma_{14}^3$,\\
                                     $\dot{\rho}_{22}=-\Gamma_{26}^1-\Gamma_{28}^1+\Gamma_{12}^3-\Gamma_{23}^3$,\\
                                     $\dot{\rho}_{33}=-\Gamma_{37}^1-\Gamma_{35}^1-\Gamma_{34}^3+\Gamma_{23}^3$,\\
                                     $\dot{\rho}_{44}=-\Gamma_{48}^1-\Gamma_{46}^1+\Gamma_{34}^3+\Gamma_{14}^3$,\\
                                     $\dot{\rho}_{55}=+\Gamma_{15}^1+\Gamma_{35}^1-\Gamma_{56}^3-\Gamma_{58}^3$,\\
                                     $\dot{\rho}_{66}=+\Gamma_{26}^1+\Gamma_{46}^1+\Gamma_{56}^3-\Gamma_{67}^3$,\\
                                     $\dot{\rho}_{77}=+\Gamma_{37}^1+\Gamma_{17}^1-\Gamma_{78}^3+\Gamma_{67}^3$,\\
                                     $\dot{\rho}_{88}=+\Gamma_{48}^1+\Gamma_{28}^1+\Gamma_{78}^3+\Gamma_{58}^3$}}     
       &\multicolumn{3}{c|}{\multirow{2}*{\makecell[c]{$\dot{\rho}_{11}=-\Gamma_{15}^{1(3)}-\Gamma_{17}^{1(3)}-\Gamma_{12}^{3(1)}-\Gamma_{14}^{3(1)},$\\
                                     $\dot{\rho}_{22}=-\Gamma_{26}^{1(3)}-\Gamma_{28}^{1(3)}+\Gamma_{12}^{3(1)}-\Gamma_{23}^{3(1)},$\\
                                     $\dot{\rho}_{33}=-\Gamma_{37}^{1(3)}-\Gamma_{35}^{1(3)}-\Gamma_{34}^{3(1)}+\Gamma_{23}^{3(1)},$\\
                                     $\dot{\rho}_{44}=-\Gamma_{48}^{1(3)}-\Gamma_{46}^{1(3)}+\Gamma_{34}^{3(1)}+\Gamma_{14}^{3(1)},$\\
                                     $\dot{\rho}_{55}=+\Gamma_{15}^{1(3)}+\Gamma_{35}^{1(3)}-\Gamma_{56}^{3(1)}-\Gamma_{58}^{3(1)},$\\
                                     $\dot{\rho}_{66}=+\Gamma_{26}^{1(3)}+\Gamma_{46}^{1(3)}+\Gamma_{56}^{3(1)}-\Gamma_{67}^{3(1)},$\\
                                     $\dot{\rho}_{77}=+\Gamma_{37}^{1(3)}+\Gamma_{17}^{1(3)}-\Gamma_{78}^{3(1)}+\Gamma_{67}^{3(1)},$\\
                                     $\dot{\rho}_{88}=+\Gamma_{48}^{1(3)}+\Gamma_{28}^{1(3)}+\Gamma_{78}^{3(1)}+\Gamma_{58}^{3(1)}$}}}
       & $\mathbb{S}_1^\perp$ & \makecell[c]{\\$\dot{\rho}_{11}=\sum_{j=1,3}(-\Gamma_{12}^j-\Gamma_{13}^j-\Gamma_{14}^j)$,\\
                                    $\dot{\rho}_{22}=\sum_{j=1,3}(+\Gamma_{12}^j-\Gamma_{23}^j-\Gamma_{24}^j)$,\\
                                    $\dot{\rho}_{33}=\sum_{j=1,3}(+\Gamma_{13}^j+\Gamma_{23}^j-\Gamma_{34}^j)$,\\
                                    $\dot{\rho}_{44}=\sum_{j=1,3}(+\Gamma_{14}^j+\Gamma_{24}^j+\Gamma_{34}^j)$\\ \quad} \\
       \cline{7-8}
        \multicolumn{2}{c|}{ }& & \multicolumn{3}{c|}{ }
       & $\mathbb{S}_2^\perp$ & \makecell[c]{\\$\dot{\rho}_{55}=\sum_{j=1,3}(-\Gamma_{56}^j-\Gamma_{57}^j-\Gamma_{58}^j)$,\\
                                    $\dot{\rho}_{66}=\sum_{j=1,3}(+\Gamma_{56}^j-\Gamma_{67}^j-\Gamma_{68}^j)$,\\
                                    $\dot{\rho}_{77}=\sum_{j=1,3}(+\Gamma_{57}^j+\Gamma_{67}^j-\Gamma_{78}^j)$,\\
                                    $\dot{\rho}_{88}=\sum_{j=1,3}(+\Gamma_{58}^j+\Gamma_{68}^j+\Gamma_{78}^j)$\\ \quad}\\
  \hline
    \multicolumn{2}{c|}{Conduction} & \multicolumn{6}{c}{Yes}\\
 \hline
 \end{tabular}}
 \end{table}
\textit{Model TTT}-This model is studied in Ref. \cite{PhysRevE.109.014142} and we won't go into details here.

In Table \ref{tableS1}, we give the dynamics when the middle spin is in TF, and it is easy to find that model TTL and model LTT have symmetric dynamics. Only when all three spins are in TF, i.e., complete transverse field case, does the non-dissipative second spin divide the system into two independent subspaces. It is worth mentioning that the non-dissipative bulk spin is not a sufficient condition for a complete transverse field to divide the system into independent subspaces. Still, it is a sufficient condition for a longitudinal field to divide the system into subspaces. 
As long as the middle spin is in TF, the system will not be in the thermal state, and the steady-state heat current will not disappear. Therefore, the key of the magnetic-controlled heat modulator is to regulate the direction of the magnetic field where the non-dissipative spin is immersed.

\end{widetext}
\bibliography{nspin-chains}

\begin{thebibliography}{122}%
\makeatletter
\providecommand \@ifxundefined [1]{%
 \@ifx{#1\undefined}
}%
\providecommand \@ifnum [1]{%
 \ifnum #1\expandafter \@firstoftwo
 \else \expandafter \@secondoftwo
 \fi
}%
\providecommand \@ifx [1]{%
 \ifx #1\expandafter \@firstoftwo
 \else \expandafter \@secondoftwo
 \fi
}%
\providecommand \natexlab [1]{#1}%
\providecommand \enquote  [1]{``#1''}%
\providecommand \bibnamefont  [1]{#1}%
\providecommand \bibfnamefont [1]{#1}%
\providecommand \citenamefont [1]{#1}%
\providecommand \href@noop [0]{\@secondoftwo}%
\providecommand \href [0]{\begingroup \@sanitize@url \@href}%
\providecommand \@href[1]{\@@startlink{#1}\@@href}%
\providecommand \@@href[1]{\endgroup#1\@@endlink}%
\providecommand \@sanitize@url [0]{\catcode `\\12\catcode `\$12\catcode
  `\&12\catcode `\#12\catcode `\^12\catcode `\_12\catcode `\%12\relax}%
\providecommand \@@startlink[1]{}%
\providecommand \@@endlink[0]{}%
\providecommand \url  [0]{\begingroup\@sanitize@url \@url }%
\providecommand \@url [1]{\endgroup\@href {#1}{\urlprefix }}%
\providecommand \urlprefix  [0]{URL }%
\providecommand \Eprint [0]{\href }%
\providecommand \doibase [0]{http://dx.doi.org/}%
\providecommand \selectlanguage [0]{\@gobble}%
\providecommand \bibinfo  [0]{\@secondoftwo}%
\providecommand \bibfield  [0]{\@secondoftwo}%
\providecommand \translation [1]{[#1]}%
\providecommand \BibitemOpen [0]{}%
\providecommand \bibitemStop [0]{}%
\providecommand \bibitemNoStop [0]{.\EOS\space}%
\providecommand \EOS [0]{\spacefactor3000\relax}%
\providecommand \BibitemShut  [1]{\csname bibitem#1\endcsname}%
\let\auto@bib@innerbib\@empty
\bibitem [{\citenamefont {Prosen}(2011)}]{PhysRevLett.106.217206}%
  \BibitemOpen
  \bibfield  {author} {\bibinfo {author} {\bibfnamefont {Toma\ifmmode
  \check{z}\else~\v{z}\fi{}}\ \bibnamefont {Prosen}},\ }\bibfield  {title}
  {\enquote {\bibinfo {title} {Open $xxz$ spin chain: Nonequilibrium steady
  state and a strict bound on ballistic transport},}\ }\href {\doibase
  10.1103/PhysRevLett.106.217206} {\bibfield  {journal} {\bibinfo  {journal}
  {Phys. Rev. Lett.}\ }\textbf {\bibinfo {volume} {106}},\ \bibinfo {pages}
  {217206} (\bibinfo {year} {2011})}\BibitemShut {NoStop}%
\bibitem [{\citenamefont {Vogl}\ \emph {et~al.}(2012)\citenamefont {Vogl},
  \citenamefont {Schaller},\ and\ \citenamefont
  {Brandes}}]{PhysRevLett.109.240402}%
  \BibitemOpen
  \bibfield  {author} {\bibinfo {author} {\bibfnamefont {Malte}\ \bibnamefont
  {Vogl}}, \bibinfo {author} {\bibfnamefont {Gernot}\ \bibnamefont {Schaller}},
  \ and\ \bibinfo {author} {\bibfnamefont {Tobias}\ \bibnamefont {Brandes}},\
  }\bibfield  {title} {\enquote {\bibinfo {title} {Criticality in transport
  through the quantum ising chain},}\ }\href {\doibase
  10.1103/PhysRevLett.109.240402} {\bibfield  {journal} {\bibinfo  {journal}
  {Phys. Rev. Lett.}\ }\textbf {\bibinfo {volume} {109}},\ \bibinfo {pages}
  {240402} (\bibinfo {year} {2012})}\BibitemShut {NoStop}%
\bibitem [{\citenamefont {Weimer}(2015)}]{PhysRevLett.114.040402}%
  \BibitemOpen
  \bibfield  {author} {\bibinfo {author} {\bibfnamefont {Hendrik}\ \bibnamefont
  {Weimer}},\ }\bibfield  {title} {\enquote {\bibinfo {title} {Variational
  principle for steady states of dissipative quantum many-body systems},}\
  }\href {\doibase 10.1103/PhysRevLett.114.040402} {\bibfield  {journal}
  {\bibinfo  {journal} {Phys. Rev. Lett.}\ }\textbf {\bibinfo {volume} {114}},\
  \bibinfo {pages} {040402} (\bibinfo {year} {2015})}\BibitemShut {NoStop}%
\bibitem [{\citenamefont {Morigi}\ \emph {et~al.}(2015)\citenamefont {Morigi},
  \citenamefont {Eschner}, \citenamefont {Cormick}, \citenamefont {Lin},
  \citenamefont {Leibfried},\ and\ \citenamefont
  {Wineland}}]{PhysRevLett.115.200502}%
  \BibitemOpen
  \bibfield  {author} {\bibinfo {author} {\bibfnamefont {Giovanna}\
  \bibnamefont {Morigi}}, \bibinfo {author} {\bibfnamefont {J\"urgen}\
  \bibnamefont {Eschner}}, \bibinfo {author} {\bibfnamefont {Cecilia}\
  \bibnamefont {Cormick}}, \bibinfo {author} {\bibfnamefont {Yiheng}\
  \bibnamefont {Lin}}, \bibinfo {author} {\bibfnamefont {Dietrich}\
  \bibnamefont {Leibfried}}, \ and\ \bibinfo {author} {\bibfnamefont
  {David~J.}\ \bibnamefont {Wineland}},\ }\bibfield  {title} {\enquote
  {\bibinfo {title} {Dissipative quantum control of a spin chain},}\ }\href
  {\doibase 10.1103/PhysRevLett.115.200502} {\bibfield  {journal} {\bibinfo
  {journal} {Phys. Rev. Lett.}\ }\textbf {\bibinfo {volume} {115}},\ \bibinfo
  {pages} {200502} (\bibinfo {year} {2015})}\BibitemShut {NoStop}%
\bibitem [{\citenamefont {Bertini}\ \emph {et~al.}(2016)\citenamefont
  {Bertini}, \citenamefont {Collura}, \citenamefont {De~Nardis},\ and\
  \citenamefont {Fagotti}}]{PhysRevLett.117.207201}%
  \BibitemOpen
  \bibfield  {author} {\bibinfo {author} {\bibfnamefont {Bruno}\ \bibnamefont
  {Bertini}}, \bibinfo {author} {\bibfnamefont {Mario}\ \bibnamefont
  {Collura}}, \bibinfo {author} {\bibfnamefont {Jacopo}\ \bibnamefont
  {De~Nardis}}, \ and\ \bibinfo {author} {\bibfnamefont {Maurizio}\
  \bibnamefont {Fagotti}},\ }\bibfield  {title} {\enquote {\bibinfo {title}
  {Transport in out-of-equilibrium $xxz$ chains: Exact profiles of charges and
  currents},}\ }\href {\doibase 10.1103/PhysRevLett.117.207201} {\bibfield
  {journal} {\bibinfo  {journal} {Phys. Rev. Lett.}\ }\textbf {\bibinfo
  {volume} {117}},\ \bibinfo {pages} {207201} (\bibinfo {year}
  {2016})}\BibitemShut {NoStop}%
\bibitem [{\citenamefont {Gopalakrishnan}\ and\ \citenamefont
  {Vasseur}(2019)}]{PhysRevLett.122.127202}%
  \BibitemOpen
  \bibfield  {author} {\bibinfo {author} {\bibfnamefont {Sarang}\ \bibnamefont
  {Gopalakrishnan}}\ and\ \bibinfo {author} {\bibfnamefont {Romain}\
  \bibnamefont {Vasseur}},\ }\bibfield  {title} {\enquote {\bibinfo {title}
  {Kinetic theory of spin diffusion and superdiffusion in $xxz$ spin chains},}\
  }\href {\doibase 10.1103/PhysRevLett.122.127202} {\bibfield  {journal}
  {\bibinfo  {journal} {Phys. Rev. Lett.}\ }\textbf {\bibinfo {volume} {122}},\
  \bibinfo {pages} {127202} (\bibinfo {year} {2019})}\BibitemShut {NoStop}%
\bibitem [{\citenamefont {Maier}\ \emph {et~al.}(2019)\citenamefont {Maier},
  \citenamefont {Brydges}, \citenamefont {Jurcevic}, \citenamefont {Trautmann},
  \citenamefont {Hempel}, \citenamefont {Lanyon}, \citenamefont {Hauke},
  \citenamefont {Blatt},\ and\ \citenamefont {Roos}}]{PhysRevLett.122.050501}%
  \BibitemOpen
  \bibfield  {author} {\bibinfo {author} {\bibfnamefont {Christine}\
  \bibnamefont {Maier}}, \bibinfo {author} {\bibfnamefont {Tiff}\ \bibnamefont
  {Brydges}}, \bibinfo {author} {\bibfnamefont {Petar}\ \bibnamefont
  {Jurcevic}}, \bibinfo {author} {\bibfnamefont {Nils}\ \bibnamefont
  {Trautmann}}, \bibinfo {author} {\bibfnamefont {Cornelius}\ \bibnamefont
  {Hempel}}, \bibinfo {author} {\bibfnamefont {Ben~P.}\ \bibnamefont {Lanyon}},
  \bibinfo {author} {\bibfnamefont {Philipp}\ \bibnamefont {Hauke}}, \bibinfo
  {author} {\bibfnamefont {Rainer}\ \bibnamefont {Blatt}}, \ and\ \bibinfo
  {author} {\bibfnamefont {Christian~F.}\ \bibnamefont {Roos}},\ }\bibfield
  {title} {\enquote {\bibinfo {title} {Environment-assisted quantum transport
  in a 10-qubit network},}\ }\href {\doibase 10.1103/PhysRevLett.122.050501}
  {\bibfield  {journal} {\bibinfo  {journal} {Phys. Rev. Lett.}\ }\textbf
  {\bibinfo {volume} {122}},\ \bibinfo {pages} {050501} (\bibinfo {year}
  {2019})}\BibitemShut {NoStop}%
\bibitem [{\citenamefont {Weimer}\ \emph {et~al.}(2021)\citenamefont {Weimer},
  \citenamefont {Kshetrimayum},\ and\ \citenamefont
  {Or\'us}}]{RevModPhys.93.015008}%
  \BibitemOpen
  \bibfield  {author} {\bibinfo {author} {\bibfnamefont {Hendrik}\ \bibnamefont
  {Weimer}}, \bibinfo {author} {\bibfnamefont {Augustine}\ \bibnamefont
  {Kshetrimayum}}, \ and\ \bibinfo {author} {\bibfnamefont {Rom\'an}\
  \bibnamefont {Or\'us}},\ }\bibfield  {title} {\enquote {\bibinfo {title}
  {Simulation methods for open quantum many-body systems},}\ }\href {\doibase
  10.1103/RevModPhys.93.015008} {\bibfield  {journal} {\bibinfo  {journal}
  {Rev. Mod. Phys.}\ }\textbf {\bibinfo {volume} {93}},\ \bibinfo {pages}
  {015008} (\bibinfo {year} {2021})}\BibitemShut {NoStop}%
\bibitem [{\citenamefont {Werlang}\ \emph {et~al.}(2022)\citenamefont
  {Werlang}, \citenamefont {Matos}, \citenamefont {Brito},\ and\ \citenamefont
  {Valente}}]{0Emergence}%
  \BibitemOpen
  \bibfield  {author} {\bibinfo {author} {\bibfnamefont {Thiago}\ \bibnamefont
  {Werlang}}, \bibinfo {author} {\bibfnamefont {Maurício}\ \bibnamefont
  {Matos}}, \bibinfo {author} {\bibfnamefont {Frederico}\ \bibnamefont
  {Brito}}, \ and\ \bibinfo {author} {\bibfnamefont {Daniel}\ \bibnamefont
  {Valente}},\ }\bibfield  {title} {\enquote {\bibinfo {title} {Emergence of
  energy-avoiding and energy-seeking behaviors in nonequilibrium dissipative
  quantum systems},}\ }\href {\doibase 10.1038/s42005-021-00780-4} {\bibfield
  {journal} {\bibinfo  {journal} {Commun. Phys.}\ }\textbf {\bibinfo {volume}
  {5}},\ \bibinfo {pages} {7} (\bibinfo {year} {2022})}\BibitemShut {NoStop}%
\bibitem [{\citenamefont {Yoo}\ \emph {et~al.}(2023)\citenamefont {Yoo},
  \citenamefont {White},\ and\ \citenamefont {Swingle}}]{PhysRevB.107.115118}%
  \BibitemOpen
  \bibfield  {author} {\bibinfo {author} {\bibfnamefont {Yongchan}\
  \bibnamefont {Yoo}}, \bibinfo {author} {\bibfnamefont {Christopher~David}\
  \bibnamefont {White}}, \ and\ \bibinfo {author} {\bibfnamefont {Brian}\
  \bibnamefont {Swingle}},\ }\bibfield  {title} {\enquote {\bibinfo {title}
  {Open-system spin transport and operator weight dissipation in spin
  chains},}\ }\href {\doibase 10.1103/PhysRevB.107.115118} {\bibfield
  {journal} {\bibinfo  {journal} {Phys. Rev. B}\ }\textbf {\bibinfo {volume}
  {107}},\ \bibinfo {pages} {115118} (\bibinfo {year} {2023})}\BibitemShut
  {NoStop}%
\bibitem [{\citenamefont {Begg}\ and\ \citenamefont
  {Hanai}(2024)}]{PhysRevLett.132.120401}%
  \BibitemOpen
  \bibfield  {author} {\bibinfo {author} {\bibfnamefont {Samuel~E.}\
  \bibnamefont {Begg}}\ and\ \bibinfo {author} {\bibfnamefont {Ryo}\
  \bibnamefont {Hanai}},\ }\bibfield  {title} {\enquote {\bibinfo {title}
  {Quantum criticality in open quantum spin chains with nonreciprocity},}\
  }\href {\doibase 10.1103/PhysRevLett.132.120401} {\bibfield  {journal}
  {\bibinfo  {journal} {Phys. Rev. Lett.}\ }\textbf {\bibinfo {volume} {132}},\
  \bibinfo {pages} {120401} (\bibinfo {year} {2024})}\BibitemShut {NoStop}%
\bibitem [{\citenamefont {Gu}\ \emph {et~al.}(2018)\citenamefont {Gu},
  \citenamefont {Wei}, \citenamefont {Yin}, \citenamefont {Li},\ and\
  \citenamefont {Yang}}]{RevModPhys.90.041002}%
  \BibitemOpen
  \bibfield  {author} {\bibinfo {author} {\bibfnamefont {Xiaokun}\ \bibnamefont
  {Gu}}, \bibinfo {author} {\bibfnamefont {Yujie}\ \bibnamefont {Wei}},
  \bibinfo {author} {\bibfnamefont {Xiaobo}\ \bibnamefont {Yin}}, \bibinfo
  {author} {\bibfnamefont {Baowen}\ \bibnamefont {Li}}, \ and\ \bibinfo
  {author} {\bibfnamefont {Ronggui}\ \bibnamefont {Yang}},\ }\bibfield  {title}
  {\enquote {\bibinfo {title} {Colloquium: Phononic thermal properties of
  two-dimensional materials},}\ }\href {\doibase 10.1103/RevModPhys.90.041002}
  {\bibfield  {journal} {\bibinfo  {journal} {Rev. Mod. Phys.}\ }\textbf
  {\bibinfo {volume} {90}},\ \bibinfo {pages} {041002} (\bibinfo {year}
  {2018})}\BibitemShut {NoStop}%
\bibitem [{\citenamefont {Landi}\ \emph {et~al.}(2022)\citenamefont {Landi},
  \citenamefont {Poletti},\ and\ \citenamefont
  {Schaller}}]{RevModPhys.94.045006}%
  \BibitemOpen
  \bibfield  {author} {\bibinfo {author} {\bibfnamefont {Gabriel~T.}\
  \bibnamefont {Landi}}, \bibinfo {author} {\bibfnamefont {Dario}\ \bibnamefont
  {Poletti}}, \ and\ \bibinfo {author} {\bibfnamefont {Gernot}\ \bibnamefont
  {Schaller}},\ }\bibfield  {title} {\enquote {\bibinfo {title} {Nonequilibrium
  boundary-driven quantum systems: Models, methods, and properties},}\ }\href
  {\doibase 10.1103/RevModPhys.94.045006} {\bibfield  {journal} {\bibinfo
  {journal} {Rev. Mod. Phys.}\ }\textbf {\bibinfo {volume} {94}},\ \bibinfo
  {pages} {045006} (\bibinfo {year} {2022})}\BibitemShut {NoStop}%
\bibitem [{\citenamefont {de~Leon}\ \emph {et~al.}(2021)\citenamefont
  {de~Leon}, \citenamefont {Itoh}, \citenamefont {Kim}, \citenamefont {Mehta},
  \citenamefont {Northup}, \citenamefont {Paik}, \citenamefont {Palmer},
  \citenamefont {Samarth}, \citenamefont {Sangtawesin},\ and\ \citenamefont
  {Steuerman}}]{doi:10.1126/science.abb2823}%
  \BibitemOpen
  \bibfield  {author} {\bibinfo {author} {\bibfnamefont {Nathalie~P.}\
  \bibnamefont {de~Leon}}, \bibinfo {author} {\bibfnamefont {Kohei~M.}\
  \bibnamefont {Itoh}}, \bibinfo {author} {\bibfnamefont {Dohun}\ \bibnamefont
  {Kim}}, \bibinfo {author} {\bibfnamefont {Karan~K.}\ \bibnamefont {Mehta}},
  \bibinfo {author} {\bibfnamefont {Tracy~E.}\ \bibnamefont {Northup}},
  \bibinfo {author} {\bibfnamefont {Hanhee}\ \bibnamefont {Paik}}, \bibinfo
  {author} {\bibfnamefont {B.~S.}\ \bibnamefont {Palmer}}, \bibinfo {author}
  {\bibfnamefont {N.}~\bibnamefont {Samarth}}, \bibinfo {author} {\bibfnamefont
  {Sorawis}\ \bibnamefont {Sangtawesin}}, \ and\ \bibinfo {author}
  {\bibfnamefont {D.~W.}\ \bibnamefont {Steuerman}},\ }\bibfield  {title}
  {\enquote {\bibinfo {title} {Materials challenges and opportunities for
  quantum computing hardware},}\ }\href {\doibase 10.1126/science.abb2823}
  {\bibfield  {journal} {\bibinfo  {journal} {Science}\ }\textbf {\bibinfo
  {volume} {372}},\ \bibinfo {pages} {2823} (\bibinfo {year}
  {2021})}\BibitemShut {NoStop}%
\bibitem [{\citenamefont {Arrachea}(2023)}]{Arrachea_2023}%
  \BibitemOpen
  \bibfield  {author} {\bibinfo {author} {\bibfnamefont {Liliana}\ \bibnamefont
  {Arrachea}},\ }\bibfield  {title} {\enquote {\bibinfo {title} {Energy
  dynamics, heat production and heat–work conversion with qubits: toward the
  development of quantum machines},}\ }\href {\doibase
  10.1088/1361-6633/acb06b} {\bibfield  {journal} {\bibinfo  {journal} {Rep.
  Prog. Phys.}\ }\textbf {\bibinfo {volume} {86}},\ \bibinfo {pages} {036501}
  (\bibinfo {year} {2023})}\BibitemShut {NoStop}%
\bibitem [{\citenamefont {Barreiro}\ \emph {et~al.}(2011)\citenamefont
  {Barreiro}, \citenamefont {M\"uller}, \citenamefont {Schindler},
  \citenamefont {Nigg}, \citenamefont {Monz}, \citenamefont {Chwalla},
  \citenamefont {Hennrich}, \citenamefont {Roos}, \citenamefont {Zoller},\ and\
  \citenamefont {Blatt}}]{Barreiro:2011fge}%
  \BibitemOpen
  \bibfield  {author} {\bibinfo {author} {\bibfnamefont {Julio~T.}\
  \bibnamefont {Barreiro}}, \bibinfo {author} {\bibfnamefont {Markus}\
  \bibnamefont {M\"uller}}, \bibinfo {author} {\bibfnamefont {Philipp}\
  \bibnamefont {Schindler}}, \bibinfo {author} {\bibfnamefont {Daniel}\
  \bibnamefont {Nigg}}, \bibinfo {author} {\bibfnamefont {Thomas}\ \bibnamefont
  {Monz}}, \bibinfo {author} {\bibfnamefont {Michael}\ \bibnamefont {Chwalla}},
  \bibinfo {author} {\bibfnamefont {Markus}\ \bibnamefont {Hennrich}}, \bibinfo
  {author} {\bibfnamefont {Christian~F.}\ \bibnamefont {Roos}}, \bibinfo
  {author} {\bibfnamefont {Peter}\ \bibnamefont {Zoller}}, \ and\ \bibinfo
  {author} {\bibfnamefont {Rainer}\ \bibnamefont {Blatt}},\ }\bibfield  {title}
  {\enquote {\bibinfo {title} {An open-system quantum simulator with trapped
  ions},}\ }\href {\doibase 10.1038/nature09801} {\bibfield  {journal}
  {\bibinfo  {journal} {Nature}\ }\textbf {\bibinfo {volume} {470}},\ \bibinfo
  {pages} {486--491} (\bibinfo {year} {2011})}\BibitemShut {NoStop}%
\bibitem [{\citenamefont {Cole}\ \emph {et~al.}(2022)\citenamefont {Cole},
  \citenamefont {Erickson}, \citenamefont {Zarantonello}, \citenamefont {Horn},
  \citenamefont {Hou}, \citenamefont {Wu}, \citenamefont {Slichter},
  \citenamefont {Reiter}, \citenamefont {Koch},\ and\ \citenamefont
  {Leibfried}}]{PhysRevLett.128.080502}%
  \BibitemOpen
  \bibfield  {author} {\bibinfo {author} {\bibfnamefont {Daniel~C.}\
  \bibnamefont {Cole}}, \bibinfo {author} {\bibfnamefont {Stephen~D.}\
  \bibnamefont {Erickson}}, \bibinfo {author} {\bibfnamefont {Giorgio}\
  \bibnamefont {Zarantonello}}, \bibinfo {author} {\bibfnamefont {Karl~P.}\
  \bibnamefont {Horn}}, \bibinfo {author} {\bibfnamefont {Pan-Yu}\ \bibnamefont
  {Hou}}, \bibinfo {author} {\bibfnamefont {Jenny~J.}\ \bibnamefont {Wu}},
  \bibinfo {author} {\bibfnamefont {Daniel~H.}\ \bibnamefont {Slichter}},
  \bibinfo {author} {\bibfnamefont {Florentin}\ \bibnamefont {Reiter}},
  \bibinfo {author} {\bibfnamefont {Christiane~P.}\ \bibnamefont {Koch}}, \
  and\ \bibinfo {author} {\bibfnamefont {Dietrich}\ \bibnamefont {Leibfried}},\
  }\bibfield  {title} {\enquote {\bibinfo {title} {Resource-efficient
  dissipative entanglement of two trapped-ion qubits},}\ }\href {\doibase
  10.1103/PhysRevLett.128.080502} {\bibfield  {journal} {\bibinfo  {journal}
  {Phys. Rev. Lett.}\ }\textbf {\bibinfo {volume} {128}},\ \bibinfo {pages}
  {080502} (\bibinfo {year} {2022})}\BibitemShut {NoStop}%
\bibitem [{\citenamefont {Devoret}\ and\ \citenamefont
  {Schoelkopf}(2013)}]{doi:10.1126/science.1231930}%
  \BibitemOpen
  \bibfield  {author} {\bibinfo {author} {\bibfnamefont {M.~H.}\ \bibnamefont
  {Devoret}}\ and\ \bibinfo {author} {\bibfnamefont {R.~J.}\ \bibnamefont
  {Schoelkopf}},\ }\bibfield  {title} {\enquote {\bibinfo {title}
  {Superconducting circuits for quantum information: An outlook},}\ }\href
  {\doibase 10.1126/science.1231930} {\bibfield  {journal} {\bibinfo  {journal}
  {Science}\ }\textbf {\bibinfo {volume} {339}},\ \bibinfo {pages} {1169--1174}
  (\bibinfo {year} {2013})}\BibitemShut {NoStop}%
\bibitem [{\citenamefont {Wendin}(2017)}]{Wendin_2017}%
  \BibitemOpen
  \bibfield  {author} {\bibinfo {author} {\bibfnamefont {G}~\bibnamefont
  {Wendin}},\ }\bibfield  {title} {\enquote {\bibinfo {title} {Quantum
  information processing with superconducting circuits: a review},}\ }\href
  {\doibase 10.1088/1361-6633/aa7e1a} {\bibfield  {journal} {\bibinfo
  {journal} {Rep. Prog. Phys.}\ }\textbf {\bibinfo {volume} {80}},\ \bibinfo
  {pages} {106001} (\bibinfo {year} {2017})}\BibitemShut {NoStop}%
\bibitem [{\citenamefont {Cai}\ \emph {et~al.}(2019)\citenamefont {Cai},
  \citenamefont {Han}, \citenamefont {Mei}, \citenamefont {Xu}, \citenamefont
  {Ma}, \citenamefont {Li}, \citenamefont {Wang}, \citenamefont {Song},
  \citenamefont {Xue}, \citenamefont {Yin}, \citenamefont {Jia},\ and\
  \citenamefont {Sun}}]{PhysRevLett.123.080501}%
  \BibitemOpen
  \bibfield  {author} {\bibinfo {author} {\bibfnamefont {W.}~\bibnamefont
  {Cai}}, \bibinfo {author} {\bibfnamefont {J.}~\bibnamefont {Han}}, \bibinfo
  {author} {\bibfnamefont {Feng}\ \bibnamefont {Mei}}, \bibinfo {author}
  {\bibfnamefont {Y.}~\bibnamefont {Xu}}, \bibinfo {author} {\bibfnamefont
  {Y.}~\bibnamefont {Ma}}, \bibinfo {author} {\bibfnamefont {X.}~\bibnamefont
  {Li}}, \bibinfo {author} {\bibfnamefont {H.}~\bibnamefont {Wang}}, \bibinfo
  {author} {\bibfnamefont {Y.~P.}\ \bibnamefont {Song}}, \bibinfo {author}
  {\bibfnamefont {Zheng-Yuan}\ \bibnamefont {Xue}}, \bibinfo {author}
  {\bibfnamefont {Zhang-Qi}\ \bibnamefont {Yin}}, \bibinfo {author}
  {\bibfnamefont {Suotang}\ \bibnamefont {Jia}}, \ and\ \bibinfo {author}
  {\bibfnamefont {Luyan}\ \bibnamefont {Sun}},\ }\bibfield  {title} {\enquote
  {\bibinfo {title} {Observation of topological magnon insulator states in a
  superconducting circuit},}\ }\href {\doibase 10.1103/PhysRevLett.123.080501}
  {\bibfield  {journal} {\bibinfo  {journal} {Phys. Rev. Lett.}\ }\textbf
  {\bibinfo {volume} {123}},\ \bibinfo {pages} {080501} (\bibinfo {year}
  {2019})}\BibitemShut {NoStop}%
\bibitem [{\citenamefont {Léger}\ \emph {et~al.}(2019)\citenamefont {Léger},
  \citenamefont {Puertas-Martínez}, \citenamefont {Bharadwaj}, \citenamefont
  {Dassonneville}, \citenamefont {Delaforce}, \citenamefont {Foroughi},
  \citenamefont {Milchakov}, \citenamefont {Planat}, \citenamefont {Buisson},
  \citenamefont {Naud}, \citenamefont {Hasch-Guichard}, \citenamefont
  {Florens}, \citenamefont {Snyman},\ and\ \citenamefont {Roch}}]{Leger2019}%
  \BibitemOpen
  \bibfield  {author} {\bibinfo {author} {\bibfnamefont {Sébastien}\
  \bibnamefont {Léger}}, \bibinfo {author} {\bibfnamefont {Javier}\
  \bibnamefont {Puertas-Martínez}}, \bibinfo {author} {\bibfnamefont
  {Karthik}\ \bibnamefont {Bharadwaj}}, \bibinfo {author} {\bibfnamefont
  {Rémy}\ \bibnamefont {Dassonneville}}, \bibinfo {author} {\bibfnamefont
  {Jovian}\ \bibnamefont {Delaforce}}, \bibinfo {author} {\bibfnamefont
  {Farshad}\ \bibnamefont {Foroughi}}, \bibinfo {author} {\bibfnamefont
  {Vladimir}\ \bibnamefont {Milchakov}}, \bibinfo {author} {\bibfnamefont
  {Luca}\ \bibnamefont {Planat}}, \bibinfo {author} {\bibfnamefont {Olivier}\
  \bibnamefont {Buisson}}, \bibinfo {author} {\bibfnamefont {Cécile}\
  \bibnamefont {Naud}}, \bibinfo {author} {\bibfnamefont {Wiebke}\ \bibnamefont
  {Hasch-Guichard}}, \bibinfo {author} {\bibfnamefont {Serge}\ \bibnamefont
  {Florens}}, \bibinfo {author} {\bibfnamefont {Izak}\ \bibnamefont {Snyman}},
  \ and\ \bibinfo {author} {\bibfnamefont {Nicolas}\ \bibnamefont {Roch}},\
  }\bibfield  {title} {\enquote {\bibinfo {title} {Observation of quantum
  many-body effects due to zero point fluctuations in superconducting
  circuits},}\ }\href {\doibase 10.1038/s41467-019-13199-x} {\bibfield
  {journal} {\bibinfo  {journal} {Nat. Commun.}\ }\textbf {\bibinfo {volume}
  {10}},\ \bibinfo {pages} {5259} (\bibinfo {year} {2019})}\BibitemShut
  {NoStop}%
\bibitem [{\citenamefont {Maillet}\ \emph {et~al.}(2020)\citenamefont
  {Maillet}, \citenamefont {Subero}, \citenamefont {Peltonen}, \citenamefont
  {Golubev},\ and\ \citenamefont {Pekola}}]{maillet2020electric}%
  \BibitemOpen
  \bibfield  {author} {\bibinfo {author} {\bibfnamefont {Olivier}\ \bibnamefont
  {Maillet}}, \bibinfo {author} {\bibfnamefont {Diego}\ \bibnamefont {Subero}},
  \bibinfo {author} {\bibfnamefont {Joonas~T}\ \bibnamefont {Peltonen}},
  \bibinfo {author} {\bibfnamefont {Dmitry~S}\ \bibnamefont {Golubev}}, \ and\
  \bibinfo {author} {\bibfnamefont {Jukka~P}\ \bibnamefont {Pekola}},\
  }\bibfield  {title} {\enquote {\bibinfo {title} {Electric field control of
  radiative heat transfer in a superconducting circuit},}\ }\href {\doibase
  10.1038/s41467-020-18163-8} {\bibfield  {journal} {\bibinfo  {journal} {Nat.
  commun.}\ }\textbf {\bibinfo {volume} {11}},\ \bibinfo {pages} {4326}
  (\bibinfo {year} {2020})}\BibitemShut {NoStop}%
\bibitem [{\citenamefont {Blais}\ \emph {et~al.}(2021)\citenamefont {Blais},
  \citenamefont {Grimsmo}, \citenamefont {Girvin},\ and\ \citenamefont
  {Wallraff}}]{RevModPhys.93.025005}%
  \BibitemOpen
  \bibfield  {author} {\bibinfo {author} {\bibfnamefont {Alexandre}\
  \bibnamefont {Blais}}, \bibinfo {author} {\bibfnamefont {Arne~L.}\
  \bibnamefont {Grimsmo}}, \bibinfo {author} {\bibfnamefont {S.~M.}\
  \bibnamefont {Girvin}}, \ and\ \bibinfo {author} {\bibfnamefont {Andreas}\
  \bibnamefont {Wallraff}},\ }\bibfield  {title} {\enquote {\bibinfo {title}
  {Circuit quantum electrodynamics},}\ }\href {\doibase
  10.1103/RevModPhys.93.025005} {\bibfield  {journal} {\bibinfo  {journal}
  {Rev. Mod. Phys.}\ }\textbf {\bibinfo {volume} {93}},\ \bibinfo {pages}
  {025005} (\bibinfo {year} {2021})}\BibitemShut {NoStop}%
\bibitem [{\citenamefont {Chowdhury}\ \emph {et~al.}(2024)\citenamefont
  {Chowdhury}, \citenamefont {Yu}, \citenamefont {Shamim}, \citenamefont
  {Kabir},\ and\ \citenamefont {Sufian}}]{PhysRevResearch.6.033107}%
  \BibitemOpen
  \bibfield  {author} {\bibinfo {author} {\bibfnamefont {Talal~Ahmed}\
  \bibnamefont {Chowdhury}}, \bibinfo {author} {\bibfnamefont {Kwangmin}\
  \bibnamefont {Yu}}, \bibinfo {author} {\bibfnamefont {Mahmud~Ashraf}\
  \bibnamefont {Shamim}}, \bibinfo {author} {\bibfnamefont {M.~L.}\
  \bibnamefont {Kabir}}, \ and\ \bibinfo {author} {\bibfnamefont {Raza~Sabbir}\
  \bibnamefont {Sufian}},\ }\bibfield  {title} {\enquote {\bibinfo {title}
  {Enhancing quantum utility: Simulating large-scale quantum spin chains on
  superconducting quantum computers},}\ }\href {\doibase
  10.1103/PhysRevResearch.6.033107} {\bibfield  {journal} {\bibinfo  {journal}
  {Phys. Rev. Res.}\ }\textbf {\bibinfo {volume} {6}},\ \bibinfo {pages}
  {033107} (\bibinfo {year} {2024})}\BibitemShut {NoStop}%
\bibitem [{\citenamefont {Lee}\ \emph {et~al.}(2011)\citenamefont {Lee},
  \citenamefont {H\"affner},\ and\ \citenamefont {Cross}}]{PhysRevA.84.031402}%
  \BibitemOpen
  \bibfield  {author} {\bibinfo {author} {\bibfnamefont {Tony~E.}\ \bibnamefont
  {Lee}}, \bibinfo {author} {\bibfnamefont {H.}~\bibnamefont {H\"affner}}, \
  and\ \bibinfo {author} {\bibfnamefont {M.~C.}\ \bibnamefont {Cross}},\
  }\bibfield  {title} {\enquote {\bibinfo {title} {Antiferromagnetic phase
  transition in a nonequilibrium lattice of rydberg atoms},}\ }\href {\doibase
  10.1103/PhysRevA.84.031402} {\bibfield  {journal} {\bibinfo  {journal} {Phys.
  Rev. A}\ }\textbf {\bibinfo {volume} {84}},\ \bibinfo {pages} {031402}
  (\bibinfo {year} {2011})}\BibitemShut {NoStop}%
\bibitem [{\citenamefont {Lee}\ and\ \citenamefont
  {Chan}(2013)}]{PhysRevA.88.063811}%
  \BibitemOpen
  \bibfield  {author} {\bibinfo {author} {\bibfnamefont {Tony~E.}\ \bibnamefont
  {Lee}}\ and\ \bibinfo {author} {\bibfnamefont {Ching-Kit}\ \bibnamefont
  {Chan}},\ }\bibfield  {title} {\enquote {\bibinfo {title} {Dissipative
  transverse-field ising model: Steady-state correlations and spin
  squeezing},}\ }\href {\doibase 10.1103/PhysRevA.88.063811} {\bibfield
  {journal} {\bibinfo  {journal} {Phys. Rev. A}\ }\textbf {\bibinfo {volume}
  {88}},\ \bibinfo {pages} {063811} (\bibinfo {year} {2013})}\BibitemShut
  {NoStop}%
\bibitem [{\citenamefont {Goldstein}\ \emph {et~al.}(2015)\citenamefont
  {Goldstein}, \citenamefont {Aron},\ and\ \citenamefont
  {Chamon}}]{PhysRevB.92.174418}%
  \BibitemOpen
  \bibfield  {author} {\bibinfo {author} {\bibfnamefont {G.}~\bibnamefont
  {Goldstein}}, \bibinfo {author} {\bibfnamefont {C.}~\bibnamefont {Aron}}, \
  and\ \bibinfo {author} {\bibfnamefont {C.}~\bibnamefont {Chamon}},\
  }\bibfield  {title} {\enquote {\bibinfo {title} {Driven-dissipative ising
  model: Mean-field solution},}\ }\href {\doibase 10.1103/PhysRevB.92.174418}
  {\bibfield  {journal} {\bibinfo  {journal} {Phys. Rev. B}\ }\textbf {\bibinfo
  {volume} {92}},\ \bibinfo {pages} {174418} (\bibinfo {year}
  {2015})}\BibitemShut {NoStop}%
\bibitem [{\citenamefont {Rose}\ \emph {et~al.}(2016)\citenamefont {Rose},
  \citenamefont {Macieszczak}, \citenamefont {Lesanovsky},\ and\ \citenamefont
  {Garrahan}}]{PhysRevE.94.052132}%
  \BibitemOpen
  \bibfield  {author} {\bibinfo {author} {\bibfnamefont {Dominic~C.}\
  \bibnamefont {Rose}}, \bibinfo {author} {\bibfnamefont {Katarzyna}\
  \bibnamefont {Macieszczak}}, \bibinfo {author} {\bibfnamefont {Igor}\
  \bibnamefont {Lesanovsky}}, \ and\ \bibinfo {author} {\bibfnamefont
  {Juan~P.}\ \bibnamefont {Garrahan}},\ }\bibfield  {title} {\enquote {\bibinfo
  {title} {Metastability in an open quantum ising model},}\ }\href {\doibase
  10.1103/PhysRevE.94.052132} {\bibfield  {journal} {\bibinfo  {journal} {Phys.
  Rev. E}\ }\textbf {\bibinfo {volume} {94}},\ \bibinfo {pages} {052132}
  (\bibinfo {year} {2016})}\BibitemShut {NoStop}%
\bibitem [{\citenamefont {Overbeck}\ \emph {et~al.}(2017)\citenamefont
  {Overbeck}, \citenamefont {Maghrebi}, \citenamefont {Gorshkov},\ and\
  \citenamefont {Weimer}}]{PhysRevA.95.042133}%
  \BibitemOpen
  \bibfield  {author} {\bibinfo {author} {\bibfnamefont {Vincent~R.}\
  \bibnamefont {Overbeck}}, \bibinfo {author} {\bibfnamefont {Mohammad~F.}\
  \bibnamefont {Maghrebi}}, \bibinfo {author} {\bibfnamefont {Alexey~V.}\
  \bibnamefont {Gorshkov}}, \ and\ \bibinfo {author} {\bibfnamefont {Hendrik}\
  \bibnamefont {Weimer}},\ }\bibfield  {title} {\enquote {\bibinfo {title}
  {Multicritical behavior in dissipative ising models},}\ }\href {\doibase
  10.1103/PhysRevA.95.042133} {\bibfield  {journal} {\bibinfo  {journal} {Phys.
  Rev. A}\ }\textbf {\bibinfo {volume} {95}},\ \bibinfo {pages} {042133}
  (\bibinfo {year} {2017})}\BibitemShut {NoStop}%
\bibitem [{\citenamefont {Buyskikh}\ \emph {et~al.}(2019)\citenamefont
  {Buyskikh}, \citenamefont {Tagliacozzo}, \citenamefont {Schuricht},
  \citenamefont {Hooley}, \citenamefont {Pekker},\ and\ \citenamefont
  {Daley}}]{PhysRevLett.123.090401}%
  \BibitemOpen
  \bibfield  {author} {\bibinfo {author} {\bibfnamefont {Anton~S.}\
  \bibnamefont {Buyskikh}}, \bibinfo {author} {\bibfnamefont {Luca}\
  \bibnamefont {Tagliacozzo}}, \bibinfo {author} {\bibfnamefont {Dirk}\
  \bibnamefont {Schuricht}}, \bibinfo {author} {\bibfnamefont {Chris~A.}\
  \bibnamefont {Hooley}}, \bibinfo {author} {\bibfnamefont {David}\
  \bibnamefont {Pekker}}, \ and\ \bibinfo {author} {\bibfnamefont {Andrew~J.}\
  \bibnamefont {Daley}},\ }\bibfield  {title} {\enquote {\bibinfo {title} {Spin
  models, dynamics, and criticality with atoms in tilted optical
  superlattices},}\ }\href {\doibase 10.1103/PhysRevLett.123.090401} {\bibfield
   {journal} {\bibinfo  {journal} {Phys. Rev. Lett.}\ }\textbf {\bibinfo
  {volume} {123}},\ \bibinfo {pages} {090401} (\bibinfo {year}
  {2019})}\BibitemShut {NoStop}%
\bibitem [{\citenamefont {Shibata}\ and\ \citenamefont
  {Katsura}(2019)}]{PhysRevB.99.224432}%
  \BibitemOpen
  \bibfield  {author} {\bibinfo {author} {\bibfnamefont {Naoyuki}\ \bibnamefont
  {Shibata}}\ and\ \bibinfo {author} {\bibfnamefont {Hosho}\ \bibnamefont
  {Katsura}},\ }\bibfield  {title} {\enquote {\bibinfo {title} {Dissipative
  quantum ising chain as a non-hermitian ashkin-teller model},}\ }\href
  {\doibase 10.1103/PhysRevB.99.224432} {\bibfield  {journal} {\bibinfo
  {journal} {Phys. Rev. B}\ }\textbf {\bibinfo {volume} {99}},\ \bibinfo
  {pages} {224432} (\bibinfo {year} {2019})}\BibitemShut {NoStop}%
\bibitem [{\citenamefont {Lang}\ and\ \citenamefont
  {B\"uchler}(2020)}]{PhysRevB.102.094204}%
  \BibitemOpen
  \bibfield  {author} {\bibinfo {author} {\bibfnamefont {Nicolai}\ \bibnamefont
  {Lang}}\ and\ \bibinfo {author} {\bibfnamefont {Hans~Peter}\ \bibnamefont
  {B\"uchler}},\ }\bibfield  {title} {\enquote {\bibinfo {title} {Entanglement
  transition in the projective transverse field ising model},}\ }\href
  {\doibase 10.1103/PhysRevB.102.094204} {\bibfield  {journal} {\bibinfo
  {journal} {Phys. Rev. B}\ }\textbf {\bibinfo {volume} {102}},\ \bibinfo
  {pages} {094204} (\bibinfo {year} {2020})}\BibitemShut {NoStop}%
\bibitem [{\citenamefont {Jin}\ \emph {et~al.}(2021)\citenamefont {Jin},
  \citenamefont {He}, \citenamefont {Iemini}, \citenamefont {Ferreira},
  \citenamefont {Wang}, \citenamefont {Chesi},\ and\ \citenamefont
  {Fazio}}]{PhysRevB.104.214301}%
  \BibitemOpen
  \bibfield  {author} {\bibinfo {author} {\bibfnamefont {Jiasen}\ \bibnamefont
  {Jin}}, \bibinfo {author} {\bibfnamefont {Wen-Bin}\ \bibnamefont {He}},
  \bibinfo {author} {\bibfnamefont {Fernando}\ \bibnamefont {Iemini}}, \bibinfo
  {author} {\bibfnamefont {Diego}\ \bibnamefont {Ferreira}}, \bibinfo {author}
  {\bibfnamefont {Ying-Dan}\ \bibnamefont {Wang}}, \bibinfo {author}
  {\bibfnamefont {Stefano}\ \bibnamefont {Chesi}}, \ and\ \bibinfo {author}
  {\bibfnamefont {Rosario}\ \bibnamefont {Fazio}},\ }\bibfield  {title}
  {\enquote {\bibinfo {title} {Determination of the critical exponents in
  dissipative phase transitions: Coherent anomaly approach},}\ }\href {\doibase
  10.1103/PhysRevB.104.214301} {\bibfield  {journal} {\bibinfo  {journal}
  {Phys. Rev. B}\ }\textbf {\bibinfo {volume} {104}},\ \bibinfo {pages}
  {214301} (\bibinfo {year} {2021})}\BibitemShut {NoStop}%
\bibitem [{\citenamefont {Paz}\ and\ \citenamefont
  {Maghrebi}(2021)}]{PhysRevA.104.023713}%
  \BibitemOpen
  \bibfield  {author} {\bibinfo {author} {\bibfnamefont {Daniel~A.}\
  \bibnamefont {Paz}}\ and\ \bibinfo {author} {\bibfnamefont {Mohammad~F.}\
  \bibnamefont {Maghrebi}},\ }\bibfield  {title} {\enquote {\bibinfo {title}
  {Driven-dissipative ising model: An exact field-theoretical analysis},}\
  }\href {\doibase 10.1103/PhysRevA.104.023713} {\bibfield  {journal} {\bibinfo
   {journal} {Phys. Rev. A}\ }\textbf {\bibinfo {volume} {104}},\ \bibinfo
  {pages} {023713} (\bibinfo {year} {2021})}\BibitemShut {NoStop}%
\bibitem [{\citenamefont {Turkeshi}\ \emph {et~al.}(2021)\citenamefont
  {Turkeshi}, \citenamefont {Biella}, \citenamefont {Fazio}, \citenamefont
  {Dalmonte},\ and\ \citenamefont {Schir\'o}}]{PhysRevB.103.224210}%
  \BibitemOpen
  \bibfield  {author} {\bibinfo {author} {\bibfnamefont {Xhek}\ \bibnamefont
  {Turkeshi}}, \bibinfo {author} {\bibfnamefont {Alberto}\ \bibnamefont
  {Biella}}, \bibinfo {author} {\bibfnamefont {Rosario}\ \bibnamefont {Fazio}},
  \bibinfo {author} {\bibfnamefont {Marcello}\ \bibnamefont {Dalmonte}}, \ and\
  \bibinfo {author} {\bibfnamefont {Marco}\ \bibnamefont {Schir\'o}},\
  }\bibfield  {title} {\enquote {\bibinfo {title} {Measurement-induced
  entanglement transitions in the quantum ising chain: From infinite to zero
  clicks},}\ }\href {\doibase 10.1103/PhysRevB.103.224210} {\bibfield
  {journal} {\bibinfo  {journal} {Phys. Rev. B}\ }\textbf {\bibinfo {volume}
  {103}},\ \bibinfo {pages} {224210} (\bibinfo {year} {2021})}\BibitemShut
  {NoStop}%
\bibitem [{\citenamefont {Paz}\ and\ \citenamefont
  {Maghrebi}(2022)}]{Paz_2021}%
  \BibitemOpen
  \bibfield  {author} {\bibinfo {author} {\bibfnamefont {Daniel~A.}\
  \bibnamefont {Paz}}\ and\ \bibinfo {author} {\bibfnamefont {Mohammad~F.}\
  \bibnamefont {Maghrebi}},\ }\bibfield  {title} {\enquote {\bibinfo {title}
  {Driven-dissipative ising model: Dynamical crossover at weak dissipation},}\
  }\href {\doibase 10.1209/0295-5075/ac33cb} {\bibfield  {journal} {\bibinfo
  {journal} {EPL}\ }\textbf {\bibinfo {volume} {136}},\ \bibinfo {pages}
  {10002} (\bibinfo {year} {2022})}\BibitemShut {NoStop}%
\bibitem [{\citenamefont {Tirrito}\ \emph {et~al.}(2023)\citenamefont
  {Tirrito}, \citenamefont {Santini}, \citenamefont {Fazio},\ and\
  \citenamefont {Collura}}]{10.21468/SciPostPhys.15.3.096}%
  \BibitemOpen
  \bibfield  {author} {\bibinfo {author} {\bibfnamefont {Emanuele}\
  \bibnamefont {Tirrito}}, \bibinfo {author} {\bibfnamefont {Alessandro}\
  \bibnamefont {Santini}}, \bibinfo {author} {\bibfnamefont {Rosario}\
  \bibnamefont {Fazio}}, \ and\ \bibinfo {author} {\bibfnamefont {Mario}\
  \bibnamefont {Collura}},\ }\bibfield  {title} {\enquote {\bibinfo {title}
  {{Full counting statistics as probe of measurement-induced transitions in the
  quantum Ising chain}},}\ }\href {\doibase 10.21468/SciPostPhys.15.3.096}
  {\bibfield  {journal} {\bibinfo  {journal} {SciPost Phys.}\ }\textbf
  {\bibinfo {volume} {15}},\ \bibinfo {pages} {096} (\bibinfo {year}
  {2023})}\BibitemShut {NoStop}%
\bibitem [{\citenamefont {Sulz}\ \emph {et~al.}(2024)\citenamefont {Sulz},
  \citenamefont {Lubich}, \citenamefont {Ceruti}, \citenamefont {Lesanovsky},\
  and\ \citenamefont {Carollo}}]{PhysRevA.109.022420}%
  \BibitemOpen
  \bibfield  {author} {\bibinfo {author} {\bibfnamefont {Dominik}\ \bibnamefont
  {Sulz}}, \bibinfo {author} {\bibfnamefont {Christian}\ \bibnamefont
  {Lubich}}, \bibinfo {author} {\bibfnamefont {Gianluca}\ \bibnamefont
  {Ceruti}}, \bibinfo {author} {\bibfnamefont {Igor}\ \bibnamefont
  {Lesanovsky}}, \ and\ \bibinfo {author} {\bibfnamefont {Federico}\
  \bibnamefont {Carollo}},\ }\bibfield  {title} {\enquote {\bibinfo {title}
  {Numerical simulation of long-range open quantum many-body dynamics with tree
  tensor networks},}\ }\href {\doibase 10.1103/PhysRevA.109.022420} {\bibfield
  {journal} {\bibinfo  {journal} {Phys. Rev. A}\ }\textbf {\bibinfo {volume}
  {109}},\ \bibinfo {pages} {022420} (\bibinfo {year} {2024})}\BibitemShut
  {NoStop}%
\bibitem [{\citenamefont {Carr}\ \emph {et~al.}(2013)\citenamefont {Carr},
  \citenamefont {Ritter}, \citenamefont {Wade}, \citenamefont {Adams},\ and\
  \citenamefont {Weatherill}}]{PhysRevLett.111.113901}%
  \BibitemOpen
  \bibfield  {author} {\bibinfo {author} {\bibfnamefont {C.}~\bibnamefont
  {Carr}}, \bibinfo {author} {\bibfnamefont {R.}~\bibnamefont {Ritter}},
  \bibinfo {author} {\bibfnamefont {C.~G.}\ \bibnamefont {Wade}}, \bibinfo
  {author} {\bibfnamefont {C.~S.}\ \bibnamefont {Adams}}, \ and\ \bibinfo
  {author} {\bibfnamefont {K.~J.}\ \bibnamefont {Weatherill}},\ }\bibfield
  {title} {\enquote {\bibinfo {title} {Nonequilibrium phase transition in a
  dilute rydberg ensemble},}\ }\href {\doibase 10.1103/PhysRevLett.111.113901}
  {\bibfield  {journal} {\bibinfo  {journal} {Phys. Rev. Lett.}\ }\textbf
  {\bibinfo {volume} {111}},\ \bibinfo {pages} {113901} (\bibinfo {year}
  {2013})}\BibitemShut {NoStop}%
\bibitem [{\citenamefont {Malossi}\ \emph {et~al.}(2014)\citenamefont
  {Malossi}, \citenamefont {Valado}, \citenamefont {Scotto}, \citenamefont
  {Huillery}, \citenamefont {Pillet}, \citenamefont {Ciampini}, \citenamefont
  {Arimondo},\ and\ \citenamefont {Morsch}}]{PhysRevLett.113.023006}%
  \BibitemOpen
  \bibfield  {author} {\bibinfo {author} {\bibfnamefont {N.}~\bibnamefont
  {Malossi}}, \bibinfo {author} {\bibfnamefont {M.~M.}\ \bibnamefont {Valado}},
  \bibinfo {author} {\bibfnamefont {S.}~\bibnamefont {Scotto}}, \bibinfo
  {author} {\bibfnamefont {P.}~\bibnamefont {Huillery}}, \bibinfo {author}
  {\bibfnamefont {P.}~\bibnamefont {Pillet}}, \bibinfo {author} {\bibfnamefont
  {D.}~\bibnamefont {Ciampini}}, \bibinfo {author} {\bibfnamefont
  {E.}~\bibnamefont {Arimondo}}, \ and\ \bibinfo {author} {\bibfnamefont
  {O.}~\bibnamefont {Morsch}},\ }\bibfield  {title} {\enquote {\bibinfo {title}
  {Full counting statistics and phase diagram of a dissipative rydberg gas},}\
  }\href {\doibase 10.1103/PhysRevLett.113.023006} {\bibfield  {journal}
  {\bibinfo  {journal} {Phys. Rev. Lett.}\ }\textbf {\bibinfo {volume} {113}},\
  \bibinfo {pages} {023006} (\bibinfo {year} {2014})}\BibitemShut {NoStop}%
\bibitem [{\citenamefont {Letscher}\ \emph {et~al.}(2017)\citenamefont
  {Letscher}, \citenamefont {Thomas}, \citenamefont {Niederpr\"um},
  \citenamefont {Fleischhauer},\ and\ \citenamefont {Ott}}]{PhysRevX.7.021020}%
  \BibitemOpen
  \bibfield  {author} {\bibinfo {author} {\bibfnamefont {F.}~\bibnamefont
  {Letscher}}, \bibinfo {author} {\bibfnamefont {O.}~\bibnamefont {Thomas}},
  \bibinfo {author} {\bibfnamefont {T.}~\bibnamefont {Niederpr\"um}}, \bibinfo
  {author} {\bibfnamefont {M.}~\bibnamefont {Fleischhauer}}, \ and\ \bibinfo
  {author} {\bibfnamefont {H.}~\bibnamefont {Ott}},\ }\bibfield  {title}
  {\enquote {\bibinfo {title} {Bistability versus metastability in driven
  dissipative rydberg gases},}\ }\href {\doibase 10.1103/PhysRevX.7.021020}
  {\bibfield  {journal} {\bibinfo  {journal} {Phys. Rev. X}\ }\textbf {\bibinfo
  {volume} {7}},\ \bibinfo {pages} {021020} (\bibinfo {year}
  {2017})}\BibitemShut {NoStop}%
\bibitem [{\citenamefont {Foss-Feig}\ \emph {et~al.}(2017)\citenamefont
  {Foss-Feig}, \citenamefont {Young}, \citenamefont {Albert}, \citenamefont
  {Gorshkov},\ and\ \citenamefont {Maghrebi}}]{PhysRevLett.119.190402}%
  \BibitemOpen
  \bibfield  {author} {\bibinfo {author} {\bibfnamefont {Michael}\ \bibnamefont
  {Foss-Feig}}, \bibinfo {author} {\bibfnamefont {Jeremy~T.}\ \bibnamefont
  {Young}}, \bibinfo {author} {\bibfnamefont {Victor~V.}\ \bibnamefont
  {Albert}}, \bibinfo {author} {\bibfnamefont {Alexey~V.}\ \bibnamefont
  {Gorshkov}}, \ and\ \bibinfo {author} {\bibfnamefont {Mohammad~F.}\
  \bibnamefont {Maghrebi}},\ }\bibfield  {title} {\enquote {\bibinfo {title}
  {Solvable family of driven-dissipative many-body systems},}\ }\href {\doibase
  10.1103/PhysRevLett.119.190402} {\bibfield  {journal} {\bibinfo  {journal}
  {Phys. Rev. Lett.}\ }\textbf {\bibinfo {volume} {119}},\ \bibinfo {pages}
  {190402} (\bibinfo {year} {2017})}\BibitemShut {NoStop}%
\bibitem [{\citenamefont {Scholl}\ \emph {et~al.}(2021)\citenamefont {Scholl},
  \citenamefont {Schuler}, \citenamefont {Williams}, \citenamefont
  {Eberharter},\ and\ \citenamefont {Browaeys}}]{2021Quantum}%
  \BibitemOpen
  \bibfield  {author} {\bibinfo {author} {\bibfnamefont {Pascal}\ \bibnamefont
  {Scholl}}, \bibinfo {author} {\bibfnamefont {Michael}\ \bibnamefont
  {Schuler}}, \bibinfo {author} {\bibfnamefont {Hannah~J.}\ \bibnamefont
  {Williams}}, \bibinfo {author} {\bibfnamefont {Alexander~A.}\ \bibnamefont
  {Eberharter}}, \ and\ \bibinfo {author} {\bibfnamefont {Antoine}\
  \bibnamefont {Browaeys}},\ }\bibfield  {title} {\enquote {\bibinfo {title}
  {Quantum simulation of 2d antiferromagnets with hundreds of rydberg atoms},}\
  }\href {\doibase 10.1038/s41586-021-03585-1} {\bibfield  {journal} {\bibinfo
  {journal} {Nature}\ }\textbf {\bibinfo {volume} {595}},\ \bibinfo {pages}
  {233--238} (\bibinfo {year} {2021})}\BibitemShut {NoStop}%
\bibitem [{\citenamefont {Liu}\ \emph {et~al.}(2024)\citenamefont {Liu},
  \citenamefont {Sun}, \citenamefont {Cabot}, \citenamefont {Carollo},
  \citenamefont {Zhang}, \citenamefont {Zhang}, \citenamefont {Zhang},
  \citenamefont {Liu}, \citenamefont {Han}, \citenamefont {Li}, \citenamefont
  {Ma}, \citenamefont {Chen}, \citenamefont {Lesanovsky}, \citenamefont
  {Ding},\ and\ \citenamefont {Shi}}]{PhysRevResearch.6.L032069}%
  \BibitemOpen
  \bibfield  {author} {\bibinfo {author} {\bibfnamefont {Zong-Kai}\
  \bibnamefont {Liu}}, \bibinfo {author} {\bibfnamefont {Kong-Hao}\
  \bibnamefont {Sun}}, \bibinfo {author} {\bibfnamefont {Albert}\ \bibnamefont
  {Cabot}}, \bibinfo {author} {\bibfnamefont {Federico}\ \bibnamefont
  {Carollo}}, \bibinfo {author} {\bibfnamefont {Jun}\ \bibnamefont {Zhang}},
  \bibinfo {author} {\bibfnamefont {Zheng-Yuan}\ \bibnamefont {Zhang}},
  \bibinfo {author} {\bibfnamefont {Li-Hua}\ \bibnamefont {Zhang}}, \bibinfo
  {author} {\bibfnamefont {Bang}\ \bibnamefont {Liu}}, \bibinfo {author}
  {\bibfnamefont {Tian-Yu}\ \bibnamefont {Han}}, \bibinfo {author}
  {\bibfnamefont {Qing}\ \bibnamefont {Li}}, \bibinfo {author} {\bibfnamefont
  {Yu}~\bibnamefont {Ma}}, \bibinfo {author} {\bibfnamefont {Han-Chao}\
  \bibnamefont {Chen}}, \bibinfo {author} {\bibfnamefont {Igor}\ \bibnamefont
  {Lesanovsky}}, \bibinfo {author} {\bibfnamefont {Dong-Sheng}\ \bibnamefont
  {Ding}}, \ and\ \bibinfo {author} {\bibfnamefont {Bao-Sen}\ \bibnamefont
  {Shi}},\ }\bibfield  {title} {\enquote {\bibinfo {title} {Emergence of
  subharmonics in a microwave driven dissipative rydberg gas},}\ }\href
  {\doibase 10.1103/PhysRevResearch.6.L032069} {\bibfield  {journal} {\bibinfo
  {journal} {Phys. Rev. Res.}\ }\textbf {\bibinfo {volume} {6}},\ \bibinfo
  {pages} {L032069} (\bibinfo {year} {2024})}\BibitemShut {NoStop}%
\bibitem [{\citenamefont {Fauseweh}(2024)}]{0Quantum}%
  \BibitemOpen
  \bibfield  {author} {\bibinfo {author} {\bibfnamefont {Benedikt}\
  \bibnamefont {Fauseweh}},\ }\bibfield  {title} {\enquote {\bibinfo {title}
  {Quantum many-body simulations on digital quantum computers: State-of-the-art
  and future challenges},}\ }\href {\doibase 10.1038/s41467-024-46402-9}
  {\bibfield  {journal} {\bibinfo  {journal} {Nat. Commun.}\ }\textbf {\bibinfo
  {volume} {15}},\ \bibinfo {pages} {2123} (\bibinfo {year}
  {2024})}\BibitemShut {NoStop}%
\bibitem [{\citenamefont {Senthil}(1998)}]{PhysRevB.57.8375}%
  \BibitemOpen
  \bibfield  {author} {\bibinfo {author} {\bibfnamefont {T.}~\bibnamefont
  {Senthil}},\ }\bibfield  {title} {\enquote {\bibinfo {title} {Properties of
  the random-field ising model in a transverse magnetic field},}\ }\href
  {\doibase 10.1103/PhysRevB.57.8375} {\bibfield  {journal} {\bibinfo
  {journal} {Phys. Rev. B}\ }\textbf {\bibinfo {volume} {57}},\ \bibinfo
  {pages} {8375--8380} (\bibinfo {year} {1998})}\BibitemShut {NoStop}%
\bibitem [{\citenamefont {{do Nascimento}}\ \emph {et~al.}(2017)\citenamefont
  {{do Nascimento}}, \citenamefont {Pacobahyba}, \citenamefont {Neto},
  \citenamefont {Salmon},\ and\ \citenamefont {Plascak}}]{DONASCIMENTO2017224}%
  \BibitemOpen
  \bibfield  {author} {\bibinfo {author} {\bibfnamefont {Denise~A.}\
  \bibnamefont {{do Nascimento}}}, \bibinfo {author} {\bibfnamefont
  {Josefa~T.}\ \bibnamefont {Pacobahyba}}, \bibinfo {author} {\bibfnamefont
  {Minos~A.}\ \bibnamefont {Neto}}, \bibinfo {author} {\bibfnamefont {Octavio
  D.~Rodriguez}\ \bibnamefont {Salmon}}, \ and\ \bibinfo {author}
  {\bibfnamefont {J.A.}\ \bibnamefont {Plascak}},\ }\bibfield  {title}
  {\enquote {\bibinfo {title} {Multicritical behavior of the two-dimensional
  transverse ising metamagnet in a longitudinal magnetic field},}\ }\href
  {\doibase https://doi.org/10.1016/j.physa.2017.01.078} {\bibfield  {journal}
  {\bibinfo  {journal} {Physica A}\ }\textbf {\bibinfo {volume} {474}},\
  \bibinfo {pages} {224--229} (\bibinfo {year} {2017})}\BibitemShut {NoStop}%
\bibitem [{\citenamefont {Kormos}\ \emph {et~al.}(2017)\citenamefont {Kormos},
  \citenamefont {Collura}, \citenamefont {Takács},\ and\ \citenamefont
  {Calabrese}}]{2016Real}%
  \BibitemOpen
  \bibfield  {author} {\bibinfo {author} {\bibfnamefont {Marton}\ \bibnamefont
  {Kormos}}, \bibinfo {author} {\bibfnamefont {Mario}\ \bibnamefont {Collura}},
  \bibinfo {author} {\bibfnamefont {Gabor}\ \bibnamefont {Takács}}, \ and\
  \bibinfo {author} {\bibfnamefont {Pasquale}\ \bibnamefont {Calabrese}},\
  }\bibfield  {title} {\enquote {\bibinfo {title} {Real time confinement
  following a quantum quench to a non-integrable model},}\ }\href {\doibase
  10.1038/nphys3934} {\bibfield  {journal} {\bibinfo  {journal} {Nat. Phys.}\
  }\textbf {\bibinfo {volume} {13}},\ \bibinfo {pages} {246--249} (\bibinfo
  {year} {2017})}\BibitemShut {NoStop}%
\bibitem [{\citenamefont {Bonfim}\ \emph {et~al.}(2019)\citenamefont {Bonfim},
  \citenamefont {Boechat},\ and\ \citenamefont
  {Florencio}}]{PhysRevE.99.012122}%
  \BibitemOpen
  \bibfield  {author} {\bibinfo {author} {\bibfnamefont {O.~F. D.~A.}\
  \bibnamefont {Bonfim}}, \bibinfo {author} {\bibfnamefont {B.}~\bibnamefont
  {Boechat}}, \ and\ \bibinfo {author} {\bibfnamefont {J.}~\bibnamefont
  {Florencio}},\ }\bibfield  {title} {\enquote {\bibinfo {title} {Ground-state
  properties of the one-dimensional transverse ising model in a longitudinal
  magnetic field},}\ }\href {\doibase 10.1103/PhysRevE.99.012122} {\bibfield
  {journal} {\bibinfo  {journal} {Phys. Rev. E}\ }\textbf {\bibinfo {volume}
  {99}},\ \bibinfo {pages} {012122} (\bibinfo {year} {2019})}\BibitemShut
  {NoStop}%
\bibitem [{\citenamefont {Yu}\ and\ \citenamefont
  {Dumke}(2019)}]{PhysRevA.100.022124}%
  \BibitemOpen
  \bibfield  {author} {\bibinfo {author} {\bibfnamefont {Deshui}\ \bibnamefont
  {Yu}}\ and\ \bibinfo {author} {\bibfnamefont {Rainer}\ \bibnamefont
  {Dumke}},\ }\bibfield  {title} {\enquote {\bibinfo {title} {Open ising model
  perturbed by classical colored noise},}\ }\href {\doibase
  10.1103/PhysRevA.100.022124} {\bibfield  {journal} {\bibinfo  {journal}
  {Phys. Rev. A}\ }\textbf {\bibinfo {volume} {100}},\ \bibinfo {pages}
  {022124} (\bibinfo {year} {2019})}\BibitemShut {NoStop}%
\bibitem [{\citenamefont {Mazza}\ \emph {et~al.}(2019)\citenamefont {Mazza},
  \citenamefont {Perfetto}, \citenamefont {Lerose}, \citenamefont {Collura},\
  and\ \citenamefont {Gambassi}}]{PhysRevB.99.180302}%
  \BibitemOpen
  \bibfield  {author} {\bibinfo {author} {\bibfnamefont {Paolo~Pietro}\
  \bibnamefont {Mazza}}, \bibinfo {author} {\bibfnamefont {Gabriele}\
  \bibnamefont {Perfetto}}, \bibinfo {author} {\bibfnamefont {Alessio}\
  \bibnamefont {Lerose}}, \bibinfo {author} {\bibfnamefont {Mario}\
  \bibnamefont {Collura}}, \ and\ \bibinfo {author} {\bibfnamefont {Andrea}\
  \bibnamefont {Gambassi}},\ }\bibfield  {title} {\enquote {\bibinfo {title}
  {Suppression of transport in nondisordered quantum spin chains due to
  confined excitations},}\ }\href {\doibase 10.1103/PhysRevB.99.180302}
  {\bibfield  {journal} {\bibinfo  {journal} {Phys. Rev. B}\ }\textbf {\bibinfo
  {volume} {99}},\ \bibinfo {pages} {180302} (\bibinfo {year}
  {2019})}\BibitemShut {NoStop}%
\bibitem [{\citenamefont {Białończyk}\ and\ \citenamefont
  {Damski}(2020)}]{Białończyk_2020}%
  \BibitemOpen
  \bibfield  {author} {\bibinfo {author} {\bibfnamefont {Michał}\ \bibnamefont
  {Białończyk}}\ and\ \bibinfo {author} {\bibfnamefont {Bogdan}\ \bibnamefont
  {Damski}},\ }\bibfield  {title} {\enquote {\bibinfo {title} {Dynamics of
  longitudinal magnetization in transverse-field quantum ising model: from
  symmetry-breaking gap to kibble–zurek mechanism},}\ }\href {\doibase
  10.1088/1742-5468/ab609a} {\bibfield  {journal} {\bibinfo  {journal} {J.
  Stat. Mech.}\ }\textbf {\bibinfo {volume} {2020}},\ \bibinfo {pages} {013108}
  (\bibinfo {year} {2020})}\BibitemShut {NoStop}%
\bibitem [{\citenamefont {Kshetrimayum}\ \emph {et~al.}(2017)\citenamefont
  {Kshetrimayum}, \citenamefont {Weimer},\ and\ \citenamefont {Orús}}]{2017A}%
  \BibitemOpen
  \bibfield  {author} {\bibinfo {author} {\bibfnamefont {Augustine}\
  \bibnamefont {Kshetrimayum}}, \bibinfo {author} {\bibfnamefont {Hendrik}\
  \bibnamefont {Weimer}}, \ and\ \bibinfo {author} {\bibfnamefont {Román}\
  \bibnamefont {Orús}},\ }\bibfield  {title} {\enquote {\bibinfo {title} {A
  simple tensor network algorithm for two-dimensional steady states},}\ }\href
  {\doibase 10.1038/s41467-017-01511-6} {\bibfield  {journal} {\bibinfo
  {journal} {Nat. Commun.}\ }\textbf {\bibinfo {volume} {8}},\ \bibinfo {pages}
  {1291} (\bibinfo {year} {2017})}\BibitemShut {NoStop}%
\bibitem [{\citenamefont {Wang}\ \emph {et~al.}(2021)\citenamefont {Wang},
  \citenamefont {Zou}, \citenamefont {H\'ods\'agi}, \citenamefont {Kormos},
  \citenamefont {Tak\'acs},\ and\ \citenamefont {Wu}}]{PhysRevB.103.235117}%
  \BibitemOpen
  \bibfield  {author} {\bibinfo {author} {\bibfnamefont {Xiao}\ \bibnamefont
  {Wang}}, \bibinfo {author} {\bibfnamefont {Haiyuan}\ \bibnamefont {Zou}},
  \bibinfo {author} {\bibfnamefont {Krist\'of}\ \bibnamefont {H\'ods\'agi}},
  \bibinfo {author} {\bibfnamefont {M\'arton}\ \bibnamefont {Kormos}}, \bibinfo
  {author} {\bibfnamefont {G\'abor}\ \bibnamefont {Tak\'acs}}, \ and\ \bibinfo
  {author} {\bibfnamefont {Jianda}\ \bibnamefont {Wu}},\ }\bibfield  {title}
  {\enquote {\bibinfo {title} {Cascade of singularities in the spin dynamics of
  a perturbed quantum critical ising chain},}\ }\href {\doibase
  10.1103/PhysRevB.103.235117} {\bibfield  {journal} {\bibinfo  {journal}
  {Phys. Rev. B}\ }\textbf {\bibinfo {volume} {103}},\ \bibinfo {pages}
  {235117} (\bibinfo {year} {2021})}\BibitemShut {NoStop}%
\bibitem [{\citenamefont {Montenegro}\ \emph {et~al.}(2021)\citenamefont
  {Montenegro}, \citenamefont {Mishra},\ and\ \citenamefont
  {Bayat}}]{PhysRevLett.126.200501}%
  \BibitemOpen
  \bibfield  {author} {\bibinfo {author} {\bibfnamefont {Victor}\ \bibnamefont
  {Montenegro}}, \bibinfo {author} {\bibfnamefont {Utkarsh}\ \bibnamefont
  {Mishra}}, \ and\ \bibinfo {author} {\bibfnamefont {Abolfazl}\ \bibnamefont
  {Bayat}},\ }\bibfield  {title} {\enquote {\bibinfo {title} {Global sensing
  and its impact for quantum many-body probes with criticality},}\ }\href
  {\doibase 10.1103/PhysRevLett.126.200501} {\bibfield  {journal} {\bibinfo
  {journal} {Phys. Rev. Lett.}\ }\textbf {\bibinfo {volume} {126}},\ \bibinfo
  {pages} {200501} (\bibinfo {year} {2021})}\BibitemShut {NoStop}%
\bibitem [{\citenamefont {Birnkammer}\ \emph {et~al.}(2022)\citenamefont
  {Birnkammer}, \citenamefont {Bastianello},\ and\ \citenamefont
  {Knap}}]{ncs41467-022-35301-6}%
  \BibitemOpen
  \bibfield  {author} {\bibinfo {author} {\bibfnamefont {Stefan}\ \bibnamefont
  {Birnkammer}}, \bibinfo {author} {\bibfnamefont {Alvise}\ \bibnamefont
  {Bastianello}}, \ and\ \bibinfo {author} {\bibfnamefont {Michael}\
  \bibnamefont {Knap}},\ }\bibfield  {title} {\enquote {\bibinfo {title}
  {Prethermalization in one-dimensional quantum many-body systems with
  confinement},}\ }\href {\doibase 10.1038/s41467-022-35301-6} {\bibfield
  {journal} {\bibinfo  {journal} {Nat. Commun.}\ }\textbf {\bibinfo {volume}
  {13}},\ \bibinfo {pages} {7663} (\bibinfo {year} {2022})}\BibitemShut
  {NoStop}%
\bibitem [{\citenamefont {Scopa}\ \emph {et~al.}(2022)\citenamefont {Scopa},
  \citenamefont {Calabrese},\ and\ \citenamefont
  {Bastianello}}]{PhysRevB.105.125413}%
  \BibitemOpen
  \bibfield  {author} {\bibinfo {author} {\bibfnamefont {Stefano}\ \bibnamefont
  {Scopa}}, \bibinfo {author} {\bibfnamefont {Pasquale}\ \bibnamefont
  {Calabrese}}, \ and\ \bibinfo {author} {\bibfnamefont {Alvise}\ \bibnamefont
  {Bastianello}},\ }\bibfield  {title} {\enquote {\bibinfo {title}
  {Entanglement dynamics in confining spin chains},}\ }\href {\doibase
  10.1103/PhysRevB.105.125413} {\bibfield  {journal} {\bibinfo  {journal}
  {Phys. Rev. B}\ }\textbf {\bibinfo {volume} {105}},\ \bibinfo {pages}
  {125413} (\bibinfo {year} {2022})}\BibitemShut {NoStop}%
\bibitem [{\citenamefont {Peng}\ and\ \citenamefont
  {Cui}(2022)}]{PhysRevB.106.214311}%
  \BibitemOpen
  \bibfield  {author} {\bibinfo {author} {\bibfnamefont {Cheng}\ \bibnamefont
  {Peng}}\ and\ \bibinfo {author} {\bibfnamefont {Xiaoling}\ \bibnamefont
  {Cui}},\ }\bibfield  {title} {\enquote {\bibinfo {title} {Bridging quantum
  many-body scars and quantum integrability in ising chains with transverse and
  longitudinal fields},}\ }\href {\doibase 10.1103/PhysRevB.106.214311}
  {\bibfield  {journal} {\bibinfo  {journal} {Phys. Rev. B}\ }\textbf {\bibinfo
  {volume} {106}},\ \bibinfo {pages} {214311} (\bibinfo {year}
  {2022})}\BibitemShut {NoStop}%
\bibitem [{\citenamefont {Roberts}\ and\ \citenamefont
  {Clerk}(2023)}]{PhysRevLett.131.190403}%
  \BibitemOpen
  \bibfield  {author} {\bibinfo {author} {\bibfnamefont {David}\ \bibnamefont
  {Roberts}}\ and\ \bibinfo {author} {\bibfnamefont {A.~A.}\ \bibnamefont
  {Clerk}},\ }\bibfield  {title} {\enquote {\bibinfo {title} {Exact solution of
  the infinite-range dissipative transverse-field ising model},}\ }\href
  {\doibase 10.1103/PhysRevLett.131.190403} {\bibfield  {journal} {\bibinfo
  {journal} {Phys. Rev. Lett.}\ }\textbf {\bibinfo {volume} {131}},\ \bibinfo
  {pages} {190403} (\bibinfo {year} {2023})}\BibitemShut {NoStop}%
\bibitem [{\citenamefont {Zhang}\ and\ \citenamefont
  {Song}(2024)}]{PhysRevB.109.104312}%
  \BibitemOpen
  \bibfield  {author} {\bibinfo {author} {\bibfnamefont {K.~L.}\ \bibnamefont
  {Zhang}}\ and\ \bibinfo {author} {\bibfnamefont {Z.}~\bibnamefont {Song}},\
  }\bibfield  {title} {\enquote {\bibinfo {title} {Magnetic bloch oscillations
  in a non-hermitian quantum ising chain},}\ }\href {\doibase
  10.1103/PhysRevB.109.104312} {\bibfield  {journal} {\bibinfo  {journal}
  {Phys. Rev. B}\ }\textbf {\bibinfo {volume} {109}},\ \bibinfo {pages}
  {104312} (\bibinfo {year} {2024})}\BibitemShut {NoStop}%
\bibitem [{\citenamefont {Narasimhan}\ \emph {et~al.}(2024)\citenamefont
  {Narasimhan}, \citenamefont {Humeniuk}, \citenamefont {Roy},\ and\
  \citenamefont {Drouin-Touchette}}]{PhysRevB.110.054432}%
  \BibitemOpen
  \bibfield  {author} {\bibinfo {author} {\bibfnamefont {Pratyankara}\
  \bibnamefont {Narasimhan}}, \bibinfo {author} {\bibfnamefont {Stephan}\
  \bibnamefont {Humeniuk}}, \bibinfo {author} {\bibfnamefont {Ananda}\
  \bibnamefont {Roy}}, \ and\ \bibinfo {author} {\bibfnamefont {Victor}\
  \bibnamefont {Drouin-Touchette}},\ }\bibfield  {title} {\enquote {\bibinfo
  {title} {Simulating the transverse-field ising model on the kagome lattice
  using a programmable quantum annealer},}\ }\href {\doibase
  10.1103/PhysRevB.110.054432} {\bibfield  {journal} {\bibinfo  {journal}
  {Phys. Rev. B}\ }\textbf {\bibinfo {volume} {110}},\ \bibinfo {pages}
  {054432} (\bibinfo {year} {2024})}\BibitemShut {NoStop}%
\bibitem [{\citenamefont {Xiao}\ \emph {et~al.}(2024)\citenamefont {Xiao},
  \citenamefont {Zhang}, \citenamefont {Shen}, \citenamefont {Bao},\ and\
  \citenamefont {Gambassi}}]{10.1007/s11128-024-04567-8}%
  \BibitemOpen
  \bibfield  {author} {\bibinfo {author} {\bibfnamefont {Wenyuan}\ \bibnamefont
  {Xiao}}, \bibinfo {author} {\bibfnamefont {Wenqiong}\ \bibnamefont {Zhang}},
  \bibinfo {author} {\bibfnamefont {Longhui}\ \bibnamefont {Shen}}, \bibinfo
  {author} {\bibfnamefont {Jia}\ \bibnamefont {Bao}}, \ and\ \bibinfo {author}
  {\bibfnamefont {Andrea}\ \bibnamefont {Gambassi}},\ }\bibfield  {title}
  {\enquote {\bibinfo {title} {Suppression of transport in nondisordered
  quantum spin chains due to confined excitations},}\ }\href {\doibase
  10.1007/s11128-024-04567-8} {\bibfield  {journal} {\bibinfo  {journal}
  {Quantum Inf. Process.}\ }\textbf {\bibinfo {volume} {23}},\ \bibinfo {pages}
  {347} (\bibinfo {year} {2024})}\BibitemShut {NoStop}%
\bibitem [{\citenamefont {Simon}\ \emph {et~al.}(2011)\citenamefont {Simon},
  \citenamefont {Bakr}, \citenamefont {Ma}, \citenamefont {Tai}, \citenamefont
  {Preiss},\ and\ \citenamefont {Greiner}}]{nature09994}%
  \BibitemOpen
  \bibfield  {author} {\bibinfo {author} {\bibfnamefont {Jonathan}\
  \bibnamefont {Simon}}, \bibinfo {author} {\bibfnamefont {Waseem~S}\
  \bibnamefont {Bakr}}, \bibinfo {author} {\bibfnamefont {Ruichao}\
  \bibnamefont {Ma}}, \bibinfo {author} {\bibfnamefont {M~Eric}\ \bibnamefont
  {Tai}}, \bibinfo {author} {\bibfnamefont {Philipp~M}\ \bibnamefont {Preiss}},
  \ and\ \bibinfo {author} {\bibfnamefont {Markus}\ \bibnamefont {Greiner}},\
  }\bibfield  {title} {\enquote {\bibinfo {title} {Quantum simulation of
  antiferromagnetic spin chains in an optical lattice},}\ }\href {\doibase
  10.1038/nature09994} {\bibfield  {journal} {\bibinfo  {journal} {Nature}\
  }\textbf {\bibinfo {volume} {472}},\ \bibinfo {pages} {307--312} (\bibinfo
  {year} {2011})}\BibitemShut {NoStop}%
\bibitem [{\citenamefont {Weber}\ \emph {et~al.}(2022)\citenamefont {Weber},
  \citenamefont {Luitz},\ and\ \citenamefont
  {Assaad}}]{PhysRevLett.129.056402}%
  \BibitemOpen
  \bibfield  {author} {\bibinfo {author} {\bibfnamefont {Manuel}\ \bibnamefont
  {Weber}}, \bibinfo {author} {\bibfnamefont {David~J.}\ \bibnamefont {Luitz}},
  \ and\ \bibinfo {author} {\bibfnamefont {Fakher~F.}\ \bibnamefont {Assaad}},\
  }\bibfield  {title} {\enquote {\bibinfo {title} {Dissipation-induced order:
  The $s=1/2$ quantum spin chain coupled to an ohmic bath},}\ }\href {\doibase
  10.1103/PhysRevLett.129.056402} {\bibfield  {journal} {\bibinfo  {journal}
  {Phys. Rev. Lett.}\ }\textbf {\bibinfo {volume} {129}},\ \bibinfo {pages}
  {056402} (\bibinfo {year} {2022})}\BibitemShut {NoStop}%
\bibitem [{\citenamefont {Min}\ \emph {et~al.}(2024)\citenamefont {Min},
  \citenamefont {Anto-Sztrikacs}, \citenamefont {Brenes},\ and\ \citenamefont
  {Segal}}]{PhysRevLett.132.266701}%
  \BibitemOpen
  \bibfield  {author} {\bibinfo {author} {\bibfnamefont {Brett}\ \bibnamefont
  {Min}}, \bibinfo {author} {\bibfnamefont {Nicholas}\ \bibnamefont
  {Anto-Sztrikacs}}, \bibinfo {author} {\bibfnamefont {Marlon}\ \bibnamefont
  {Brenes}}, \ and\ \bibinfo {author} {\bibfnamefont {Dvira}\ \bibnamefont
  {Segal}},\ }\bibfield  {title} {\enquote {\bibinfo {title} {Bath-engineering
  magnetic order in quantum spin chains: An analytic mapping approach},}\
  }\href {\doibase 10.1103/PhysRevLett.132.266701} {\bibfield  {journal}
  {\bibinfo  {journal} {Phys. Rev. Lett.}\ }\textbf {\bibinfo {volume} {132}},\
  \bibinfo {pages} {266701} (\bibinfo {year} {2024})}\BibitemShut {NoStop}%
\bibitem [{\citenamefont {Cavalcante}\ \emph {et~al.}(2024)\citenamefont
  {Cavalcante}, \citenamefont {Bonan\ifmmode~\mbox{\c{c}}\else \c{c}\fi{}a},
  \citenamefont {Miranda},\ and\ \citenamefont
  {Deffner}}]{PhysRevB.110.064304}%
  \BibitemOpen
  \bibfield  {author} {\bibinfo {author} {\bibfnamefont {Moallison~F.}\
  \bibnamefont {Cavalcante}}, \bibinfo {author} {\bibfnamefont {Marcus V.~S.}\
  \bibnamefont {Bonan\ifmmode~\mbox{\c{c}}\else \c{c}\fi{}a}}, \bibinfo
  {author} {\bibfnamefont {Eduardo}\ \bibnamefont {Miranda}}, \ and\ \bibinfo
  {author} {\bibfnamefont {Sebastian}\ \bibnamefont {Deffner}},\ }\bibfield
  {title} {\enquote {\bibinfo {title} {Nanowelding of quantum
  spin-$\frac{1}{2}$ chains at minimal dissipation},}\ }\href {\doibase
  10.1103/PhysRevB.110.064304} {\bibfield  {journal} {\bibinfo  {journal}
  {Phys. Rev. B}\ }\textbf {\bibinfo {volume} {110}},\ \bibinfo {pages}
  {064304} (\bibinfo {year} {2024})}\BibitemShut {NoStop}%
\bibitem [{\citenamefont {Fazio}\ \emph {et~al.}(2024)\citenamefont {Fazio},
  \citenamefont {Keeling}, \citenamefont {Mazza},\ and\ \citenamefont
  {Schirò}}]{fazio2024manybodyopenquantumsystems}%
  \BibitemOpen
  \bibfield  {author} {\bibinfo {author} {\bibfnamefont {Rosario}\ \bibnamefont
  {Fazio}}, \bibinfo {author} {\bibfnamefont {Jonathan}\ \bibnamefont
  {Keeling}}, \bibinfo {author} {\bibfnamefont {Leonardo}\ \bibnamefont
  {Mazza}}, \ and\ \bibinfo {author} {\bibfnamefont {Marco}\ \bibnamefont
  {Schirò}},\ }\href {https://arxiv.org/abs/2409.10300} {\enquote {\bibinfo
  {title} {Many-body open quantum systems},}\ } (\bibinfo {year} {2024}),\
  \Eprint {http://arxiv.org/abs/2409.10300} {arXiv:2409.10300 [quant-ph]}
  \BibitemShut {NoStop}%
\bibitem [{\citenamefont {Minganti}\ and\ \citenamefont
  {Biella}(2024)}]{minganti2024openquantumsystems}%
  \BibitemOpen
  \bibfield  {author} {\bibinfo {author} {\bibfnamefont {Fabrizio}\
  \bibnamefont {Minganti}}\ and\ \bibinfo {author} {\bibfnamefont {Alberto}\
  \bibnamefont {Biella}},\ }\href {https://arxiv.org/abs/2407.16855} {\enquote
  {\bibinfo {title} {Open quantum systems -- a brief introduction},}\ }
  (\bibinfo {year} {2024}),\ \Eprint {http://arxiv.org/abs/2407.16855}
  {arXiv:2407.16855 [quant-ph]} \BibitemShut {NoStop}%
\bibitem [{\citenamefont {Sperstad}\ \emph {et~al.}(2010)\citenamefont
  {Sperstad}, \citenamefont {Stiansen},\ and\ \citenamefont
  {Sudb\o{}}}]{PhysRevB.81.104302}%
  \BibitemOpen
  \bibfield  {author} {\bibinfo {author} {\bibfnamefont {Iver~Bakken}\
  \bibnamefont {Sperstad}}, \bibinfo {author} {\bibfnamefont {Einar~B.}\
  \bibnamefont {Stiansen}}, \ and\ \bibinfo {author} {\bibfnamefont {Asle}\
  \bibnamefont {Sudb\o{}}},\ }\bibfield  {title} {\enquote {\bibinfo {title}
  {Monte carlo simulations of dissipative quantum ising models},}\ }\href
  {\doibase 10.1103/PhysRevB.81.104302} {\bibfield  {journal} {\bibinfo
  {journal} {Phys. Rev. B}\ }\textbf {\bibinfo {volume} {81}},\ \bibinfo
  {pages} {104302} (\bibinfo {year} {2010})}\BibitemShut {NoStop}%
\bibitem [{\citenamefont {Koziol}\ \emph {et~al.}(2021)\citenamefont {Koziol},
  \citenamefont {Langheld}, \citenamefont {Kapfer},\ and\ \citenamefont
  {Schmidt}}]{PhysRevB.103.245135}%
  \BibitemOpen
  \bibfield  {author} {\bibinfo {author} {\bibfnamefont {Jan~Alexander}\
  \bibnamefont {Koziol}}, \bibinfo {author} {\bibfnamefont {Anja}\ \bibnamefont
  {Langheld}}, \bibinfo {author} {\bibfnamefont {Sebastian~C.}\ \bibnamefont
  {Kapfer}}, \ and\ \bibinfo {author} {\bibfnamefont {Kai~Phillip}\
  \bibnamefont {Schmidt}},\ }\bibfield  {title} {\enquote {\bibinfo {title}
  {Quantum-critical properties of the long-range transverse-field ising model
  from quantum monte carlo simulations},}\ }\href {\doibase
  10.1103/PhysRevB.103.245135} {\bibfield  {journal} {\bibinfo  {journal}
  {Phys. Rev. B}\ }\textbf {\bibinfo {volume} {103}},\ \bibinfo {pages}
  {245135} (\bibinfo {year} {2021})}\BibitemShut {NoStop}%
\bibitem [{\citenamefont {Kiss}\ \emph {et~al.}(2024)\citenamefont {Kiss},
  \citenamefont {Zar\'and},\ and\ \citenamefont {Lovas}}]{PhysRevB.109.024431}%
  \BibitemOpen
  \bibfield  {author} {\bibinfo {author} {\bibfnamefont {Annam\'aria}\
  \bibnamefont {Kiss}}, \bibinfo {author} {\bibfnamefont {Gergely}\
  \bibnamefont {Zar\'and}}, \ and\ \bibinfo {author} {\bibfnamefont {Izabella}\
  \bibnamefont {Lovas}},\ }\bibfield  {title} {\enquote {\bibinfo {title}
  {Complete replica solution for the transverse field sherrington-kirkpatrick
  spin glass model with continuous-time quantum monte carlo method},}\ }\href
  {\doibase 10.1103/PhysRevB.109.024431} {\bibfield  {journal} {\bibinfo
  {journal} {Phys. Rev. B}\ }\textbf {\bibinfo {volume} {109}},\ \bibinfo
  {pages} {024431} (\bibinfo {year} {2024})}\BibitemShut {NoStop}%
\bibitem [{\citenamefont {Humeniuk}(2020)}]{Humeniuk_2020}%
  \BibitemOpen
  \bibfield  {author} {\bibinfo {author} {\bibfnamefont {Stephan}\ \bibnamefont
  {Humeniuk}},\ }\bibfield  {title} {\enquote {\bibinfo {title} {Thermal
  kosterlitz–thouless transitions in the 1/r2 long-range ferromagnetic
  quantum ising chain revisited},}\ }\href {\doibase 10.1088/1742-5468/ab900c}
  {\bibfield  {journal} {\bibinfo  {journal} {J. Stat. Mech.}\ }\textbf
  {\bibinfo {volume} {2020}},\ \bibinfo {pages} {063105} (\bibinfo {year}
  {2020})}\BibitemShut {NoStop}%
\bibitem [{\citenamefont {Laurell}\ \emph {et~al.}(2023)\citenamefont
  {Laurell}, \citenamefont {Alvarez},\ and\ \citenamefont
  {Dagotto}}]{PhysRevB.107.104414}%
  \BibitemOpen
  \bibfield  {author} {\bibinfo {author} {\bibfnamefont {Pontus}\ \bibnamefont
  {Laurell}}, \bibinfo {author} {\bibfnamefont {Gonzalo}\ \bibnamefont
  {Alvarez}}, \ and\ \bibinfo {author} {\bibfnamefont {Elbio}\ \bibnamefont
  {Dagotto}},\ }\bibfield  {title} {\enquote {\bibinfo {title} {Spin dynamics
  of the generalized quantum spin compass chain},}\ }\href {\doibase
  10.1103/PhysRevB.107.104414} {\bibfield  {journal} {\bibinfo  {journal}
  {Phys. Rev. B}\ }\textbf {\bibinfo {volume} {107}},\ \bibinfo {pages}
  {104414} (\bibinfo {year} {2023})}\BibitemShut {NoStop}%
\bibitem [{\citenamefont {Martin}\ and\ \citenamefont
  {Grover}(2023)}]{PhysRevResearch.5.043270}%
  \BibitemOpen
  \bibfield  {author} {\bibinfo {author} {\bibfnamefont {Simon}\ \bibnamefont
  {Martin}}\ and\ \bibinfo {author} {\bibfnamefont {Tarun}\ \bibnamefont
  {Grover}},\ }\bibfield  {title} {\enquote {\bibinfo {title} {Critical phase
  induced by berry phase and dissipation in a spin chain},}\ }\href {\doibase
  10.1103/PhysRevResearch.5.043270} {\bibfield  {journal} {\bibinfo  {journal}
  {Phys. Rev. Res.}\ }\textbf {\bibinfo {volume} {5}},\ \bibinfo {pages}
  {043270} (\bibinfo {year} {2023})}\BibitemShut {NoStop}%
\bibitem [{\citenamefont {Schrder}\ \emph {et~al.}(2019)\citenamefont
  {Schrder}, \citenamefont {Turban}, \citenamefont {Musser}, \citenamefont
  {Hine},\ and\ \citenamefont {Chin}}]{2019Tensor}%
  \BibitemOpen
  \bibfield  {author} {\bibinfo {author} {\bibfnamefont {Florian}\ \bibnamefont
  {Schrder}}, \bibinfo {author} {\bibfnamefont {David}\ \bibnamefont {Turban}},
  \bibinfo {author} {\bibfnamefont {Andrew}\ \bibnamefont {Musser}}, \bibinfo
  {author} {\bibfnamefont {Nicholas}\ \bibnamefont {Hine}}, \ and\ \bibinfo
  {author} {\bibfnamefont {Alex}\ \bibnamefont {Chin}},\ }\bibfield  {title}
  {\enquote {\bibinfo {title} {Tensor network simulation of multi-environmental
  open quantum dynamics via machine learning and entanglement
  renormalisation},}\ }\href {\doibase 10.1038/s41467-019-09039-7} {\bibfield
  {journal} {\bibinfo  {journal} {Nat. Commun.}\ }\textbf {\bibinfo {volume}
  {10}},\ \bibinfo {pages} {1062} (\bibinfo {year} {2019})}\BibitemShut
  {NoStop}%
\bibitem [{\citenamefont {Jin}\ \emph {et~al.}(2016)\citenamefont {Jin},
  \citenamefont {Biella}, \citenamefont {Viyuela}, \citenamefont {Mazza},
  \citenamefont {Keeling}, \citenamefont {Fazio},\ and\ \citenamefont
  {Rossini}}]{PhysRevX.6.031011}%
  \BibitemOpen
  \bibfield  {author} {\bibinfo {author} {\bibfnamefont {Jiasen}\ \bibnamefont
  {Jin}}, \bibinfo {author} {\bibfnamefont {Alberto}\ \bibnamefont {Biella}},
  \bibinfo {author} {\bibfnamefont {Oscar}\ \bibnamefont {Viyuela}}, \bibinfo
  {author} {\bibfnamefont {Leonardo}\ \bibnamefont {Mazza}}, \bibinfo {author}
  {\bibfnamefont {Jonathan}\ \bibnamefont {Keeling}}, \bibinfo {author}
  {\bibfnamefont {Rosario}\ \bibnamefont {Fazio}}, \ and\ \bibinfo {author}
  {\bibfnamefont {Davide}\ \bibnamefont {Rossini}},\ }\bibfield  {title}
  {\enquote {\bibinfo {title} {Cluster mean-field approach to the steady-state
  phase diagram of dissipative spin systems},}\ }\href {\doibase
  10.1103/PhysRevX.6.031011} {\bibfield  {journal} {\bibinfo  {journal} {Phys.
  Rev. X}\ }\textbf {\bibinfo {volume} {6}},\ \bibinfo {pages} {031011}
  (\bibinfo {year} {2016})}\BibitemShut {NoStop}%
\bibitem [{\citenamefont {Slagle}\ \emph {et~al.}(2022)\citenamefont {Slagle},
  \citenamefont {Liu}, \citenamefont {Aasen}, \citenamefont {Pichler},
  \citenamefont {Mong}, \citenamefont {Chen}, \citenamefont {Endres},\ and\
  \citenamefont {Alicea}}]{PhysRevB.106.115122}%
  \BibitemOpen
  \bibfield  {author} {\bibinfo {author} {\bibfnamefont {Kevin}\ \bibnamefont
  {Slagle}}, \bibinfo {author} {\bibfnamefont {Yue}\ \bibnamefont {Liu}},
  \bibinfo {author} {\bibfnamefont {David}\ \bibnamefont {Aasen}}, \bibinfo
  {author} {\bibfnamefont {Hannes}\ \bibnamefont {Pichler}}, \bibinfo {author}
  {\bibfnamefont {Roger S.~K.}\ \bibnamefont {Mong}}, \bibinfo {author}
  {\bibfnamefont {Xie}\ \bibnamefont {Chen}}, \bibinfo {author} {\bibfnamefont
  {Manuel}\ \bibnamefont {Endres}}, \ and\ \bibinfo {author} {\bibfnamefont
  {Jason}\ \bibnamefont {Alicea}},\ }\bibfield  {title} {\enquote {\bibinfo
  {title} {Quantum spin liquids bootstrapped from ising criticality in rydberg
  arrays},}\ }\href {\doibase 10.1103/PhysRevB.106.115122} {\bibfield
  {journal} {\bibinfo  {journal} {Phys. Rev. B}\ }\textbf {\bibinfo {volume}
  {106}},\ \bibinfo {pages} {115122} (\bibinfo {year} {2022})}\BibitemShut
  {NoStop}%
\bibitem [{\citenamefont {Jin}\ \emph {et~al.}(2018)\citenamefont {Jin},
  \citenamefont {Biella}, \citenamefont {Viyuela}, \citenamefont {Ciuti},
  \citenamefont {Fazio},\ and\ \citenamefont {Rossini}}]{PhysRevB.98.241108}%
  \BibitemOpen
  \bibfield  {author} {\bibinfo {author} {\bibfnamefont {Jiasen}\ \bibnamefont
  {Jin}}, \bibinfo {author} {\bibfnamefont {Alberto}\ \bibnamefont {Biella}},
  \bibinfo {author} {\bibfnamefont {Oscar}\ \bibnamefont {Viyuela}}, \bibinfo
  {author} {\bibfnamefont {Cristiano}\ \bibnamefont {Ciuti}}, \bibinfo {author}
  {\bibfnamefont {Rosario}\ \bibnamefont {Fazio}}, \ and\ \bibinfo {author}
  {\bibfnamefont {Davide}\ \bibnamefont {Rossini}},\ }\bibfield  {title}
  {\enquote {\bibinfo {title} {Phase diagram of the dissipative quantum ising
  model on a square lattice},}\ }\href {\doibase 10.1103/PhysRevB.98.241108}
  {\bibfield  {journal} {\bibinfo  {journal} {Phys. Rev. B}\ }\textbf {\bibinfo
  {volume} {98}},\ \bibinfo {pages} {241108(R)} (\bibinfo {year}
  {2018})}\BibitemShut {NoStop}%
\bibitem [{\citenamefont {Nathan}\ and\ \citenamefont
  {Rudner}(2020)}]{PhysRevB.102.115109}%
  \BibitemOpen
  \bibfield  {author} {\bibinfo {author} {\bibfnamefont {Frederik}\
  \bibnamefont {Nathan}}\ and\ \bibinfo {author} {\bibfnamefont {Mark~S.}\
  \bibnamefont {Rudner}},\ }\bibfield  {title} {\enquote {\bibinfo {title}
  {Universal lindblad equation for open quantum systems},}\ }\href {\doibase
  10.1103/PhysRevB.102.115109} {\bibfield  {journal} {\bibinfo  {journal}
  {Phys. Rev. B}\ }\textbf {\bibinfo {volume} {102}},\ \bibinfo {pages}
  {115109} (\bibinfo {year} {2020})}\BibitemShut {NoStop}%
\bibitem [{\citenamefont {Breuer}\ and\ \citenamefont
  {Petruccione}(2002)}]{breuer2002theory}%
  \BibitemOpen
  \bibfield  {author} {\bibinfo {author} {\bibfnamefont {Heinz-Peter}\
  \bibnamefont {Breuer}}\ and\ \bibinfo {author} {\bibfnamefont {Francesco}\
  \bibnamefont {Petruccione}},\ }\href@noop {} {\emph {\bibinfo {title} {The
  theory of open quantum systems}}}\ (\bibinfo  {publisher} {Oxford University
  Press on Demand},\ \bibinfo {year} {2002})\BibitemShut {NoStop}%
\bibitem [{\citenamefont {Manzano}(2020)}]{10.1063/1.5115323}%
  \BibitemOpen
  \bibfield  {author} {\bibinfo {author} {\bibfnamefont {Daniel}\ \bibnamefont
  {Manzano}},\ }\bibfield  {title} {\enquote {\bibinfo {title} {{A short
  introduction to the Lindblad master equation}},}\ }\href {\doibase
  10.1063/1.5115323} {\bibfield  {journal} {\bibinfo  {journal} {AIP Advances}\
  }\textbf {\bibinfo {volume} {10}},\ \bibinfo {pages} {025106} (\bibinfo
  {year} {2020})}\BibitemShut {NoStop}%
\bibitem [{\citenamefont {Levy}\ and\ \citenamefont
  {Kosloff}(2014)}]{Levy_2014}%
  \BibitemOpen
  \bibfield  {author} {\bibinfo {author} {\bibfnamefont {Amikam}\ \bibnamefont
  {Levy}}\ and\ \bibinfo {author} {\bibfnamefont {Ronnie}\ \bibnamefont
  {Kosloff}},\ }\bibfield  {title} {\enquote {\bibinfo {title} {The local
  approach to quantum transport may violate the second law of
  thermodynamics},}\ }\href {\doibase 10.1209/0295-5075/107/20004} {\bibfield
  {journal} {\bibinfo  {journal} {EPL}\ }\textbf {\bibinfo {volume} {107}},\
  \bibinfo {pages} {20004} (\bibinfo {year} {2014})}\BibitemShut {NoStop}%
\bibitem [{\citenamefont {Hofer}\ \emph {et~al.}(2017)\citenamefont {Hofer},
  \citenamefont {Perarnau-Llobet}, \citenamefont {Miranda}, \citenamefont
  {Haack}, \citenamefont {Silva}, \citenamefont {Brask},\ and\ \citenamefont
  {Brunner}}]{Hofer_2017}%
  \BibitemOpen
  \bibfield  {author} {\bibinfo {author} {\bibfnamefont {Patrick~P}\
  \bibnamefont {Hofer}}, \bibinfo {author} {\bibfnamefont {Martí}\
  \bibnamefont {Perarnau-Llobet}}, \bibinfo {author} {\bibfnamefont
  {L~David~M}\ \bibnamefont {Miranda}}, \bibinfo {author} {\bibfnamefont
  {Géraldine}\ \bibnamefont {Haack}}, \bibinfo {author} {\bibfnamefont
  {Ralph}\ \bibnamefont {Silva}}, \bibinfo {author} {\bibfnamefont
  {Jonatan~Bohr}\ \bibnamefont {Brask}}, \ and\ \bibinfo {author}
  {\bibfnamefont {Nicolas}\ \bibnamefont {Brunner}},\ }\bibfield  {title}
  {\enquote {\bibinfo {title} {Markovian master equations for quantum thermal
  machines: local versus global approach},}\ }\href {\doibase
  10.1088/1367-2630/aa964f} {\bibfield  {journal} {\bibinfo  {journal} {New J.
  Phys.}\ }\textbf {\bibinfo {volume} {19}},\ \bibinfo {pages} {123037}
  (\bibinfo {year} {2017})}\BibitemShut {NoStop}%
\bibitem [{\citenamefont {Balachandran}\ \emph {et~al.}(2018)\citenamefont
  {Balachandran}, \citenamefont {Benenti}, \citenamefont {Pereira},
  \citenamefont {Casati},\ and\ \citenamefont
  {Poletti}}]{PhysRevLett.120.200603}%
  \BibitemOpen
  \bibfield  {author} {\bibinfo {author} {\bibfnamefont {Vinitha}\ \bibnamefont
  {Balachandran}}, \bibinfo {author} {\bibfnamefont {Giuliano}\ \bibnamefont
  {Benenti}}, \bibinfo {author} {\bibfnamefont {Emmanuel}\ \bibnamefont
  {Pereira}}, \bibinfo {author} {\bibfnamefont {Giulio}\ \bibnamefont
  {Casati}}, \ and\ \bibinfo {author} {\bibfnamefont {Dario}\ \bibnamefont
  {Poletti}},\ }\bibfield  {title} {\enquote {\bibinfo {title} {Perfect diode
  in quantum spin chains},}\ }\href {\doibase 10.1103/PhysRevLett.120.200603}
  {\bibfield  {journal} {\bibinfo  {journal} {Phys. Rev. Lett.}\ }\textbf
  {\bibinfo {volume} {120}},\ \bibinfo {pages} {200603} (\bibinfo {year}
  {2018})}\BibitemShut {NoStop}%
\bibitem [{\citenamefont {Cattaneo}\ \emph {et~al.}(2019)\citenamefont
  {Cattaneo}, \citenamefont {Giorgi}, \citenamefont {Maniscalco},\ and\
  \citenamefont {Zambrini}}]{Cattaneo_2019}%
  \BibitemOpen
  \bibfield  {author} {\bibinfo {author} {\bibfnamefont {Marco}\ \bibnamefont
  {Cattaneo}}, \bibinfo {author} {\bibfnamefont {Gian~Luca}\ \bibnamefont
  {Giorgi}}, \bibinfo {author} {\bibfnamefont {Sabrina}\ \bibnamefont
  {Maniscalco}}, \ and\ \bibinfo {author} {\bibfnamefont {Roberta}\
  \bibnamefont {Zambrini}},\ }\bibfield  {title} {\enquote {\bibinfo {title}
  {Local versus global master equation with common and separate baths:
  superiority of the global approach in partial secular approximation},}\
  }\href {\doibase 10.1088/1367-2630/ab54ac} {\bibfield  {journal} {\bibinfo
  {journal} {New J. Phys.}\ }\textbf {\bibinfo {volume} {21}},\ \bibinfo
  {pages} {113045} (\bibinfo {year} {2019})}\BibitemShut {NoStop}%
\bibitem [{\citenamefont {Hewgill}\ \emph {et~al.}(2021)\citenamefont
  {Hewgill}, \citenamefont {De~Chiara},\ and\ \citenamefont
  {Imparato}}]{hewgill2021quantum}%
  \BibitemOpen
  \bibfield  {author} {\bibinfo {author} {\bibfnamefont {Adam}\ \bibnamefont
  {Hewgill}}, \bibinfo {author} {\bibfnamefont {Gabriele}\ \bibnamefont
  {De~Chiara}}, \ and\ \bibinfo {author} {\bibfnamefont {Alberto}\ \bibnamefont
  {Imparato}},\ }\bibfield  {title} {\enquote {\bibinfo {title} {Quantum
  thermodynamically consistent local master equations},}\ }\href {\doibase
  10.1103/PhysRevResearch.3.013165} {\bibfield  {journal} {\bibinfo  {journal}
  {Phys. Rev. Research}\ }\textbf {\bibinfo {volume} {3}},\ \bibinfo {pages}
  {013165} (\bibinfo {year} {2021})}\BibitemShut {NoStop}%
\bibitem [{\citenamefont {Lee}\ \emph {et~al.}(2022)\citenamefont {Lee},
  \citenamefont {Balachandran}, \citenamefont {Guo},\ and\ \citenamefont
  {Poletti}}]{PhysRevE.105.024120}%
  \BibitemOpen
  \bibfield  {author} {\bibinfo {author} {\bibfnamefont {Kang~Hao}\
  \bibnamefont {Lee}}, \bibinfo {author} {\bibfnamefont {Vinitha}\ \bibnamefont
  {Balachandran}}, \bibinfo {author} {\bibfnamefont {Chu}\ \bibnamefont {Guo}},
  \ and\ \bibinfo {author} {\bibfnamefont {Dario}\ \bibnamefont {Poletti}},\
  }\bibfield  {title} {\enquote {\bibinfo {title} {Transport and spectral
  properties of the $xx+xxz$ diode and stability to dephasing},}\ }\href
  {\doibase 10.1103/PhysRevE.105.024120} {\bibfield  {journal} {\bibinfo
  {journal} {Phys. Rev. E}\ }\textbf {\bibinfo {volume} {105}},\ \bibinfo
  {pages} {024120} (\bibinfo {year} {2022})}\BibitemShut {NoStop}%
\bibitem [{\citenamefont {Pereira}(2019)}]{PhysRevE.99.032116}%
  \BibitemOpen
  \bibfield  {author} {\bibinfo {author} {\bibfnamefont {Emmanuel}\
  \bibnamefont {Pereira}},\ }\bibfield  {title} {\enquote {\bibinfo {title}
  {Perfect thermal rectification in a many-body quantum ising model},}\ }\href
  {\doibase 10.1103/PhysRevE.99.032116} {\bibfield  {journal} {\bibinfo
  {journal} {Phys. Rev. E}\ }\textbf {\bibinfo {volume} {99}},\ \bibinfo
  {pages} {032116} (\bibinfo {year} {2019})}\BibitemShut {NoStop}%
\bibitem [{\citenamefont {Joulain}\ \emph {et~al.}(2016)\citenamefont
  {Joulain}, \citenamefont {Drevillon}, \citenamefont {Ezzahri},\ and\
  \citenamefont {Ordonez-Miranda}}]{PhysRevLett.116.200601}%
  \BibitemOpen
  \bibfield  {author} {\bibinfo {author} {\bibfnamefont {Karl}\ \bibnamefont
  {Joulain}}, \bibinfo {author} {\bibfnamefont {J\'er\'emie}\ \bibnamefont
  {Drevillon}}, \bibinfo {author} {\bibfnamefont {Youn\`es}\ \bibnamefont
  {Ezzahri}}, \ and\ \bibinfo {author} {\bibfnamefont {Jose}\ \bibnamefont
  {Ordonez-Miranda}},\ }\bibfield  {title} {\enquote {\bibinfo {title} {Quantum
  thermal transistor},}\ }\href {\doibase 10.1103/PhysRevLett.116.200601}
  {\bibfield  {journal} {\bibinfo  {journal} {Phys. Rev. Lett.}\ }\textbf
  {\bibinfo {volume} {116}},\ \bibinfo {pages} {200601} (\bibinfo {year}
  {2016})}\BibitemShut {NoStop}%
\bibitem [{\citenamefont {Wijesekara}\ \emph {et~al.}(2020)\citenamefont
  {Wijesekara}, \citenamefont {Gunapala}, \citenamefont {Stockman},\ and\
  \citenamefont {Premaratne}}]{PhysRevB.101.245402}%
  \BibitemOpen
  \bibfield  {author} {\bibinfo {author} {\bibfnamefont {Ravi~T.}\ \bibnamefont
  {Wijesekara}}, \bibinfo {author} {\bibfnamefont {Sarath~D.}\ \bibnamefont
  {Gunapala}}, \bibinfo {author} {\bibfnamefont {Mark~I.}\ \bibnamefont
  {Stockman}}, \ and\ \bibinfo {author} {\bibfnamefont {Malin}\ \bibnamefont
  {Premaratne}},\ }\bibfield  {title} {\enquote {\bibinfo {title} {Optically
  controlled quantum thermal gate},}\ }\href {\doibase
  10.1103/PhysRevB.101.245402} {\bibfield  {journal} {\bibinfo  {journal}
  {Phys. Rev. B}\ }\textbf {\bibinfo {volume} {101}},\ \bibinfo {pages}
  {245402} (\bibinfo {year} {2020})}\BibitemShut {NoStop}%
\bibitem [{\citenamefont {Ghosh}\ \emph {et~al.}(2021)\citenamefont {Ghosh},
  \citenamefont {Ghoshal},\ and\ \citenamefont {Sen}}]{PhysRevA.103.052613}%
  \BibitemOpen
  \bibfield  {author} {\bibinfo {author} {\bibfnamefont {Riddhi}\ \bibnamefont
  {Ghosh}}, \bibinfo {author} {\bibfnamefont {Ahana}\ \bibnamefont {Ghoshal}},
  \ and\ \bibinfo {author} {\bibfnamefont {Ujjwal}\ \bibnamefont {Sen}},\
  }\bibfield  {title} {\enquote {\bibinfo {title} {Quantum thermal transistors:
  Operation characteristics in steady state versus transient regimes},}\ }\href
  {\doibase 10.1103/PhysRevA.103.052613} {\bibfield  {journal} {\bibinfo
  {journal} {Phys. Rev. A}\ }\textbf {\bibinfo {volume} {103}},\ \bibinfo
  {pages} {052613} (\bibinfo {year} {2021})}\BibitemShut {NoStop}%
\bibitem [{\citenamefont {Caldeira}\ and\ \citenamefont
  {Leggett}(1981)}]{PhysRevLett.46.211}%
  \BibitemOpen
  \bibfield  {author} {\bibinfo {author} {\bibfnamefont {A.~O.}\ \bibnamefont
  {Caldeira}}\ and\ \bibinfo {author} {\bibfnamefont {A.~J.}\ \bibnamefont
  {Leggett}},\ }\bibfield  {title} {\enquote {\bibinfo {title} {Influence of
  dissipation on quantum tunneling in macroscopic systems},}\ }\href {\doibase
  10.1103/PhysRevLett.46.211} {\bibfield  {journal} {\bibinfo  {journal} {Phys.
  Rev. Lett.}\ }\textbf {\bibinfo {volume} {46}},\ \bibinfo {pages} {211--214}
  (\bibinfo {year} {1981})}\BibitemShut {NoStop}%
\bibitem [{\citenamefont {Leggett}\ \emph {et~al.}(1987)\citenamefont
  {Leggett}, \citenamefont {Chakravarty}, \citenamefont {Dorsey}, \citenamefont
  {Fisher}, \citenamefont {Garg},\ and\ \citenamefont
  {Zwerger}}]{RevModPhys.59.1}%
  \BibitemOpen
  \bibfield  {author} {\bibinfo {author} {\bibfnamefont {A.~J.}\ \bibnamefont
  {Leggett}}, \bibinfo {author} {\bibfnamefont {S.}~\bibnamefont
  {Chakravarty}}, \bibinfo {author} {\bibfnamefont {A.~T.}\ \bibnamefont
  {Dorsey}}, \bibinfo {author} {\bibfnamefont {Matthew P.~A.}\ \bibnamefont
  {Fisher}}, \bibinfo {author} {\bibfnamefont {Anupam}\ \bibnamefont {Garg}}, \
  and\ \bibinfo {author} {\bibfnamefont {W.}~\bibnamefont {Zwerger}},\
  }\bibfield  {title} {\enquote {\bibinfo {title} {Dynamics of the dissipative
  two-state system},}\ }\href {\doibase 10.1103/RevModPhys.59.1} {\bibfield
  {journal} {\bibinfo  {journal} {Rev. Mod. Phys.}\ }\textbf {\bibinfo {volume}
  {59}},\ \bibinfo {pages} {1--85} (\bibinfo {year} {1987})}\BibitemShut
  {NoStop}%
\bibitem [{\citenamefont {Gonz\'{a}lez}\ \emph {et~al.}(2017)\citenamefont
  {Gonz\'{a}lez}, \citenamefont {Correa}, \citenamefont {Nocerino},
  \citenamefont {Palao}, \citenamefont {Alonso},\ and\ \citenamefont
  {Adesso}}]{doi:10.1142/S1230161217400108}%
  \BibitemOpen
  \bibfield  {author} {\bibinfo {author} {\bibfnamefont {J.~Onam}\ \bibnamefont
  {Gonz\'{a}lez}}, \bibinfo {author} {\bibfnamefont {Luis~A.}\ \bibnamefont
  {Correa}}, \bibinfo {author} {\bibfnamefont {Giorgio}\ \bibnamefont
  {Nocerino}}, \bibinfo {author} {\bibfnamefont {Jos\'{e}~P.}\ \bibnamefont
  {Palao}}, \bibinfo {author} {\bibfnamefont {Daniel}\ \bibnamefont {Alonso}},
  \ and\ \bibinfo {author} {\bibfnamefont {Gerardo}\ \bibnamefont {Adesso}},\
  }\bibfield  {title} {\enquote {\bibinfo {title} {Testing the validity of the
  ‘local’ and ‘global’ gkls master equations on an exactly solvable
  model},}\ }\href {\doibase 10.1142/S1230161217400108} {\bibfield  {journal}
  {\bibinfo  {journal} {Open Sys. Information Dyn.}\ }\textbf {\bibinfo
  {volume} {24}},\ \bibinfo {pages} {1740010} (\bibinfo {year}
  {2017})}\BibitemShut {NoStop}%
\bibitem [{\citenamefont {Upadhyay}\ \emph {et~al.}(2024)\citenamefont
  {Upadhyay}, \citenamefont {Tahir~Naseem}, \citenamefont {Müstecaplıoğlu},\
  and\ \citenamefont {Marathe}}]{Upadhyay_2024}%
  \BibitemOpen
  \bibfield  {author} {\bibinfo {author} {\bibfnamefont {Vipul}\ \bibnamefont
  {Upadhyay}}, \bibinfo {author} {\bibfnamefont {M}~\bibnamefont
  {Tahir~Naseem}}, \bibinfo {author} {\bibfnamefont {Özgür~E}\ \bibnamefont
  {Müstecaplıoğlu}}, \ and\ \bibinfo {author} {\bibfnamefont {Rahul}\
  \bibnamefont {Marathe}},\ }\bibfield  {title} {\enquote {\bibinfo {title}
  {Signature of topology via heat transfer analysis in the
  su–schrieffer–heeger (ssh) model},}\ }\href {\doibase
  10.1088/1367-2630/ad19aa} {\bibfield  {journal} {\bibinfo  {journal} {New J.
  Phys.}\ }\textbf {\bibinfo {volume} {26}},\ \bibinfo {pages} {013014}
  (\bibinfo {year} {2024})}\BibitemShut {NoStop}%
\bibitem [{\citenamefont {Wichterich}\ \emph {et~al.}(2007)\citenamefont
  {Wichterich}, \citenamefont {Henrich}, \citenamefont {Breuer}, \citenamefont
  {Gemmer},\ and\ \citenamefont {Michel}}]{PhysRevE.76.031115}%
  \BibitemOpen
  \bibfield  {author} {\bibinfo {author} {\bibfnamefont {Hannu}\ \bibnamefont
  {Wichterich}}, \bibinfo {author} {\bibfnamefont {Markus~J.}\ \bibnamefont
  {Henrich}}, \bibinfo {author} {\bibfnamefont {Heinz-Peter}\ \bibnamefont
  {Breuer}}, \bibinfo {author} {\bibfnamefont {Jochen}\ \bibnamefont {Gemmer}},
  \ and\ \bibinfo {author} {\bibfnamefont {Mathias}\ \bibnamefont {Michel}},\
  }\bibfield  {title} {\enquote {\bibinfo {title} {Modeling heat transport
  through completely positive maps},}\ }\href {\doibase
  10.1103/PhysRevE.76.031115} {\bibfield  {journal} {\bibinfo  {journal} {Phys.
  Rev. E}\ }\textbf {\bibinfo {volume} {76}},\ \bibinfo {pages} {031115}
  (\bibinfo {year} {2007})}\BibitemShut {NoStop}%
\bibitem [{\citenamefont {Yang}\ \emph {et~al.}(2023)\citenamefont {Yang},
  \citenamefont {Liu},\ and\ \citenamefont {Yu}}]{PhysRevE.107.064125}%
  \BibitemOpen
  \bibfield  {author} {\bibinfo {author} {\bibfnamefont {Yi-jia}\ \bibnamefont
  {Yang}}, \bibinfo {author} {\bibfnamefont {Yu-qiang}\ \bibnamefont {Liu}}, \
  and\ \bibinfo {author} {\bibfnamefont {Chang-shui}\ \bibnamefont {Yu}},\
  }\bibfield  {title} {\enquote {\bibinfo {title} {Quantum thermal diode
  dominated by pure classical correlation via three triangular-coupled
  qubits},}\ }\href {\doibase 10.1103/PhysRevE.107.064125} {\bibfield
  {journal} {\bibinfo  {journal} {Phys. Rev. E}\ }\textbf {\bibinfo {volume}
  {107}},\ \bibinfo {pages} {064125} (\bibinfo {year} {2023})}\BibitemShut
  {NoStop}%
\bibitem [{\citenamefont {Majland}\ \emph {et~al.}(2020)\citenamefont
  {Majland}, \citenamefont {Christensen},\ and\ \citenamefont
  {Zinner}}]{PhysRevB.101.184510}%
  \BibitemOpen
  \bibfield  {author} {\bibinfo {author} {\bibfnamefont {Marco}\ \bibnamefont
  {Majland}}, \bibinfo {author} {\bibfnamefont {Kasper~Sangild}\ \bibnamefont
  {Christensen}}, \ and\ \bibinfo {author} {\bibfnamefont {Nikolaj~Thomas}\
  \bibnamefont {Zinner}},\ }\bibfield  {title} {\enquote {\bibinfo {title}
  {Quantum thermal transistor in superconducting circuits},}\ }\href {\doibase
  10.1103/PhysRevB.101.184510} {\bibfield  {journal} {\bibinfo  {journal}
  {Phys. Rev. B}\ }\textbf {\bibinfo {volume} {101}},\ \bibinfo {pages}
  {184510} (\bibinfo {year} {2020})}\BibitemShut {NoStop}%
\bibitem [{\citenamefont {Yamamoto}\ and\ \citenamefont
  {Kato}(2021)}]{Yamamoto_2021}%
  \BibitemOpen
  \bibfield  {author} {\bibinfo {author} {\bibfnamefont {Tsuyoshi}\
  \bibnamefont {Yamamoto}}\ and\ \bibinfo {author} {\bibfnamefont {Takeo}\
  \bibnamefont {Kato}},\ }\bibfield  {title} {\enquote {\bibinfo {title} {Heat
  transport through a two-level system embedded between two harmonic
  resonators},}\ }\href {\doibase 10.1088/1361-648X/ac1281} {\bibfield
  {journal} {\bibinfo  {journal} {J. Phys.: Condens. Matter}\ }\textbf
  {\bibinfo {volume} {33}},\ \bibinfo {pages} {395303} (\bibinfo {year}
  {2021})}\BibitemShut {NoStop}%
\bibitem [{\citenamefont {Gubaydullin}\ \emph {et~al.}(2022)\citenamefont
  {Gubaydullin}, \citenamefont {Thomas}, \citenamefont {Golubev}, \citenamefont
  {Lvov}, \citenamefont {Peltonen},\ and\ \citenamefont
  {Pekola}}]{Gubaydullin2022}%
  \BibitemOpen
  \bibfield  {author} {\bibinfo {author} {\bibfnamefont {Azat}\ \bibnamefont
  {Gubaydullin}}, \bibinfo {author} {\bibfnamefont {George}\ \bibnamefont
  {Thomas}}, \bibinfo {author} {\bibfnamefont {Dmitry~S.}\ \bibnamefont
  {Golubev}}, \bibinfo {author} {\bibfnamefont {Dmitrii}\ \bibnamefont {Lvov}},
  \bibinfo {author} {\bibfnamefont {Joonas~T.}\ \bibnamefont {Peltonen}}, \
  and\ \bibinfo {author} {\bibfnamefont {Jukka~P.}\ \bibnamefont {Pekola}},\
  }\bibfield  {title} {\enquote {\bibinfo {title} {Photonic heat transport in
  three terminal superconducting circuit},}\ }\href {\doibase
  10.1038/s41467-022-29078-x} {\bibfield  {journal} {\bibinfo  {journal} {Nat.
  Commun.}\ }\textbf {\bibinfo {volume} {13}},\ \bibinfo {pages} {1552}
  (\bibinfo {year} {2022})}\BibitemShut {NoStop}%
\bibitem [{\citenamefont {Sánchez}\ \emph {et~al.}(2017)\citenamefont
  {Sánchez}, \citenamefont {Thierschmann},\ and\ \citenamefont
  {Molenkamp}}]{Sanchez_2017}%
  \BibitemOpen
  \bibfield  {author} {\bibinfo {author} {\bibfnamefont {Rafael}\ \bibnamefont
  {Sánchez}}, \bibinfo {author} {\bibfnamefont {Holger}\ \bibnamefont
  {Thierschmann}}, \ and\ \bibinfo {author} {\bibfnamefont {Laurens~W}\
  \bibnamefont {Molenkamp}},\ }\bibfield  {title} {\enquote {\bibinfo {title}
  {Single-electron thermal devices coupled to a mesoscopic gate},}\ }\href
  {\doibase 10.1088/1367-2630/aa8b94} {\bibfield  {journal} {\bibinfo
  {journal} {New J. Phys.}\ }\textbf {\bibinfo {volume} {19}},\ \bibinfo
  {pages} {113040} (\bibinfo {year} {2017})}\BibitemShut {NoStop}%
\bibitem [{\citenamefont {Ordonez-Miranda}\ \emph {et~al.}(2017)\citenamefont
  {Ordonez-Miranda}, \citenamefont {Joulain}, \citenamefont {De~Sousa~Meneses},
  \citenamefont {Ezzahri},\ and\ \citenamefont
  {Drevillon}}]{10.1063/1.4991516}%
  \BibitemOpen
  \bibfield  {author} {\bibinfo {author} {\bibfnamefont {Jose}\ \bibnamefont
  {Ordonez-Miranda}}, \bibinfo {author} {\bibfnamefont {Karl}\ \bibnamefont
  {Joulain}}, \bibinfo {author} {\bibfnamefont {Domingos}\ \bibnamefont
  {De~Sousa~Meneses}}, \bibinfo {author} {\bibfnamefont {Younès}\ \bibnamefont
  {Ezzahri}}, \ and\ \bibinfo {author} {\bibfnamefont {Jérémie}\ \bibnamefont
  {Drevillon}},\ }\bibfield  {title} {\enquote {\bibinfo {title} {{Photonic
  thermal diode based on superconductors}},}\ }\href {\doibase
  10.1063/1.4991516} {\bibfield  {journal} {\bibinfo  {journal} {J. Appl.
  Phys.}\ }\textbf {\bibinfo {volume} {122}},\ \bibinfo {pages} {093105}
  (\bibinfo {year} {2017})}\BibitemShut {NoStop}%
\bibitem [{\citenamefont {Senior}\ \emph {et~al.}(2020)\citenamefont {Senior},
  \citenamefont {Gubaydullin}, \citenamefont {Karimi}, \citenamefont
  {Peltonen}, \citenamefont {Ankerhold},\ and\ \citenamefont
  {Pekola}}]{senior2020heat}%
  \BibitemOpen
  \bibfield  {author} {\bibinfo {author} {\bibfnamefont {Jorden}\ \bibnamefont
  {Senior}}, \bibinfo {author} {\bibfnamefont {Azat}\ \bibnamefont
  {Gubaydullin}}, \bibinfo {author} {\bibfnamefont {Bayan}\ \bibnamefont
  {Karimi}}, \bibinfo {author} {\bibfnamefont {Joonas~T}\ \bibnamefont
  {Peltonen}}, \bibinfo {author} {\bibfnamefont {Joachim}\ \bibnamefont
  {Ankerhold}}, \ and\ \bibinfo {author} {\bibfnamefont {Jukka~P}\ \bibnamefont
  {Pekola}},\ }\bibfield  {title} {\enquote {\bibinfo {title} {Heat
  rectification via a superconducting artificial atom},}\ }\href {\doibase
  10.1038/s42005-020-0307-5} {\bibfield  {journal} {\bibinfo  {journal}
  {Commun. Phys.}\ }\textbf {\bibinfo {volume} {3}},\ \bibinfo {pages} {40}
  (\bibinfo {year} {2020})}\BibitemShut {NoStop}%
\bibitem [{\citenamefont {Díaz}\ and\ \citenamefont
  {Sánchez}(2021)}]{Diaz_2021}%
  \BibitemOpen
  \bibfield  {author} {\bibinfo {author} {\bibfnamefont {Israel}\ \bibnamefont
  {Díaz}}\ and\ \bibinfo {author} {\bibfnamefont {Rafael}\ \bibnamefont
  {Sánchez}},\ }\bibfield  {title} {\enquote {\bibinfo {title} {The qutrit as
  a heat diode and circulator},}\ }\href {\doibase 10.1088/1367-2630/ac4211}
  {\bibfield  {journal} {\bibinfo  {journal} {New J. Phys.}\ }\textbf {\bibinfo
  {volume} {23}},\ \bibinfo {pages} {125006} (\bibinfo {year}
  {2021})}\BibitemShut {NoStop}%
\bibitem [{\citenamefont {Liu}\ \emph {et~al.}(2023)\citenamefont {Liu},
  \citenamefont {Yang}, \citenamefont {Ma},\ and\ \citenamefont
  {Yu}}]{10.1063/5.0160675}%
  \BibitemOpen
  \bibfield  {author} {\bibinfo {author} {\bibfnamefont {Yu-qiang}\
  \bibnamefont {Liu}}, \bibinfo {author} {\bibfnamefont {Yi-jia}\ \bibnamefont
  {Yang}}, \bibinfo {author} {\bibfnamefont {Ting-ting}\ \bibnamefont {Ma}}, \
  and\ \bibinfo {author} {\bibfnamefont {Chang-shui}\ \bibnamefont {Yu}},\
  }\bibfield  {title} {\enquote {\bibinfo {title} {Quantum heat valve and
  entanglement in superconducting lc resonators},}\ }\href {\doibase
  10.1063/5.0160675} {\bibfield  {journal} {\bibinfo  {journal} {Appl. Phys.
  Lett.}\ }\textbf {\bibinfo {volume} {123}},\ \bibinfo {pages} {144002}
  (\bibinfo {year} {2023})}\BibitemShut {NoStop}%
\bibitem [{\citenamefont {Gu}\ \emph {et~al.}(2017)\citenamefont {Gu},
  \citenamefont {Kockum}, \citenamefont {Miranowicz}, \citenamefont {xi~Liu},\
  and\ \citenamefont {Nori}}]{GU20171}%
  \BibitemOpen
  \bibfield  {author} {\bibinfo {author} {\bibfnamefont {Xiu}\ \bibnamefont
  {Gu}}, \bibinfo {author} {\bibfnamefont {Anton~Frisk}\ \bibnamefont
  {Kockum}}, \bibinfo {author} {\bibfnamefont {Adam}\ \bibnamefont
  {Miranowicz}}, \bibinfo {author} {\bibfnamefont {Yu}~\bibnamefont {xi~Liu}},
  \ and\ \bibinfo {author} {\bibfnamefont {Franco}\ \bibnamefont {Nori}},\
  }\bibfield  {title} {\enquote {\bibinfo {title} {Microwave photonics with
  superconducting quantum circuits},}\ }\href {\doibase
  https://doi.org/10.1016/j.physrep.2017.10.002} {\bibfield  {journal}
  {\bibinfo  {journal} {Phys. Rep.}\ }\textbf {\bibinfo {volume} {718-719}},\
  \bibinfo {pages} {1--102} (\bibinfo {year} {2017})},\ \bibinfo {note}
  {microwave photonics with superconducting quantum circuits}\BibitemShut
  {NoStop}%
\bibitem [{\citenamefont {Makhlin}\ \emph {et~al.}(2001)\citenamefont
  {Makhlin}, \citenamefont {Sch\"on},\ and\ \citenamefont
  {Shnirman}}]{RevModPhys.73.357}%
  \BibitemOpen
  \bibfield  {author} {\bibinfo {author} {\bibfnamefont {Yuriy}\ \bibnamefont
  {Makhlin}}, \bibinfo {author} {\bibfnamefont {Gerd}\ \bibnamefont {Sch\"on}},
  \ and\ \bibinfo {author} {\bibfnamefont {Alexander}\ \bibnamefont
  {Shnirman}},\ }\bibfield  {title} {\enquote {\bibinfo {title} {Quantum-state
  engineering with josephson-junction devices},}\ }\href {\doibase
  10.1103/RevModPhys.73.357} {\bibfield  {journal} {\bibinfo  {journal} {Rev.
  Mod. Phys.}\ }\textbf {\bibinfo {volume} {73}},\ \bibinfo {pages} {357--400}
  (\bibinfo {year} {2001})}\BibitemShut {NoStop}%
\bibitem [{\citenamefont {Mooij}\ \emph {et~al.}(1999)\citenamefont {Mooij},
  \citenamefont {Orlando}, \citenamefont {Levitov}, \citenamefont {Tian},
  \citenamefont {van~der Wal},\ and\ \citenamefont
  {Lloyd}}]{doi:10.1126/science.285.5430.1036}%
  \BibitemOpen
  \bibfield  {author} {\bibinfo {author} {\bibfnamefont {J.~E.}\ \bibnamefont
  {Mooij}}, \bibinfo {author} {\bibfnamefont {T.~P.}\ \bibnamefont {Orlando}},
  \bibinfo {author} {\bibfnamefont {L.}~\bibnamefont {Levitov}}, \bibinfo
  {author} {\bibfnamefont {Lin}\ \bibnamefont {Tian}}, \bibinfo {author}
  {\bibfnamefont {Caspar~H.}\ \bibnamefont {van~der Wal}}, \ and\ \bibinfo
  {author} {\bibfnamefont {Seth}\ \bibnamefont {Lloyd}},\ }\bibfield  {title}
  {\enquote {\bibinfo {title} {Josephson persistent-current qubit},}\ }\href
  {\doibase 10.1126/science.285.5430.1036} {\bibfield  {journal} {\bibinfo
  {journal} {Science}\ }\textbf {\bibinfo {volume} {285}},\ \bibinfo {pages}
  {1036--1039} (\bibinfo {year} {1999})}\BibitemShut {NoStop}%
\bibitem [{\citenamefont {Koch}\ \emph {et~al.}(2007)\citenamefont {Koch},
  \citenamefont {Yu}, \citenamefont {Gambetta}, \citenamefont {Houck},
  \citenamefont {Schuster}, \citenamefont {Majer}, \citenamefont {Blais},
  \citenamefont {Devoret}, \citenamefont {Girvin},\ and\ \citenamefont
  {Schoelkopf}}]{PhysRevA.76.042319}%
  \BibitemOpen
  \bibfield  {author} {\bibinfo {author} {\bibfnamefont {Jens}\ \bibnamefont
  {Koch}}, \bibinfo {author} {\bibfnamefont {Terri~M.}\ \bibnamefont {Yu}},
  \bibinfo {author} {\bibfnamefont {Jay}\ \bibnamefont {Gambetta}}, \bibinfo
  {author} {\bibfnamefont {A.~A.}\ \bibnamefont {Houck}}, \bibinfo {author}
  {\bibfnamefont {D.~I.}\ \bibnamefont {Schuster}}, \bibinfo {author}
  {\bibfnamefont {J.}~\bibnamefont {Majer}}, \bibinfo {author} {\bibfnamefont
  {Alexandre}\ \bibnamefont {Blais}}, \bibinfo {author} {\bibfnamefont {M.~H.}\
  \bibnamefont {Devoret}}, \bibinfo {author} {\bibfnamefont {S.~M.}\
  \bibnamefont {Girvin}}, \ and\ \bibinfo {author} {\bibfnamefont {R.~J.}\
  \bibnamefont {Schoelkopf}},\ }\bibfield  {title} {\enquote {\bibinfo {title}
  {Charge-insensitive qubit design derived from the cooper pair box},}\ }\href
  {\doibase 10.1103/PhysRevA.76.042319} {\bibfield  {journal} {\bibinfo
  {journal} {Phys. Rev. A}\ }\textbf {\bibinfo {volume} {76}},\ \bibinfo
  {pages} {042319} (\bibinfo {year} {2007})}\BibitemShut {NoStop}%
\bibitem [{\citenamefont {Guthrie}\ \emph {et~al.}(2022)\citenamefont
  {Guthrie}, \citenamefont {Satrya}, \citenamefont {Chang}, \citenamefont
  {Menczel}, \citenamefont {Nori},\ and\ \citenamefont
  {Pekola}}]{PhysRevApplied.17.064022}%
  \BibitemOpen
  \bibfield  {author} {\bibinfo {author} {\bibfnamefont {Andrew}\ \bibnamefont
  {Guthrie}}, \bibinfo {author} {\bibfnamefont {Christoforus~Dimas}\
  \bibnamefont {Satrya}}, \bibinfo {author} {\bibfnamefont {Yu-Cheng}\
  \bibnamefont {Chang}}, \bibinfo {author} {\bibfnamefont {Paul}\ \bibnamefont
  {Menczel}}, \bibinfo {author} {\bibfnamefont {Franco}\ \bibnamefont {Nori}},
  \ and\ \bibinfo {author} {\bibfnamefont {Jukka~P.}\ \bibnamefont {Pekola}},\
  }\bibfield  {title} {\enquote {\bibinfo {title} {Cooper-pair box coupled to
  two resonators: An architecture for a quantum refrigerator},}\ }\href
  {\doibase 10.1103/PhysRevApplied.17.064022} {\bibfield  {journal} {\bibinfo
  {journal} {Phys. Rev. Appl.}\ }\textbf {\bibinfo {volume} {17}},\ \bibinfo
  {pages} {064022} (\bibinfo {year} {2022})}\BibitemShut {NoStop}%
\bibitem [{\citenamefont {Kerman}(2019)}]{Kerman_2019}%
  \BibitemOpen
  \bibfield  {author} {\bibinfo {author} {\bibfnamefont {Andrew~J}\
  \bibnamefont {Kerman}},\ }\bibfield  {title} {\enquote {\bibinfo {title}
  {Superconducting qubit circuit emulation of a vector spin-1/2},}\ }\href
  {\doibase 10.1088/1367-2630/ab2ee7} {\bibfield  {journal} {\bibinfo
  {journal} {New J. Phys.}\ }\textbf {\bibinfo {volume} {21}},\ \bibinfo
  {pages} {073030} (\bibinfo {year} {2019})}\BibitemShut {NoStop}%
\bibitem [{\citenamefont {Ozfidan}\ \emph {et~al.}(2020)\citenamefont
  {Ozfidan}, \citenamefont {Deng}, \citenamefont {Smirnov}, \citenamefont
  {Lanting}, \citenamefont {Harris}, \citenamefont {Swenson}, \citenamefont
  {Whittaker}, \citenamefont {Altomare}, \citenamefont {Babcock}, \citenamefont
  {Baron}, \citenamefont {Berkley}, \citenamefont {Boothby}, \citenamefont
  {Christiani}, \citenamefont {Bunyk}, \citenamefont {Enderud}, \citenamefont
  {Evert}, \citenamefont {Hager}, \citenamefont {Hajda}, \citenamefont
  {Hilton}, \citenamefont {Huang}, \citenamefont {Hoskinson}, \citenamefont
  {Johnson}, \citenamefont {Jooya}, \citenamefont {Ladizinsky}, \citenamefont
  {Ladizinsky}, \citenamefont {Li}, \citenamefont {MacDonald}, \citenamefont
  {Marsden}, \citenamefont {Marsden}, \citenamefont {Medina}, \citenamefont
  {Molavi}, \citenamefont {Neufeld}, \citenamefont {Nissen}, \citenamefont
  {Norouzpour}, \citenamefont {Oh}, \citenamefont {Pavlov}, \citenamefont
  {Perminov}, \citenamefont {Poulin-Lamarre}, \citenamefont {Reis},
  \citenamefont {Prescott}, \citenamefont {Rich}, \citenamefont {Sato},
  \citenamefont {Sterling}, \citenamefont {Tsai}, \citenamefont {Volkmann},
  \citenamefont {Wilkinson}, \citenamefont {Yao},\ and\ \citenamefont
  {Amin}}]{PhysRevApplied.13.034037}%
  \BibitemOpen
  \bibfield  {author} {\bibinfo {author} {\bibfnamefont {I.}~\bibnamefont
  {Ozfidan}}, \bibinfo {author} {\bibfnamefont {C.}~\bibnamefont {Deng}},
  \bibinfo {author} {\bibfnamefont {A.Y.}\ \bibnamefont {Smirnov}}, \bibinfo
  {author} {\bibfnamefont {T.}~\bibnamefont {Lanting}}, \bibinfo {author}
  {\bibfnamefont {R.}~\bibnamefont {Harris}}, \bibinfo {author} {\bibfnamefont
  {L.}~\bibnamefont {Swenson}}, \bibinfo {author} {\bibfnamefont
  {J.}~\bibnamefont {Whittaker}}, \bibinfo {author} {\bibfnamefont
  {F.}~\bibnamefont {Altomare}}, \bibinfo {author} {\bibfnamefont
  {M.}~\bibnamefont {Babcock}}, \bibinfo {author} {\bibfnamefont
  {C.}~\bibnamefont {Baron}}, \bibinfo {author} {\bibfnamefont {A.J.}\
  \bibnamefont {Berkley}}, \bibinfo {author} {\bibfnamefont {K.}~\bibnamefont
  {Boothby}}, \bibinfo {author} {\bibfnamefont {H.}~\bibnamefont {Christiani}},
  \bibinfo {author} {\bibfnamefont {P.}~\bibnamefont {Bunyk}}, \bibinfo
  {author} {\bibfnamefont {C.}~\bibnamefont {Enderud}}, \bibinfo {author}
  {\bibfnamefont {B.}~\bibnamefont {Evert}}, \bibinfo {author} {\bibfnamefont
  {M.}~\bibnamefont {Hager}}, \bibinfo {author} {\bibfnamefont
  {A.}~\bibnamefont {Hajda}}, \bibinfo {author} {\bibfnamefont
  {J.}~\bibnamefont {Hilton}}, \bibinfo {author} {\bibfnamefont
  {S.}~\bibnamefont {Huang}}, \bibinfo {author} {\bibfnamefont
  {E.}~\bibnamefont {Hoskinson}}, \bibinfo {author} {\bibfnamefont {M.W.}\
  \bibnamefont {Johnson}}, \bibinfo {author} {\bibfnamefont {K.}~\bibnamefont
  {Jooya}}, \bibinfo {author} {\bibfnamefont {E.}~\bibnamefont {Ladizinsky}},
  \bibinfo {author} {\bibfnamefont {N.}~\bibnamefont {Ladizinsky}}, \bibinfo
  {author} {\bibfnamefont {R.}~\bibnamefont {Li}}, \bibinfo {author}
  {\bibfnamefont {A.}~\bibnamefont {MacDonald}}, \bibinfo {author}
  {\bibfnamefont {D.}~\bibnamefont {Marsden}}, \bibinfo {author} {\bibfnamefont
  {G.}~\bibnamefont {Marsden}}, \bibinfo {author} {\bibfnamefont
  {T.}~\bibnamefont {Medina}}, \bibinfo {author} {\bibfnamefont
  {R.}~\bibnamefont {Molavi}}, \bibinfo {author} {\bibfnamefont
  {R.}~\bibnamefont {Neufeld}}, \bibinfo {author} {\bibfnamefont
  {M.}~\bibnamefont {Nissen}}, \bibinfo {author} {\bibfnamefont
  {M.}~\bibnamefont {Norouzpour}}, \bibinfo {author} {\bibfnamefont
  {T.}~\bibnamefont {Oh}}, \bibinfo {author} {\bibfnamefont {I.}~\bibnamefont
  {Pavlov}}, \bibinfo {author} {\bibfnamefont {I.}~\bibnamefont {Perminov}},
  \bibinfo {author} {\bibfnamefont {G.}~\bibnamefont {Poulin-Lamarre}},
  \bibinfo {author} {\bibfnamefont {M.}~\bibnamefont {Reis}}, \bibinfo {author}
  {\bibfnamefont {T.}~\bibnamefont {Prescott}}, \bibinfo {author}
  {\bibfnamefont {C.}~\bibnamefont {Rich}}, \bibinfo {author} {\bibfnamefont
  {Y.}~\bibnamefont {Sato}}, \bibinfo {author} {\bibfnamefont {G.}~\bibnamefont
  {Sterling}}, \bibinfo {author} {\bibfnamefont {N.}~\bibnamefont {Tsai}},
  \bibinfo {author} {\bibfnamefont {M.}~\bibnamefont {Volkmann}}, \bibinfo
  {author} {\bibfnamefont {W.}~\bibnamefont {Wilkinson}}, \bibinfo {author}
  {\bibfnamefont {J.}~\bibnamefont {Yao}}, \ and\ \bibinfo {author}
  {\bibfnamefont {M.H.}\ \bibnamefont {Amin}},\ }\bibfield  {title} {\enquote
  {\bibinfo {title} {Demonstration of a nonstoquastic hamiltonian in coupled
  superconducting flux qubits},}\ }\href {\doibase
  10.1103/PhysRevApplied.13.034037} {\bibfield  {journal} {\bibinfo  {journal}
  {Phys. Rev. Appl.}\ }\textbf {\bibinfo {volume} {13}},\ \bibinfo {pages}
  {034037} (\bibinfo {year} {2020})}\BibitemShut {NoStop}%
\bibitem [{\citenamefont {Hita-Pérez}\ \emph {et~al.}(2021)\citenamefont
  {Hita-Pérez}, \citenamefont {Jaumà}, \citenamefont {Pino},\ and\
  \citenamefont {García-Ripoll}}]{10.1063/5.0069530}%
  \BibitemOpen
  \bibfield  {author} {\bibinfo {author} {\bibfnamefont {María}\ \bibnamefont
  {Hita-Pérez}}, \bibinfo {author} {\bibfnamefont {Gabriel}\ \bibnamefont
  {Jaumà}}, \bibinfo {author} {\bibfnamefont {Manuel}\ \bibnamefont {Pino}}, \
  and\ \bibinfo {author} {\bibfnamefont {Juan~José}\ \bibnamefont
  {García-Ripoll}},\ }\bibfield  {title} {\enquote {\bibinfo {title}
  {Three-josephson junctions flux qubit couplings},}\ }\href {\doibase
  10.1063/5.0069530} {\bibfield  {journal} {\bibinfo  {journal} {Appl. Phys.
  Lett.}\ }\textbf {\bibinfo {volume} {119}},\ \bibinfo {pages} {222601}
  (\bibinfo {year} {2021})}\BibitemShut {NoStop}%
\bibitem [{\citenamefont {Majer}\ \emph {et~al.}(2005)\citenamefont {Majer},
  \citenamefont {Paauw}, \citenamefont {ter Haar}, \citenamefont {Harmans},\
  and\ \citenamefont {Mooij}}]{PhysRevLett.94.090501}%
  \BibitemOpen
  \bibfield  {author} {\bibinfo {author} {\bibfnamefont {J.~B.}\ \bibnamefont
  {Majer}}, \bibinfo {author} {\bibfnamefont {F.~G.}\ \bibnamefont {Paauw}},
  \bibinfo {author} {\bibfnamefont {A.~C.~J.}\ \bibnamefont {ter Haar}},
  \bibinfo {author} {\bibfnamefont {C.~J. P.~M.}\ \bibnamefont {Harmans}}, \
  and\ \bibinfo {author} {\bibfnamefont {J.~E.}\ \bibnamefont {Mooij}},\
  }\bibfield  {title} {\enquote {\bibinfo {title} {Spectroscopy on two coupled
  superconducting flux qubits},}\ }\href {\doibase
  10.1103/PhysRevLett.94.090501} {\bibfield  {journal} {\bibinfo  {journal}
  {Phys. Rev. Lett.}\ }\textbf {\bibinfo {volume} {94}},\ \bibinfo {pages}
  {090501} (\bibinfo {year} {2005})}\BibitemShut {NoStop}%
\bibitem [{\citenamefont {Harrington}\ \emph {et~al.}(2022)\citenamefont
  {Harrington}, \citenamefont {Mueller},\ and\ \citenamefont
  {Murch}}]{Harrington2022}%
  \BibitemOpen
  \bibfield  {author} {\bibinfo {author} {\bibfnamefont {Patrick~M.}\
  \bibnamefont {Harrington}}, \bibinfo {author} {\bibfnamefont {Erich~J.}\
  \bibnamefont {Mueller}}, \ and\ \bibinfo {author} {\bibfnamefont {Kater~W.}\
  \bibnamefont {Murch}},\ }\bibfield  {title} {\enquote {\bibinfo {title}
  {Engineered dissipation for quantum information science},}\ }\href {\doibase
  10.1038/s42254-022-00494-8} {\bibfield  {journal} {\bibinfo  {journal} {Nat.
  Rev. Phys.}\ }\textbf {\bibinfo {volume} {4}},\ \bibinfo {pages} {660--671}
  (\bibinfo {year} {2022})}\BibitemShut {NoStop}%
\bibitem [{\citenamefont {H\"anni}\ \emph {et~al.}(2021)\citenamefont
  {H\"anni}, \citenamefont {Sheptyakov}, \citenamefont {Mena}, \citenamefont
  {Hirtenlechner}, \citenamefont {Keller}, \citenamefont {Stuhr}, \citenamefont
  {Regnault}, \citenamefont {Medarde}, \citenamefont {Cervellino},
  \citenamefont {R\"uegg}, \citenamefont {Normand},\ and\ \citenamefont
  {Kr\"amer}}]{PhysRevB.103.094424}%
  \BibitemOpen
  \bibfield  {author} {\bibinfo {author} {\bibfnamefont {N.~P.}\ \bibnamefont
  {H\"anni}}, \bibinfo {author} {\bibfnamefont {D.}~\bibnamefont {Sheptyakov}},
  \bibinfo {author} {\bibfnamefont {M.}~\bibnamefont {Mena}}, \bibinfo {author}
  {\bibfnamefont {E.}~\bibnamefont {Hirtenlechner}}, \bibinfo {author}
  {\bibfnamefont {L.}~\bibnamefont {Keller}}, \bibinfo {author} {\bibfnamefont
  {U.}~\bibnamefont {Stuhr}}, \bibinfo {author} {\bibfnamefont {L.-P.}\
  \bibnamefont {Regnault}}, \bibinfo {author} {\bibfnamefont {M.}~\bibnamefont
  {Medarde}}, \bibinfo {author} {\bibfnamefont {A.}~\bibnamefont {Cervellino}},
  \bibinfo {author} {\bibfnamefont {Ch.}\ \bibnamefont {R\"uegg}}, \bibinfo
  {author} {\bibfnamefont {B.}~\bibnamefont {Normand}}, \ and\ \bibinfo
  {author} {\bibfnamefont {K.~W.}\ \bibnamefont {Kr\"amer}},\ }\bibfield
  {title} {\enquote {\bibinfo {title} {Magnetic order in the
  quasi-one-dimensional ising system $\mathrm{RbCo}{\mathrm{cl}}_{3}$},}\
  }\href {\doibase 10.1103/PhysRevB.103.094424} {\bibfield  {journal} {\bibinfo
   {journal} {Phys. Rev. B}\ }\textbf {\bibinfo {volume} {103}},\ \bibinfo
  {pages} {094424} (\bibinfo {year} {2021})}\BibitemShut {NoStop}%
\bibitem [{\citenamefont {Fang}\ \emph {et~al.}(2021)\citenamefont {Fang},
  \citenamefont {Huang},\ and\ \citenamefont {Ruan}}]{PhysRevLett.127.043902}%
  \BibitemOpen
  \bibfield  {author} {\bibinfo {author} {\bibfnamefont {Yisheng}\ \bibnamefont
  {Fang}}, \bibinfo {author} {\bibfnamefont {Junyi}\ \bibnamefont {Huang}}, \
  and\ \bibinfo {author} {\bibfnamefont {Zhichao}\ \bibnamefont {Ruan}},\
  }\bibfield  {title} {\enquote {\bibinfo {title} {Experimental observation of
  phase transitions in spatial photonic ising machine},}\ }\href {\doibase
  10.1103/PhysRevLett.127.043902} {\bibfield  {journal} {\bibinfo  {journal}
  {Phys. Rev. Lett.}\ }\textbf {\bibinfo {volume} {127}},\ \bibinfo {pages}
  {043902} (\bibinfo {year} {2021})}\BibitemShut {NoStop}%
\bibitem [{\citenamefont {Zhang}\ and\ \citenamefont
  {Song}(2021)}]{PhysRevLett.126.116401}%
  \BibitemOpen
  \bibfield  {author} {\bibinfo {author} {\bibfnamefont {K.~L.}\ \bibnamefont
  {Zhang}}\ and\ \bibinfo {author} {\bibfnamefont {Z.}~\bibnamefont {Song}},\
  }\bibfield  {title} {\enquote {\bibinfo {title} {Quantum phase transition in
  a quantum ising chain at nonzero temperatures},}\ }\href {\doibase
  10.1103/PhysRevLett.126.116401} {\bibfield  {journal} {\bibinfo  {journal}
  {Phys. Rev. Lett.}\ }\textbf {\bibinfo {volume} {126}},\ \bibinfo {pages}
  {116401} (\bibinfo {year} {2021})}\BibitemShut {NoStop}%
\bibitem [{\citenamefont {Guo}\ \emph {et~al.}(2022)\citenamefont {Guo},
  \citenamefont {Yu}, \citenamefont {Hu},\ and\ \citenamefont
  {Li}}]{PhysRevA.105.053311}%
  \BibitemOpen
  \bibfield  {author} {\bibinfo {author} {\bibfnamefont {Zheng-Xin}\
  \bibnamefont {Guo}}, \bibinfo {author} {\bibfnamefont {Xue-Jia}\ \bibnamefont
  {Yu}}, \bibinfo {author} {\bibfnamefont {Xi-Dan}\ \bibnamefont {Hu}}, \ and\
  \bibinfo {author} {\bibfnamefont {Zhi}\ \bibnamefont {Li}},\ }\bibfield
  {title} {\enquote {\bibinfo {title} {Emergent phase transitions in a cluster
  ising model with dissipation},}\ }\href {\doibase
  10.1103/PhysRevA.105.053311} {\bibfield  {journal} {\bibinfo  {journal}
  {Phys. Rev. A}\ }\textbf {\bibinfo {volume} {105}},\ \bibinfo {pages}
  {053311} (\bibinfo {year} {2022})}\BibitemShut {NoStop}%
\bibitem [{\citenamefont {Fallas~Padilla}\ \emph {et~al.}(2022)\citenamefont
  {Fallas~Padilla}, \citenamefont {Pu}, \citenamefont {Cheng},\ and\
  \citenamefont {Zhang}}]{PhysRevLett.129.183602}%
  \BibitemOpen
  \bibfield  {author} {\bibinfo {author} {\bibfnamefont {Diego}\ \bibnamefont
  {Fallas~Padilla}}, \bibinfo {author} {\bibfnamefont {Han}\ \bibnamefont
  {Pu}}, \bibinfo {author} {\bibfnamefont {Guo-Jing}\ \bibnamefont {Cheng}}, \
  and\ \bibinfo {author} {\bibfnamefont {Yu-Yu}\ \bibnamefont {Zhang}},\
  }\bibfield  {title} {\enquote {\bibinfo {title} {Understanding the quantum
  rabi ring using analogies to quantum magnetism},}\ }\href {\doibase
  10.1103/PhysRevLett.129.183602} {\bibfield  {journal} {\bibinfo  {journal}
  {Phys. Rev. Lett.}\ }\textbf {\bibinfo {volume} {129}},\ \bibinfo {pages}
  {183602} (\bibinfo {year} {2022})}\BibitemShut {NoStop}%
\bibitem [{\citenamefont {Takesue}\ \emph {et~al.}(2023)\citenamefont
  {Takesue}, \citenamefont {Yamada}, \citenamefont {Inaba}, \citenamefont
  {Ikuta}, \citenamefont {Yonezu}, \citenamefont {Inagaki}, \citenamefont
  {Honjo}, \citenamefont {Kazama}, \citenamefont {Enbutsu}, \citenamefont
  {Umeki},\ and\ \citenamefont {Kasahara}}]{PhysRevApplied.19.L031001}%
  \BibitemOpen
  \bibfield  {author} {\bibinfo {author} {\bibfnamefont {Hiroki}\ \bibnamefont
  {Takesue}}, \bibinfo {author} {\bibfnamefont {Yasuhiro}\ \bibnamefont
  {Yamada}}, \bibinfo {author} {\bibfnamefont {Kensuke}\ \bibnamefont {Inaba}},
  \bibinfo {author} {\bibfnamefont {Takuya}\ \bibnamefont {Ikuta}}, \bibinfo
  {author} {\bibfnamefont {Yuya}\ \bibnamefont {Yonezu}}, \bibinfo {author}
  {\bibfnamefont {Takahiro}\ \bibnamefont {Inagaki}}, \bibinfo {author}
  {\bibfnamefont {Toshimori}\ \bibnamefont {Honjo}}, \bibinfo {author}
  {\bibfnamefont {Takushi}\ \bibnamefont {Kazama}}, \bibinfo {author}
  {\bibfnamefont {Koji}\ \bibnamefont {Enbutsu}}, \bibinfo {author}
  {\bibfnamefont {Takeshi}\ \bibnamefont {Umeki}}, \ and\ \bibinfo {author}
  {\bibfnamefont {Ryoichi}\ \bibnamefont {Kasahara}},\ }\bibfield  {title}
  {\enquote {\bibinfo {title} {Observing a phase transition in a coherent ising
  machine},}\ }\href {\doibase 10.1103/PhysRevApplied.19.L031001} {\bibfield
  {journal} {\bibinfo  {journal} {Phys. Rev. Appl.}\ }\textbf {\bibinfo
  {volume} {19}},\ \bibinfo {pages} {L031001} (\bibinfo {year}
  {2023})}\BibitemShut {NoStop}%
\bibitem [{\citenamefont {Yang}\ \emph {et~al.}(2024)\citenamefont {Yang},
  \citenamefont {Liu}, \citenamefont {Liu},\ and\ \citenamefont
  {Yu}}]{PhysRevE.109.014142}%
  \BibitemOpen
  \bibfield  {author} {\bibinfo {author} {\bibfnamefont {Yi-jia}\ \bibnamefont
  {Yang}}, \bibinfo {author} {\bibfnamefont {Yu-qiang}\ \bibnamefont {Liu}},
  \bibinfo {author} {\bibfnamefont {Zheng}\ \bibnamefont {Liu}}, \ and\
  \bibinfo {author} {\bibfnamefont {Chang-shui}\ \bibnamefont {Yu}},\
  }\bibfield  {title} {\enquote {\bibinfo {title} {Magnetically controlled
  quantum thermal devices via three nearest-neighbor coupled spin-1/2
  systems},}\ }\href {\doibase 10.1103/PhysRevE.109.014142} {\bibfield
  {journal} {\bibinfo  {journal} {Phys. Rev. E}\ }\textbf {\bibinfo {volume}
  {109}},\ \bibinfo {pages} {014142} (\bibinfo {year} {2024})}\BibitemShut
  {NoStop}%
\end{thebibliography}%

\end{document}